\documentclass[10pt,epsfig]{article}

\usepackage{enumerate}
\usepackage{float}
\usepackage{subfig}
\usepackage{color}
\usepackage[table]{xcolor}
\usepackage{slashed}
\usepackage{multirow}
\usepackage{cite}

\let\counterwithin\relax
\usepackage{chngcntr}
\usepackage{amssymb, amsmath,mathrsfs}

\usepackage{graphics}
\usepackage{graphicx}
\usepackage{epsf}
\usepackage{epsfig}
\usepackage{float}
\usepackage{makecell}
\usepackage{multirow}
\usepackage{color}
\usepackage{xcolor}
\usepackage{simplewick}

\usepackage[utf8]{inputenc}
\usepackage{amsmath}
\usepackage{amssymb}
\usepackage{subfig}
\usepackage[normalem]{ulem}

\usepackage{mathtools}
\usepackage{amsmath}
\usepackage{diagbox}

\newcommand\undermat[2]{% http://tex.stackexchange.com/a/102468/5764
	\makebox[0.5pt][l]{$\smash{\underbrace{\phantom{%
					\begin{matrix}#2\end{matrix}}}_{ \let\scriptstyle\textstyle\text{\large $#1$}}}$}#2}
\newcommand\overmat[2]{% http://tex.stackexchange.com/a/102468/5764
	\makebox[-1pt][l]{$\smash{\overbrace{\phantom{%
					\begin{matrix}#2\end{matrix}}}^{ \let\scriptstyle\textstyle\text{\large $#1$}}}$}#2}    
\usepackage{tikz}
\usepackage[vcentermath]{youngtab}
\usepackage{slashed} 
\usepackage[font=small]{caption} 
\usepackage{times} 

\usepackage{bm}       
\usepackage{bbm} 

\usepackage{relsize}  

\usepackage{autobreak}  

\usepackage[colorlinks=true,
linkcolor=blue,
urlcolor=red,
citecolor=red]{hyperref}

\usepackage{makeidx} 
\usepackage{bbding} 
\usepackage{listings} 
\usepackage{ytableau}
\usepackage[utf8]{inputenc}
\usepackage[T1]{fontenc}
\usepackage{lmodern}

\ytableausetup
{boxsize=1.0em}

\usepackage{mmacells}

\mmaDefineMathReplacement[≤]{<=}{\leq}
\mmaDefineMathReplacement[≥]{>=}{\geq}
\mmaDefineMathReplacement[≠]{!=}{\neq}
\mmaDefineMathReplacement[→]{->}{\to}[2]
\mmaDefineMathReplacement[⧴]{:>}{:\hspace{-.2em}\to}[2]
\mmaDefineMathReplacement{∈}{\in}
\mmaDefineMathReplacement{∉}{\notin}
\mmaDefineMathReplacement{∞}{\infty}
\mmaDefineMathReplacement{†}{\dagger}
\mmaDefineMathReplacement{ψ}{\psi}
\mmaDefineMathReplacement{⊗}{\otimes}
\mmaDefineMathReplacement{τ}{\tau}
\mmaDefineMathReplacement{λ}{\lambda}
\mmaDefineMathReplacement{Γ}{\Gamma}

\mmaSet{
  morefv={gobble=2},
  linklocaluri=mma/symbol/definition:#1,
  morecellgraphics={yoffset=1.9ex}
}

\long\def\rpl#1!!#2!!{\textcolor{red}{#1} \textcolor{blue}{#2}}

\usepackage[top=1in, left=0.95in, bottom=1.1in, right=0.95in]{geometry}
\def\baselinestretch{1.27}
\usepackage[toc]{appendix}

\newcommand{\Rd}[1] { {\color{red} #1}}

\newcommand{\beq}{\begin {equation}}  
\newcommand{\eeq}{\end   {equation}} 
\newcommand{\bea}{\begin {eqnarray}} 
\newcommand{\eea}{\end   {eqnarray}}  
\newcommand{\baa}{\begin {array}   } 
\newcommand{\eaa}{\end   {array}   }     
\newcommand{\bit}{\begin {itemize} }
\newcommand{\eit}{\end   {itemize} }
\newcommand{\be }{\begin {equation}} 
\newcommand{\ee }{\end   {equation}}
\newcommand{\nn }{\nonumber        }

\newcommand{\mc}[1]{\mathcal{#1}}

\newcommand{\ket}[1]{| #1 \rangle}
\newcommand{\bra}[1]{\langle #1 |}
\newcommand{\vev}[1]{ \left\langle {#1}  \right\rangle }

\newcommand{\ie}{{\text{i.e.}}~}

\newcommand{\eq}[1]{\begin{equation}\begin{split} #1 \end{split}\end{equation}}

\newcommand{\4}{\!\!\!\!/\,}
\newcommand{\comment}[1]{}

\newcolumntype{M}[1]{>{\centering\arraybackslash}m{#1}}
\newcolumntype{N}{@{}m{0pt}@{}}

\allowdisplaybreaks

\begin{document}

\begin{center}

%{\Large \textbf  {A Package for Listing Operator Basis in Effective Field Theory}}\\[10mm]

{\Large \textbf  {Operators For Generic Effective Field Theory at any Dimension: On-shell Amplitude Basis Construction}}\\[10mm]

Hao-Lin Li$^{a, c}$\footnote{haolin.li@uclouvain.be}, Zhe Ren$^{a, b}$\footnote{renzhe@itp.ac.cn}, Ming-Lei Xiao$^{a, d, e}$\footnote{minglei.xiao@northwestern.edu}, Jiang-Hao Yu$^{a, b, f, g, h}$\footnote{jhyu@itp.ac.cn}, Yu-Hui Zheng$^{a, b}$\footnote{zhengyuhui@itp.ac.cn}\\[10mm]

\noindent 
$^a${\em \small CAS Key Laboratory of Theoretical Physics, Institute of Theoretical Physics, Chinese Academy of Sciences,    \\ Beijing 100190, P. R. China}  \\
$^b${\em \small School of Physical Sciences, University of Chinese Academy of Sciences,   Beijing 100049, P.R. China}   \\
$^c${\em \small Centre for Cosmology, Particle Physics and Phenomenology (CP3), Universite Catholique de Louvain,\\
Chem. du Cyclotron 2, 1348, Louvain-la-neuve, Belgium}\\
$^d${\em \small Department of Physics and Astronomy, Northwestern University, Evanston, Illinois 60208, USA}\\
$^e${\em \small High Energy Physics Division, Argonne National Laboratory, Lemont, Illinois 60439, USA}\\
$^f${\em \small Center for High Energy Physics, Peking University, Beijing 100871, China} \\
$^g${\em \small School of Fundamental Physics and Mathematical Sciences, Hangzhou Institute for Advanced Study, UCAS, Hangzhou 310024, China} \\
$^h${\em \small International Centre for Theoretical Physics Asia-Pacific, Beijing/Hangzhou, China}\\[10mm]

\date{\today}   
          
\end{center}

\begin{abstract}

We describe a general procedure to construct the independent and complete operator bases for generic Lorentz invariant effective field theories, given any kind of gauge symmetry and field content, up to any mass dimension. By considering the operator as contact on-shell amplitude, the so-called amplitude operator correspondence, we provide a unified construction of the Lorentz and gauge and flavor structures by Young Tableau tensor. 
Several bases are constructed to emphasize different aspects: independence (y-basis and m-basis), repeated fields with flavors (p-basis), and conserved quantum numbers (j-basis). 
We also provide new algorithms for finding the m-basis by defining inner products for group factors and the p-basis by constructing the matrix representations of the Young symmetrizers from group generators.
The on-shell amplitude basis gives us a systematic way to convert any operator into such basis, so that the conversions between any other operator bases can be easily done by linear algebra. All of these are implemented in a \texttt{Mathematica} package: \href{https://abc4eft.hepforge.org/}{ABC4EFT} ({\textbf A}mplitude {\textbf B}asis {\textbf C}onstruction for {\textbf E}ffective {\textbf F}ield {\textbf T}heories).
%
%needs to be converted Regarding to the repeated field with flavor indices, the p-basis is constructed from .  For the UV origin 
%Various operator bases are systematically constructed to emphasize different aspects: operator independence (y-basis), flavor relation (p-basis) and conserved quantum number (j-basis). 
%It is necessary to convert the on-shell amplitude y-basis to the p-basis for the repeated field with flavor indices, and the j-basis for identifying its UV origin.
%

\end{abstract}

\newpage

\setcounter{tocdepth}{4}
\setcounter{secnumdepth}{4}

\tableofcontents

\setcounter{footnote}{0}

\def\baselinestretch{1.5}
\counterwithin{equation}{section}

\newpage

\section{Introduction}

The Standard Model (SM) has been acknowledged as the most successful model in particle physics, yet several experimental facts indicated that new physics exists beyond the SM. Under the circumstances that no signal of new physics below the TeV scale is found at the Large Hadron Collider (LHC), the effective field theory (EFT) framework provides a systematical parametrization of all kinds of new physics at an energy scale below the new physics. 
The EFT framework has been applied to various models at different energy scales, such as the standard model effective field theory (SMEFT)~\cite{Weinberg:1979sa,Buchmuller:1985jz,Grzadkowski:2010es,Lehman:2014jma,Li:2020gnx,Murphy:2020rsh,Li:2020xlh,Liao:2020jmn,Liao:2016hru}, the low-energy effective field theory (LEFT)~\cite{Jenkins:2017jig,Liao:2020zyx,Li:2020tsi,Murphy:2020cly}, the standard model effective field theory with right-handed neutrinos ($\nu$SMEFT)~\cite{delAguila:2008ir,Aparici:2009fh,Bhattacharya:2015vja,Liao:2016qyd,Li:2021tsq}, the low-energy effective field theory with right-handed neutrinos ($\nu$LEFT)~\cite{Chala:2020vqp,Li:2020lba,Li:2021tsq} and so on.

As a successful paradigm to understand particle physics at different scales, theories of EFT beyond the leading order attract more and more attention. As of the SMEFT, the first SMEFT operator was written down in 1979 by Weinberg on the mass dimension 5~\cite{Weinberg:1979sa}, and then in 1986, Buchmuler and Wyler wrote the dimension six operators~\cite{Buchmuller:1985jz}. Since then, they have been writing down the SMEFT operators getting more and more attention, especially after the LHC finished its first run. According to the power counting in the SMEFT, the lower dimension, the more dominant contributions. However, there are also many cases that a higher dimension operator dominates the physical processes. Therefore, it is necessary to write down the complete set of operators at higher dimensions.

When certain EFT is applied to study a specific physical process, it is crucial to obtain a complete and independent operator basis in order to find all independent operators related to the process without miscounting and redundancies. In the SMEFT, it takes more than 20 years to obtain the complete and independent basis, the Warsaw basis~\cite{Grzadkowski:2010es}, the widely-used operator basis at dimension 6. However, when it comes to a higher dimension, although the number of independent operators at certain dimension can be fully determined once the model is known, there are still difficulties to write down explicit operators: such as the equation of motion, covariant derivative commutator, the Bianchi identities, the Fierz identities, total derivatives, and repeated fields, etc. Furthermore, choices of the complete and independent operator basis can be multifarious due to the huge number of possible operators and the freedom to define redundant operators among them. 

Recent developments of the on-shell method have greatly reduced the difficulties for the Lorentz sector of effective operators \cite{Shadmi:2018xan,Ma:2019gtx,Durieux:2019siw,AccettulliHuber:2021uoa,Balkin:2021dko,Durieux:2019eor,Durieux:2020gip,Dong:2021yak}, where independent operators are enumerated in terms of their corresponding on-shell amplitude basis.
An ultimate algorithm that systematically deal with all the redundancy relations was proposed in \cite{Li:2020gnx,Li:2020xlh,Li:2020zfq}. In particular, the Lorentz sector is represented by a Young tensor component of the group $SL(2,C) \times SU(N)$, the former being the Lorentz group, and $N$ is the number of external particles in the on-shell amplitude. 
% We consider the operator as the on-shell amplitude by rewriting the operator with only the spinor indices and treating the symmetric part of it as the Young tensor under the $SL(2,C) \times SU(N)$ product group. 
% With the semi-standard Young tableau identified, we can write the Young tensor as the on-shell contact amplitude. 
Additionally, gauge group tensors and repeated field issues are tackled carefully to guarantee the independence among the flavor-specified operators, with flavor relations automatically derived and encoded in the final expressions of the operators.
% Based on such amplitude-operator correspondence, we can write down a complete and independent on-shell basis for any type of operator. This on-shell operator basis is guaranteed to be independent and complete by the Young tableau method. 
Furthermore, it has the advantage that any operator could be expanded on this on-shell basis, rendering a coordinate as unique identifier of the operator. 
The Young tensor method has been applied to the SMEFT~\cite{Li:2020gnx,Li:2020xlh}, the LEFT~\cite{Li:2020tsi}, $\nu$SMEFT and $\nu$LEFT~\cite{Li:2021tsq} to obtain the on-shell EFT operator bases. 
Note that we are always assuming massless particles in this algorithm; massive scalars and fermions do not make a difference as shown in~\cite{Li:2021tsq}, but operators involving higher-spin massive particles/fields require more sophisticated algorithm, which are investigated in \cite{Durieux:2019eor,Durieux:2020gip,Dong:2021yak}.
% Alternatively, the on-shell EFT operator bases can also be constructed based on the kinematic structures~\cite{Shadmi:2018xan,Ma:2019gtx,Durieux:2019siw,AccettulliHuber:2021uoa,Balkin:2021dko} in the spinor-helicity notation. 

The above Lorentz part of an operator can be applied to any generic EFT with Lorentz invariance. On the other hand, the gauge structure of the operator depends on the model. For the general $SU(N)$ gauge group, we still use the Young tableau to obtain the complete set of the gauge structure. In this case, we invent a Littlewood-Richardson method at the Young tableau level, instead of the Young diagram level traditionally, using only the fundamental indices under the $SU(N)$ group. Thus our Young tensor method could apply to any Lorentz invariant EFT with any gauge symmetry, such as SU(5), left-right symmetry, and so on.

After getting the Lorentz and gauge structure, one can take the direct product of two Young Tableau, which obtains the on-shell basis for a type of operator called the y-basis and m-basis. Suppose there are repeated fields with flavor indices in the operator. In that case, the flavor tensor will obey a certain symmetry structure: the flavor tensor can be decomposed via the Sn symmetry according to the Schur-Weyl theorem. With the permutation symmetry, we could re-write the operators, ad re-organize these operators according to the flavor symmetry, which is the permutation basis (p-basis) flavor specified operator. Given the EFT operators, one question to ask is what is in the UV for such operators. We perform the partial wave expansion on the operator with the Pauli-Lubanski and Casimir action to decompose the ones to several j-basis operators, which have a certain spin and gauge quantum numbers. These constitute different bases of a type of operator: y-basis, m-basis, p-basis, and j-basis.

We not only provide a systematical way to obtain the complete and independent operator basis of EFT that can be applied to generic models, but also present a systematical method to write any operator basis in terms of our operator basis modulo equation of motion and commutator of covariant derivative. In the traditional approach, the basis conversion is quite challenging: how to convert any operator into a standard basis in a systematic way? In this work, due to the advantage of the Young tensor basis, we can systematically convert any operator into the on-shell basis using the reduced rule we provide. All of these are presented in a Mathematica package ABC4EFT, which is publicly available in the website \href{https://abc4eft.hepforge.org}{HEPForge}.

On the technical side, we summarize several improvements and new features in the package compared to our previous work~\cite{Li:2020gnx,Li:2020xlh}:
\begin{itemize}
\item We propose a new efficient algorithm to find the independent m-basis gauge factors, which simultaneously provides a metric tensor for finding the coordinate of arbitrary gauge factors.
\item In our previous work~\cite{Li:2020xlh}, the coordinate of p-basis operator with definite flavor permutation symmetry is obtained by constructing the irreducible representation of flavor permutation group with the Clebsch-Gordon coefficients, while in this work, such information is extracted from the representation of the corresponding Young symmetrizers. A new method is also more straightforward to obtain the reduced p-basis (p'-basis) than the one in the previous work, where a de-symmetrization algorithm was implemented.
\item We generalize the Poincare partial wave expansion to the operator j-basis given particular partitions, and provide a unified Casimir method to obtain both the Lorentz and gauge j-basis.
\end{itemize}

In this paper, in section~\ref{sec:2} we introduce the building blocks used to build operator basis in any EFT and the fact that these building blocks are connected with spinor variables by amplitude-operator correspondence. In section~\ref{sec:3}, we illustrate the algorithm to obtain various operator bases, including y-basis, m-basis, p-basis, and j-basis. In section~\ref{sec:4}, we show that we are able to convert operators among different bases and reduce an over-complete basis to a complete one. In section~\ref{sec:5}, we introduce how to define a model in ABC4EFT and the functions to obtain operator bases in such a model. We conclude in section~\ref{sec:6}.

\section{Building Blocks and Amplitude-operator Correspondence}\label{sec:2}

In this paper, the building blocks are the covariant derivatives acting on fields $\Psi$s which are irreducible representations $\left( j_r,j_l \right)$ of the Lorentz group and irreducible representations $\mathbf{r}$ of gauge groups, written as $D^{\omega} \Psi$ explicitly for each field $\Psi$, where $\omega \geq 0$. If not specified, the indices on the covariant derivatives can either be the Lorentz indices or the $SL(2,\mathbb{C})$ indices depending on the context. The covariant derivative under the $SL(2,\mathbb{C})$ is defined as
\eq{
	D_{\alpha\dot\alpha} = D_{\mu}\sigma^{\mu}_{\alpha\dot\alpha} \in (1/2,1/2),
}
and here we present all kinds of fields as the irreducible representations of the $SL(2,\mathbb{C})$:
\bea
&&\phi \in (0,0), \quad \psi_{\alpha }  \in (1/2,0), \quad \psi^{\dagger}_{\dot\alpha } \in (0,1/2), \\
% &&H \in \left(0,0\right), \quad H^{\dagger} \in \left(0,0\right), \\
&&F_{{\rm L}\alpha\beta} = \frac{i}{2}F_{\mu\nu}\sigma^{\mu\nu}_{\alpha\beta}\in (1,0),  \quad F_{{\rm R} \dot\alpha\dot\beta} = -\frac{i}{2}F_{\mu\nu}\bar\sigma^{\mu\nu}_{\dot\alpha\dot\beta}\in(0,1), \\
&&C_{{\rm L}\alpha\beta\gamma\delta} = C_{\mu\nu\rho\lambda} \sigma^{\mu\nu}_{\alpha\beta} \sigma^{\rho\lambda}_{\gamma\delta}\in (2,0), \quad C_{{\rm R} \dot\alpha\dot\beta\dot\gamma\dot\delta} =  C_{\mu\nu\rho\lambda} \bar\sigma^{\mu\nu}_{\dot\alpha\dot\beta} \bar\sigma^{\rho\lambda}_{\dot\gamma\dot\delta}\in (0,2).
\eea
The covariant derivatives in our notation should be understood as acting on the nearest field on the right, and the $SL(2,\mathbb{C})$ indices and gauge indices on the covariant derivatives, and that field should be understood as the indices of the whole building block. For example, consider the case that two covariant derivatives acting on the SM field $Q$, the left-handed quark,
\eq{
	D_{\alpha^{(1)}\dot\alpha^{(1)}}D_{\alpha^{(2)}\dot\alpha^{(2)}}Q_{\alpha^{(3)}ai}=(D^2Q)_{\alpha^{(1)}\alpha^{(2)}\alpha^{(3)}\dot\alpha^{(1)}\dot\alpha^{(2)}}{}_{{ai}}.
}
Where $a$ is the index of the fundamental representation of the $SU(3)$ gauge group, and $i$ is the index of the fundamental representation of the $SU(2)$ gauge group.

In general, the building blocks $D^{\omega} \Psi$ are the reducible representations of the $SL(2,\mathbb{C})$ group and can be decomposed as a direct sum of the irreducible representations,
\begin{equation}\label{eq:LorDecom}
    D^{w} \Psi \in \left(j_{l}+\frac{w}{2}, j_{r}+\frac{w}{2}\right) \oplus \text{lower weights}.
\end{equation}

Consider the case where all kind of particle state with $| \Psi \rangle$ generated by the fields $\Psi$ are massless with the helicity $h$. The spinor helicity variables are defined as $P_{\mu} \sigma^{\mu}_{\alpha\dot{\alpha}} = \lambda_{\alpha} \tilde{\lambda}_{\dot{\alpha}}$ up to the little group transformation $\lambda \to e^{-i\varphi/2}\lambda$, \ $\tilde\lambda \to e^{i\varphi/2}\tilde\lambda$. The amplitude basis of $N$ particles, denote by $\mathcal{B}\left(h_{1}, \ldots, h_{N}\right)$, should transforms as $\mc{B} \to e^{ih_i\varphi}\mc{B}$ under the little group transformation of the $i$th particle. Thus we can apply the following amplitude-operator correspondence~\cite{Li:2020gnx,Li:2020xlh},
\beq\label{eq:AmpOpeCor1}
\left\langle 0\left|D^{w} \Psi\right| \Psi\right\rangle \sim \lambda^{2 j_{l}+w} \tilde{\lambda}^{2 j_{r}+w},
\eeq
where $(j_l,j_r) = (-h,0)$ for particle states with helicity $h \leq 0$ and $(j_l,j_r) = (0,h)$ for particle states with helicity $h \geq 0$. This formula will give a correct phase $e^{ih\varphi}$ under the little group transformation of the particle $| \Psi \rangle$. The total symmetries of $\lambda$s and $\tilde{\lambda}$s indicate that only the highest weight of the irreducible representations of $SL(2,\mathbb{C})$ group in eq.~(\ref{eq:LorDecom}) should be kept. In fact, the lower weight of the irreducible representations in eq.~(\ref{eq:LorDecom}) can be converted into other fields by equation of motion (EOM) and the corvariant derivative commutator $i[D,D]=F$, and thus are understood as redundancies in EFT operators perspective as follows,
\eq{
	& D_{[\alpha\dot\alpha}D_{\beta]\dot\beta} = D_{\mu}D_{\nu}\sigma^{\mu}_{[\alpha\dot\alpha}\sigma^{\nu}_{\beta]\dot\beta} = -{\color{red} D^2 }\epsilon_{\alpha\beta}\epsilon_{\dot\alpha\dot\beta} + \frac{i}{2}[D_{\mu},D_{\nu}]\epsilon_{\alpha\beta}\bar\sigma^{\mu\nu}_{\dot\alpha\dot\beta}, \\
	& D_{[\alpha\dot\alpha}\psi_{\beta]} = D_{\mu}\sigma^{\mu}_{[\alpha\dot\alpha}\psi_{\beta]} = -\epsilon_{\alpha\beta}{\color{red} (D\4\psi)_{\dot\alpha} }, \\
	& D_{[\alpha\dot\alpha}F_{{\rm L} \beta]\gamma} = D_{\mu}F_{\nu\rho} \sigma^{\mu}_{[\alpha\dot\alpha}\sigma^{\nu\rho}_{\beta]\gamma} = 2{\color{red} D^{\mu}F_{\mu\nu} } \epsilon_{\alpha\beta}\sigma^{\nu}_{\gamma\dot\alpha}, \\
	& D_{\left[\alpha\dot{\alpha}\right.} C_{\rm L}{}_{\left.\beta\right]\gamma\delta\epsilon} = D_{\mu} C_{\nu\rho\sigma\lambda} \sigma^{\mu}_{\left[\alpha\dot{\alpha}\right.} \sigma^{\nu\rho}_{\left.\beta\right]\gamma} \sigma^{\sigma\lambda}_{\delta\epsilon} = i \epsilon_{\alpha\beta} (\sigma^{\nu})_{\gamma\dot{\alpha}} (\sigma^{\sigma\lambda})_{\delta\epsilon} {\color{red} D^{\mu} C_{\mu\nu\sigma\lambda}}.
}
The above discussion can be generalized to the massive scalars and fermions as presented in Ref.~\cite{Li:2020tsi}.

Our field building blocks are not only representations of the Lorentz group, but also representations of gauge groups as well, so we need an invariant tensor $T^{a_1,\dots,a_N}$ as a group factor to contract $N$ field building block $D^{w_i}\Psi_{i,a_i}$ in the operator to form a gauge singlet, where the $a_i$s are indices of representation $\mathbf{r}_i$ of the gauge group.
If one only has fundamental, anti-fundamental, and adjoint representation of $SU(M)$ group in the model~\footnote{Here we use $SU(M)$ for the gauge group, because the notation $SU(N)$ is occupied for the $N$-particle symmetry group. }, then the invariant tensor $T^{a_1,\dots,a_N}$ must be able to construct with the following basic invariant tensors:
\bea
&& SU(M) : \quad f^{ABC},\ d^{ABC},\ \delta^{AB},\ ({ \rm T}^A)_a^b,\ \epsilon_{i_1\dots i_M},\ \epsilon^{i_1\dots i_M},\ \delta^a_b.
\eea
where letters in lowercase denote the indices of (anti)fundamental representations and letters in uppercase denote the indices of adjoint representations of the corresponding gauge group, ${  {\rm T}}^A$'s are the generators for the fundamental representation, and
$f^{ABC}$ and $d^{ABC}$ are defined by $-2i{\rm Tr}\left([{\rm T}^A,{\rm T}^B]{\rm T}^C\right)$ and $2{\rm Tr}\left(\{{\rm T}^A,{\rm T}^B\}{\rm T}^C\right)$, where square and curly brackets represents the commutator and anti-commutator, respectively. Concretely, in the SMEFT, we have the basic tensors for the $SU(2)$ and $SU(3)$ :
\bea\label{eq:SU3tensors}
&& SU(3) : \quad f^{ABC},d^{ABC},\delta^{AB},(\lambda^A)_a^b,\epsilon_{abc},\epsilon^{abc}, \delta^a_b\\
&& SU(2) : \quad \epsilon^{IJK},\delta^{IJ},(\tau^I)_i^j,\epsilon_{ij},\epsilon^{ij},\delta^i_j,\label{eq:SU2tensors}
\eea
where $\tau$ and $\lambda$ are Pauli and Gell-Mann matrices respectively.
If one needs a field carrying other irreducible representations, then additional invariant tensors may be needed. For example, if one adds a $SU(2)$ quartet $\Delta^I$ with a single index $I$ to the model, then at least one additional invariant tensor $\Gamma^I_{abc}$ is needed to convert the fields to the one with only fundamental indices with definite permutation symmetry --- $\Delta_{abc} = \Gamma^I_{abc}\Delta^I$. The reader can refer to appendix~\ref{app:groupprofile} for how to register a new invariant tensor in the group profile file. On the other hand, if one expresses the fields directly with tensors of only fundamental indices for $SU(M)$ group, such as $\epsilon$ and $\delta$, and leaves the symmetry among the indices implicit, then no other invariant tensor is needed.

Combining the aforementioned group factors and the Lorentz structures the operators involving $N$ fields with helicity $\{h_1,h_2,\cdots,h_N\}$ at a certain dimension $d$ can be formally expressed as,
\beq\label{eq:OpeForm}
\mathcal{O}_N^{(d)}=T^{a_{1}, \ldots, a_{N}} \epsilon^{n} \tilde{\epsilon}^{\tilde{n}} \prod_{i=1}^{N} D^{w_i} \Psi_{i, a_{i}},
\eeq
where $\tilde{n}+n \equiv r =\sum_{i} \left(\omega_{i}+|h_i| \right) \equiv \sum_{i} r_i$ and $\tilde{n}-n=\sum_{i} h_{i}$ so that the $\epsilon$'s and $\tilde{\epsilon}$'s contract all spinor indices of $su(2)_l$ and $su(2)_r$ of the building blocks. An interesting observation suggests that the following relation is correct
\beq
r = d-N
\eeq
for operators only involving fields with helicity $|h_i|<2$.
The gauge group factor $T$ can be factorized to the product of invariant tensor for each gauge group $T=\prod_G T_G$. Taking the SMEFT as example where we only have the $SU(3)$ and $SU(2)$ non-Abelian gauge group, therefore $T$ can be written as $T=T_{SU(3)}T_{SU(2)}$, and $T_{SU(3)}$ and $T_{SU(2)}$ can be the combination of the elements defined in eq.~(\ref{eq:SU3tensors}) and (\ref{eq:SU2tensors}).

%where letters in lowercase denote the indices of (anti)fundamental representations and letters in uppercase denote the indices of adjoint representations of the corresponding gauge group.
%The generalization to the $SU(N)$ group is straightforward, invariant tensors are basically the same with the $SU(3)$ listed above with following replacements:  $\lambda^A$ to the corresponding $SU(N)$ generators $T^A$, the 3rd rank $\epsilon$ tensor to the $n$th rank $\epsilon$ tensor, $f^{ABC}$ and $d^{ABC}$ to $-2i{\rm Tr}\{[T^A,T^B]T^C\}$ and $2{\rm Tr}\{\{T^A,T^B\}T^C\}$ respectively. This setup is enough for the model with only fundamental, anti-fundamental and adjoint representations, if one need field of other irreducible representation, then depending on how one express the fields additional invariant tensors may be needed. For example if one add a $SU(2)$ quartet $\Delta^I$ with single index $I$ to the model, then at least one additional invariant tensor $\Gamma^I_{abc}$ is needed to convert the fields to the one with only fundamental indices with definite permutation symmetry --- $\Delta_{abc} = \Gamma^I_{abc}\Delta^I$. Reader can refer to appendix~\ref{app:groupprofile} for the method to register a new invariant tensor in the group profile file. the  On the other hand, if one expresses the fields with only fundamental indices and leaves the symmetry among the indices implicit, then no other invariant tensor is needed.

Given the operator in the form of eq.~\ref{eq:OpeForm}, the full amplitude-operator correspondence combining everything together is
\bea\label{eq:AmpOpeCor2}
\int d^{4} x\left\langle 0\left|\mathcal{O}_N^{(d)}(x)\right| \Psi_{1}^{a_{1}}, \ldots, \Psi_{N}^{a_{N}}\right\rangle
\quad &\sim& \mc{M}^{(d)}\left( \phi_1^{a_1}(p_1),\dots,\phi_N^{a_N}(p_N) \right) \delta^{(4)}\left(\sum_{i=1}^{N} \lambda_{i} \tilde{\lambda}_{i}\right), \\
\mc{M}^{(d)}\left( \phi_1^{a_1}(p_1),\dots,\phi_N^{a_N}(p_N) \right) &=& T_G^{a_1,\dots,a_N} \mc{B}^{(d)}(h_1,\dots,h_N),
\eea
where $\phi_i$s are the external particles with momenta $p_i$, and $a_i$ are collections of the group indices for them. $\mathcal{B}=\langle\cdot\rangle^{n}[\cdot]^{\tilde{n}}$ is the kinematic part of the amplitude in which the $su(2)_l$ and $su(2)_r$ indices are contracted with the $\epsilon$'s and $\tilde{\epsilon}$'s in eq.~(\ref{eq:OpeForm}) accordingly and the notations $\langle ij \rangle =  \lambda_i^{\alpha} \lambda_{j\alpha}$ and $[ ij ] =  \tilde{\lambda}_{i\dot{\alpha}} \tilde{\lambda}_{j}^{\dot{\alpha}}$ are applied. The correspondence between the EFT operators and amplitudes fully takes care of the redundancies in the EFT operators since operators which differ by the EOM or the $i[D,D]=F$ correspond to the same amplitude according to eq.~(\ref{eq:AmpOpeCor1}) and the momentum conservation in eq.~(\ref{eq:AmpOpeCor2}) further eliminates the IBP redundancy. So after finding all independent amplitudes, which can be done in a systematically way, the complete and independent operator basis can be obtained by making use of the amplitude-operator correspondence.

Before moving further, we would like to clarify some terminology that will be used in the following. 
\begin{itemize}
    \item Class: The Lorentz irreducible representations presented by a set of abstract fields with helicities $\{h_1,h_2,\cdots,h_N\}$ and their covariant derivatives that can be Lorentz-invariant form a Lorentz class.
    \item Type: For a specific model, substitution of fields of the model into the abstract fields of each Lorentz class that can form gauge invariant is a type.
    \item Term: For each type, the spin-statistics of the $m$ repeated fields in the type constrains flavor indices of operators transforming under certain representation of the symmetric group $S_m$. The decomposition of the representation of $S_m$ of all the Lorentz and gauge invariant tensors in the type into irreducible representations of $S_m$ gives the terms in that type.
    \item Operator: Due to the Schur-Weyl duality, the irreducible representations of $S_m$ of a term are also irreducible representations of $SU(n_f)$, where $n_f$ is the flavor number of the repeated fields. Specifying each flavor index with $n_f$ according to the semi-standard Young tableau (SSYT) gives the independent (flavor-specified) operators in a term.
\end{itemize}

Table.~\ref{tab:Dim6Cl}, \ref{tab:Dim7Cl} and \ref{tab:Dim8Cl} list all possible non-vanishing classes involving the massless spin 0, 1/2 and 1 fields at mass dimension 6, 7 and 8 respectively. At and beyond mass dimension 6, classes involving only two fields vanish since these must contain EOMs of the fields, and classes involving only three fields all vanish except the $F^3_{\rm L}$ and $F^3_{\rm R}$ since these involve at least one derivative and can be converted to classes involving more fields due to the EOM. From the perspective of the on-shell amplitude, the above conclusions correspond that two-point on-shell amplitudes vanish and three-point on-shell amplitudes satisfy special kinematics for massless particles.

\begin{table}[h]
	\centering
	\begin{tabular}{|c|c|c|c|c|}
		\hline
		\diagbox{$\bar{n}$}{$n$} & 0 & 1 & 2 & 3 \\
		\hline
		0 & $\phi^6$ & $\psi^2 \phi^3$ & $F_{\rm L}^2 \phi^2$, $F_{\rm L} \psi^2 \phi$, $\psi^4$ & $F_{\rm L}^3$ \\
		\hline
		1 & $\psi^{\dagger}{}^2 \phi^3$ & $\psi^{\dagger}{} \psi \phi^2 D$, $\psi^{\dagger}{}^2 \psi^2$, $\phi^4 D^2$ & &  \\
		\hline
		2 & $F_{\rm R}^2 \phi^2$, $F_{\rm R} \psi^{\dagger}{}^2 \phi$, $\psi^{\dagger}{}^4$ & &  &  \\
		\hline
		3 & $F_{\rm R}^3$ &  &  &  \\
		\hline
	\end{tabular}
	\caption{All possible non-vanishing classes involving spin 0, 1/2, 1 fields at the mass dimension 6, where $n=\sum_i\left(\frac12\omega_i-\min(0,h_i)\right)$ and $\tilde{n}=\sum_i\left(\frac12\omega_i+\max(0,h_i)\right)$ are the number of $\epsilon$'s and $\tilde{\epsilon}$'s contraction.}
	\label{tab:Dim6Cl}
\end{table}

\begin{table}[h]
	\centering
	\begin{tabular}{|c|c|c|c|c|}
		\hline
		\diagbox{$\bar{n}$}{$n$} & 0 & 1 & 2 & 3 \\
		\hline
		0 & $\phi^7$ & $\psi^2 \phi^4$ & $\psi^4 \phi$, $F_{\rm L} \psi^2 \phi^2$, $F_{\rm L}^2 \phi^3$ & $F_{\rm L}^2 \psi^2$, $F_{\rm L}^3 \phi$ \\
		\hline
		1 & $\psi^{\dagger}{}^2 \phi^4$ & $\psi^{\dagger}{} \psi \phi^3 D$, $\psi^{\dagger}{}^2 \psi^2 \phi$, $\phi^5 D^2$ & \makecell[c]{$F_{\rm L}^2 \psi^{\dagger}{}^2$, $\psi^{\dagger}{} \psi^3 D$, $F_{\rm L} \psi^{\dagger}{} \psi \phi D$, \\ $\psi^2 \phi^2 D^2$, $F_{\rm L} \phi^3 D^2$} &  \\
		\hline
		2 & $\psi^{\dagger}{}^4 \phi$, $F_{\rm R} \psi^{\dagger}{}^2 \phi^2$, $F_{\rm R}^2 \phi^3$ & \makecell[c]{$F_{\rm R}^2 \psi^2$, $\psi^{\dagger}{}^3 \psi D$, $F_{\rm R} \psi^{\dagger}{} \psi \phi D$, \\ $\psi^{\dagger}{}^2 \phi^2 D^2$, $F_{\rm R} \phi^3 D^2$} &  &  \\
		\hline
		3 & $F_{\rm R}^2 \psi^{\dagger}{}^2$, $F_{\rm R}^3 \phi$ &  &  &  \\
		\hline
	\end{tabular}
	\caption{All possible non-vanishing classes involving spin 0, 1/2, 1 fields at the mass dimension 7.}
	\label{tab:Dim7Cl}
\end{table}

\begin{table}[h]
	\centering
	\begin{tabular}{|c|c|c|c|c|c|}
		\hline
		\diagbox{$\bar{n}$}{$n$} & 0 & 1 & 2 & 3 & 4 \\
		\hline
		0 & $\phi^8$ & $\psi^2 \phi^5$ & \makecell[c]{$\psi^4 \phi^2$, $F_{\rm L} \psi^2 \phi^3$, \\ $F_{\rm L}^2 \phi^4$} & \makecell[c]{$F_{\rm L} \psi^4$, $F_{\rm L}^2 \psi^2 \phi$, \\ $F_{\rm L}^3 \phi^2$} & $F_{\rm L}^4$\\
		\hline
		1 & $\psi^{\dagger}{}^2 \phi^5$ & \makecell[c]{$\psi^{\dagger}{}^2 \psi^2 \phi^2$, $\psi^{\dagger}{} \psi \phi^4 D$, \\ $\phi^6 D^2$} & \makecell[c]{$F_{\rm L} \psi^{\dagger}{}^2 \psi^2$, $F_{\rm L}^2 \psi^{\dagger}{}^2 \phi$, \\$\psi^{\dagger}{} \psi^3 \phi D$, $F_{\rm L} \psi^{\dagger}{} \psi \phi^2 D$, \\ $\psi^2 \phi^3 D^2$, $F_{\rm L} \phi^4 D^2$ } & \makecell[c]{$F_{\rm L}^2 \psi^{\dagger}{} \psi D$, $\psi^4 D^2$, \\ $F_{\rm L} \psi^2 \phi D^2$,  $F_{\rm L}^2 \phi^2 D^2$} &  \\
		\hline
		2 & \makecell[c]{$\psi^{\dagger}{}^4 \phi^2$, $F_{\rm R} \psi^{\dagger}{}^2 \phi^3$, \\ $F_{\rm R}^2 \phi^4$} & \makecell[c]{$F_{\rm R} \psi^{\dagger}{}^2 \psi^2$, $F_{\rm R}^2 \psi^2 \phi$, \\$\psi^{\dagger}{}^3 \psi \phi D$, $F_{\rm R} \psi^{\dagger}{} \psi \phi^2 D$, \\ $\psi^{\dagger}{}^2 \phi^3 D^2$, $F_{\rm R} \phi^4 D^2$ } & \makecell[c]{$F_{\rm R}^2 F_{\rm L}^2 $, $F_{\rm R} F_{\rm L} \psi^{\dagger}{} \psi D$, \\ $\psi^{\dagger}{}^2 \psi^2 D^2$, $F_{\rm R} \psi^2 \phi D^2$, \\ $F_{\rm L} \psi^{\dagger}{}^2 \phi D^2$, $F_{\rm R} F_{\rm L} \phi^2 D^2$, \\ $\phi^4 D^4$, $\psi^{\dagger}{} \psi \phi^2 D^3$} &  &  \\
		\hline
		3 & \makecell[c]{$F_{\rm R} \psi^{\dagger}{}^4$, $F_{\rm R}^2 \psi^{\dagger}{}^2 \phi$, \\ $F_{\rm R}^3 \phi^2$} & \makecell[c]{$F_{\rm R}^2 \psi^{\dagger}{} \psi D$, $\psi^{\dagger}{}^4 D^2$, \\ $F_{\rm R} \psi^{\dagger}{}^2 \phi D^2$,  $F_{\rm R}^2 \phi^2 D^2$} &  &  &  \\
		\hline
		4 & $F_{\rm R}^4$ &  &  &  &  \\
		\hline
	\end{tabular}
	\caption{All possible non-vanishing classes involving spin 0, 1/2, 1 fields at mass dimension 8.}
	\label{tab:Dim8Cl}
\end{table}

\section{Operator Basis}\label{sec:3}

In this section, we introduce the main function of the package, namely to construct an independent operator basis. We start by assuming that all the constituting fields are distinguishable, and the operators in this sense are the so-called \textit{flavor-blind} operators,
%A simple example is $F_{\mu\nu}\partial^\mu\phi \partial^\nu\phi$, which vanishes due to not only the anti-symmetry of $F_{\mu\nu}$, but also the identity between the two $\phi$ fields. 
By \textit{flavor-blind}, we mean to treat repeated fields with different flavor indices as formally different field objects; for those that does not have flavor indices, we also add unique flavor indices to them to distinguish them as if they have additional flavor structures.
For example, without additional labeling, the operator $F_{\mu\nu}\partial^\mu\phi \partial^\nu\phi$  vanishes not only because of the anti-symmetry of $F_{\mu\nu}$, but also because of the identity between the two $\phi$ fields. 
Hence, after the labeling the operator becomes $F_{p\mu\nu}\partial^\mu\phi_r \partial^\nu\phi_s$, which does not necessarily vanish. In particular, if we have two different $\phi$ as in, for example, the Two Higgs Doublet Model, and the indices $r,s$ can be chosen from $\{1,2\}$, then it is a perfectly valid operator. %if we set the flavor number of $\phi$ to be 2 at the end, it will be a perfectly valid operator 
Working with \textit{flavor-blind} operator is easier because repeated fields do not introduce extra redundancy relations, and we can investigate its Lorentz structures and gauge factors separately. We will construct the first complete and independent basis of \textit{flavor-blind} operators, called y-basis. Then we will deal with the repeated fields, treating flavor labels as tensor indices that can take different values (if there are no flavor degrees of freedom, then the indices take identical values), so that an operator as flavor tensor have many tensor components called flavor-specified operators. We use permutation symmetries of the flavor tensors to determine the additional constraints among the flavor-specified operators. 

\subsection{Y-basis}\label{sec:Ybasis}

In this subsection, we consider the flavor-blind operators and construct the operator basis using Young tableaux; that is why the basis obtained this way is called y-basis. As mentioned in eq.~(\ref{eq:AmpOpeCor2}), a type of local amplitudes can be decomposed as the kinematic factor $\mathcal{B}$ that describes the energy dependence and the angular distribution and the gauge factor $T_G$ that describes the gauge structure of external particles. For a given type, where the helicity $h_i$ and gauge group representation $\mathbf{r}_i$ of each external particle are known, the kinematic factors $\{\mc{B}_i\}$ span a linear space of dimension $\mc{N}_{\mc{B}}$ and the gauge factors $\{T_G{}_j\}$ span a linear space of dimension $\mc{N}_G$ for each gauge group $G$, $i=1,\cdots, \mc{N}_{\mc{B}}$ and $j=1,\cdots, \mc{N}_G$. The linear space of amplitude basis thus is the direct product of $\{\mc{B}_i\}$ and all sets of $\{T_G{}_j\}$ with dimension $\mc{N}=\mc{N}_{\mc{B}} \times \prod_G \mc{N}_{G}$.

The kinematic factors can be obtained by utilizing an auxiliary $SU(N)$ transformation introduced in \cite{Henning:2019enq,Li:2020gnx,Li:2020xlh}, where $N$ is the number of particles in a given class and the indices are raised as $SU(N)$ indices. The $SU(N)$ transformation acts on the kinematic variables $(\lambda_i,\tilde{\lambda}_i)$ and keep the total momentum invariant. Now $\lambda_{i\alpha}$ and $\tilde{\lambda}_{i\dot{\alpha}}$ transform as $\mathbf{2}\times\mathbf{N}$ and $\mathbf{\bar{2}}\times\mathbf{\bar{N}}$ representation of the $SL(2,\mathbb{C})\times SU(N)$ group respectively. For a given class with the tuple $(N,n,\tilde{n})$, where $n$ and $\tilde{n}$ denote half the number of left-handed spinor indices and right-handed spinor indices in the class, representations of $\lambda$s and $\tilde{\lambda}$s under $SU(N)$ group can be presented by the Young diagrams
\eq{
	\lambda^{\otimes n} = \underbrace{\yng(1,1)\ ...\ \yng(1,1)}_{\let\scriptstyle\textstyle\text{\large $n$}} \quad , \qquad \tilde\lambda^{\otimes \tilde{n}} = \rotatebox[]{90}{\text{$N-2$}}  \left\{
    \begin{array}{lll}
        \yng(1,1) &\ldots{}& \yng(1,1) \\
        \quad\vdotswithin{}& & \quad \vdotswithin{}\\
        \undermat{\tilde{n}}{\yng(1,1) &\ldots{}& \yng(1,1)}
    \end{array}
\right.. \\ \nn
}
Then the representation of $\mc{B}$ under the $SU(N)$ is the inner product of $\lambda$'s and $\tilde{\lambda}$'s, which can be reduced to a direct sum of irreducible representations of $SU(N)$ using the Littlewood-Richardson rule. It was proved in Ref.~\cite{Henning:2019enq} that all irreducible representations denoted by Young diagrams in the reduction vanish due to the momentum conservation except the primary Young diagram
\begin{eqnarray}\label{eq:primary_YD}
\\ \nn
Y_{N,n,\tilde{n}} \quad = \quad \arraycolsep=0pt\def\arraystretch{1}
\rotatebox[]{90}{\text{$N-2$}} \left\{
\begin{array}{cccccc}
\yng(1,1) &\ \ldots{}&\ \yng(1,1)& \overmat{n}{\yng(1,1)&\ \ldots{}\  &\yng(1,1)} \\
\vdotswithin{}& & \vdotswithin{}&&&\\
\undermat{\tilde{n}}{\yng(1,1)\ &\ldots{}&\ \yng(1,1)} &&&
\end{array}
\right..
\\
\nn 
\end{eqnarray}

\begin{table}[h]
		\centering
		\begin{tabular}{|c|c|c|c|c|c|}
			\hline
			\diagbox{$\tilde{n}$}{$n$} & 0 & 1 & 2 & 3 & 4 \\
			\hline
		0   &
		    & \begin{minipage}{2.5cm}\vspace{3mm} \centering \ydiagram{1,1} \vspace{3mm} \end{minipage} 
			& \begin{minipage}{2.5cm}\vspace{3mm} \centering \ydiagram{2,2} \vspace{3mm} \end{minipage} 
			& \begin{minipage}{2.5cm}\vspace{3mm} \centering \ydiagram{3,3} \vspace{3mm} \end{minipage} 
			& \begin{minipage}{2.5cm}\vspace{3mm} \centering \ydiagram{4,4} \vspace{3mm} \end{minipage} \\
			\hline
		1   & \begin{minipage}{2.5cm}\vspace{2mm} \centering \ydiagram[*(gray)]{1,1,1,1,1} \vspace{2mm} \end{minipage} 
		    & \begin{minipage}{2.5cm}\vspace{2mm} \centering \ydiagram[*(white) ]{1+1,1+1,0,0}*[*(gray)]{2,2,1,1} \vspace{2mm} \end{minipage}
		    & \begin{minipage}{2.5cm}\vspace{3mm} \centering \ydiagram[*(white) ]{1+2,1+2,0}*[*(gray)]{3,3,1} \vspace{3mm} \end{minipage}
		    & \begin{minipage}{2.5cm}\vspace{3mm} \centering \ydiagram[*(white) ]{1+3,1+3}*[*(gray)]{4,4} \end{minipage}
		    & \\
			\hline
		2   & \begin{minipage}{2.5cm}\vspace{3mm} \centering \ydiagram[*(gray)]{2,2,2,2} \vspace{3mm} \end{minipage}
		    & \begin{minipage}{2.5cm}\vspace{3mm} \centering \ydiagram[*(white) ]{2+1,2+1,0}*[*(gray)]{3,3,2} \vspace{3mm} \end{minipage}
		    & \begin{minipage}{2.5cm}\vspace{3mm} \centering \ydiagram[*(white) ]{2+2,2+2}*[*(gray)]{4,4} \vspace{3mm} \end{minipage} 
		    & & \\
			\hline
		3   & \begin{minipage}{2.5cm}\vspace{3mm} \centering \ydiagram[*(gray)]{3,3,3} \vspace{3mm} \end{minipage}
		    & \begin{minipage}{2.5cm}\vspace{3mm} \centering \ydiagram[*(white) ]{3+1,3+1}*[*(gray)]{4,4} \vspace{3mm} \end{minipage} 
		    & & & \\
		    \hline
	    4   & \begin{minipage}{2.5cm}\vspace{3mm} \centering \ydiagram[*(gray)]{4,4} \vspace{3mm} \end{minipage}
	        & & & & \\
			\hline
		\end{tabular}
		\caption{All possible non-vanishing classes involving spin 0, 1/2, 1 fields presented by the $SU(N)$ Young diagram at the mass dimension 8, where $\omega = N-h$, $\tilde{\omega} = N+h$. $N$ is the number of fields in the class and $h$ represents the sum of helicities of all fields in the class. The gray boxes denote the $SU(N)$ representation of $\tilde{\lambda}$s and the white boxes denote the $SU(N)$ representation of $\lambda$s.}
		\label{tab:Dim8ClY}
	\end{table}
	
However, although the primary Young diagram is in one-to-one correspondence with the tuple $(N,n,\tilde{n})$, different classes may share the same primary Young diagram since the tuple $(N,n,\tilde{n})$ does not include all information of a class $\{\{h_1,h_2,\cdots,h_N\},\omega\}$, where $\omega$ is the number of covariant derivatives in the class. Given a set of labels $\{1,\cdots,N\}$ corresponding to each particle in the class and the number of labels to be filled in the primary Young diagram
\eq{\label{eq:numoflabels}
	\#i = \tilde{n} - 2h_i, \quad i=1,\cdots,N.
}
All the SSYTs obtained by filling the labels $\{\overbrace{1,\dots,1}^{\#1},\overbrace{2,\dots,2}^{\#2},\dots,\overbrace{N,\dots,N}^{\#N}  \} $ in the primary Young diagram span the space of all amplitudes in the class, and the amplitudes corresponding to the SSYTs are the independent basis vectors of the space, that is, $\{\mc{B}^{(y)}_i\}$. The SSYTs can be translated to amplitudes column by column with the following rules
\eq{\label{eq:YT_translate}
	\begin{array}{c} \young({{k_1}},{{k_2}}) \\ \vdots \\ \young({{k_{N-3}}},{{k_{N-2}}}) \end{array} \sim \mc{E}^{k_1\dots k_{N-2}ij}[ij], \qquad \young(i,j) \sim \vev{ij},
}
where the $\mc{E}$ is the Levi-Civita tensor of the $SU(N)$ group.

Here we give an example of class $F_{\rm L}^2 \phi^2 D^2$. The class is denoted by the helicity of each field and the number of derivatives in the class, which in this case is $\{\{-1,-1,0,0\},2\}$. Then we can conclude that the primary Young diagram corresponding to the class is
\begin{eqnarray}\label{eq:egPYD}
\yng(4,4)
\end{eqnarray}
since $N=4$, $n=3$ and $\tilde{n}=1$. The number of each labels $\{1,2,3,4\}$ to be filled in the primary Young diagram (\ref{eq:egPYD}) can be deduced from eq.~(\ref{eq:numoflabels}), $\#1=3$, $\#2=3$, $\#3=1$ and $\#4=1$. Thus filling in all labels will give SSYTs as follows,
\beq
\young(1113,2224) \quad \young(1112,2234) \ ,
\eeq
and the SSYTs can be translated to amplitudes using eq.~(\ref{eq:YT_translate}),
\begin{eqnarray}\label{eq:W2H2D2lory}
\vev{12}^2 \vev{34}[34] = \mc{B}^{(y)}_1, \quad
\vev{12}\vev{13}\vev{24} [34] = \mc{B}^{(y)}_2.
\end{eqnarray}
The amplitudes can be further translated to operators using the amplitude-operator correspondence eq.~(\ref{eq:AmpOpeCor1}) and eq.~(\ref{eq:AmpOpeCor2}), which are
\begin{eqnarray}\label{eq:y-basisEGLorf}
\mc{B}^{(y)}_1 &=& F_{\rm L}{}_1{}^{\alpha\beta} F_{\rm L}{}_2{}_{\alpha\beta} (D\phi_3)^{\gamma}{}_{\dot{\alpha}} (D\phi_4)_{\gamma}{}^{\dot{\alpha}}, \\
\mc{B}^{(y)}_2 &=& F_{\rm L}{}_1{}^{\alpha\beta} F_{\rm L}{}_2{}_{\alpha}{}^{\gamma} (D\phi_3)_{\beta\dot{\alpha}} (D\phi_4)_{\gamma}{}^{\dot{\alpha}}.\label{eq:y-basisEGLorl}
\end{eqnarray}
The label $(y)$ on $\mc{B}$ indicates the kinematic factor $\mc{B}^{(y)}$ is obtained as the y-basis, we keep the same notation below.

%As for the gauge factor $T$, we only consider unitary groups in this paper and only $SU(3)$, $SU(2)$ and $U(1)$ group profiles are included in our code. A guide of how to include more unitary groups in our code is given in Appendix.

As for the gauge factor $T$, we proposed in our previous work~\cite{Li:2020gnx,Li:2020xlh} an algorithm to find all the independent $T$ expressed in terms of products of $M$th-rank Levi-Civita tensors given that all the fields are expressed with fundamental indices only. The algorithm is to apply the generalized Littlewood-Richardson rules to construct the singlet Young tableaux from the set of Young tableaux corresponding to each field. Therefore to implement the algorithm, one needs to convert all the fields with non-fundamental indices to the ones with fundamental indices only such that they have a direct correspondence to the Young tableaux. For the most interesting case, a field of adjoint representation of $SU(M)$ gauge group ${\cal F}^A$, such a conversion can be made by contracting the field with the $\epsilon$ tensor and generator ${\rm T}^A$ of fundamental representation in the following form:
\begin{eqnarray}
    \epsilon_{a_1\dots a_M} ({\rm T}^A)_i^{a_M} {\cal F}^A\equiv {\cal F}_{a_1\dots a_{M-1}i}\sim 
    \rotatebox[]{90}{\text{$M-1$}}  \left\{
    \begin{array}{l}\ytableausetup{centertableaux, boxsize=2em}
        \begin{ytableau}
            {{a_1}} & {{i}}\\
            {{a_2}}
        \end{ytableau}\\
        \quad\vdotswithin{}\\
        \begin{ytableau}
            \scriptstyle a_{M-1}
        \end{ytableau}
        %\young({{a_{M-2}}},{{a_{M-1}}}) 
    \end{array}
\right.,
\end{eqnarray}
Specifically, for the gauge boson $W$ and $G$ for $SU(2)$ and $SU(3)$ gauge groups in the SMEFT, we have the following correspondence:
%We provide some examples of how to manipulate fundamental representation indices in the SM. The $W$ amd $G$ bosons in the SM are adjoint representations of $SU(2)$ and $SU(3)$ respectively, and can be presented in fundamental representation indices as
\eq{\label{eq:example_T_gluon}
	&\epsilon_{jk}\left(\tau^I\right){}_i^k W^I = W_{ij}\sim \young(ij) ,\\
	&\epsilon_{acd}\left(\lambda^A\right){}_b^d G^A = G_{abc} \sim \young(ab,c).
}
For the fields of other representations, new invariant tensors may be needed to facilitate such conversion as discussed in appendix \ref{app:groupprofile}.

In order to illustrate this algorithm, let us take dimension-8 type $W^2 H H^{\dagger} D^2$ in the SMEFT as an example. The $SU(2)$ group factor can be obtained by constructing the singlet Young diagrams of the $SU(2)$, and the $SU(2)$ structure in this type contains $W$, $H$ and $H^{\dagger}$, which transforms as the adjoint, fundamental and anti-fundamental representations of the $SU(2)$ respectively. The correspondence between each field and its Young tableau under the $SU(2)$ is
denoted as
\begin{eqnarray}
    &(\tau^{I_1}){}_{i_1j_1} W_1^{I_1} \equiv \epsilon_{j_1 m_1}(\tau^{I_1}){}_{i_1}^{m_1} W_1^{I_1} \sim \young({{i_1}}{{j_1}}) \ , \quad
(\tau^{I_2}){}_{i_2j_2} W_2^{I_2} \equiv \epsilon_{j_2 m_2}(\tau^{I_2}){}_{i_2}^{m_2} W_2^{I_2} \sim \young({{i_2}}{{j_2}}) \ , \\
&H_{3i_3} \sim \young({{i_3}}) \ , \quad
\epsilon_{i_4j_4} H_4^{\dagger j_4} \sim \young({{i_4}}) \ .
\end{eqnarray}
The singlet Young diagram of $SU(2)$ can be constructed with the Littlewood-Richardson rule in the following ways:
\begin{eqnarray}\label{eq:y-basisEGGauf}
&\Yboxdim15pt \young({{i_1}}{{j_1}})\xrightarrow{\young({{i_2}}{{j_2}})}\young({{i_1}}{{j_1}},{{i_2}}{{j_2}})\xrightarrow{\young({{i_3}})}\young({{i_1}}{{j_1}}{{i_3}},{{i_2}}{{j_2}})\xrightarrow{\young({{i_4}})}\young({{i_1}}{{j_1}}{{i_3}},{{i_2}}{{j_2}}{{i_4}}) \nn \\ &\sim \epsilon^{i_1i_2}\epsilon^{j_1j_2}\epsilon^{i_3i_4}\epsilon_{i_4j_4}(\tau^{I_1}){}_{i_1j_1} (\tau^{I_2}){}_{i_2j_2}= T^{(y)}_{SU2,1},\\
&\Yboxdim15pt \young({{i_1}}{{j_1}})\xrightarrow{\young({{i_2}}{{j_2}})}\young({{i_1}}{{j_1}}{{i_2}},{{j_2}})\xrightarrow{\young({{i_3}})}\young({{i_1}}{{j_1}}{{i_2}},{{j_2}}{{i_3}})\xrightarrow{\young({{i_4}})}\young({{i_1}}{{j_1}}{{i_2}},{{j_2}}{{i_3}}{{i_4}}) \nn \\ &\sim \epsilon^{i_1j_2}\epsilon^{j_1i_3}\epsilon^{i_2i_4}\epsilon_{i_4j_4}(\tau^{I_1}){}_{i_1j_1} (\tau^{I_2}){}_{i_2j_2} = T^{(y)}_{SU2,2}.\label{eq:y-basisEGGaul}
\end{eqnarray} 
The tensors eq.~(\ref{eq:y-basisEGGauf}) and eq.~(\ref{eq:y-basisEGGaul}) can be further simpified to
\begin{eqnarray}\label{eq:gyb}
    T^{(y)}_{SU2,1} = 2 \delta^{I_1I_2} \delta^{i_3}_{j_4}, \quad 
    T^{(y)}_{SU2,2} = \delta^{I_1I_2} \delta^{i_3}_{j_4} - i \epsilon^{I_1I_2J} (\tau^{J})^{i_3}_{j_4}.
\end{eqnarray}
The construction of the $SU(3)$ group factor $T_{SU3}$ is trivial since $W$, $H$ and $H^{\dagger}$ transform as the singlet under the $SU(3)$ group, $T^{(y)}_{SU3,1}=1$.

The complete y-basis of the type $W^2 H H^{\dagger} D^2$ is then the direct product of the Lorentz and gauge factors. Define $\mc{T}^{(y)}_{ijk} \equiv T^{(y)}_{SU3,i} T^{(y)}_{SU2,j} \mc{B}^{(y)}_k$ and the basis vectors of y-basis of type $W^4_{\rm L}$ are
\begin{eqnarray}\label{eq:y-basisEGf}
\mathcal{O}^{(y)}_{W_{\rm L}^2 H H^{\dagger} D^2,1} &= \mc{T}^{(y)}_{111} &= 2 \delta^{I_1I_2} \delta^{i_3}_{j_4} \vev{12}^2\vev{34}[34] \nn \\
&&= 2 \delta^{I_1I_2} \delta^{i_3}_{j_4} W_{\rm L}{}^{I_1}_1{}^{\alpha\beta} W_{\rm L}{}^{I_2}_2{}_{\alpha\beta} (DH_3)_{i_3}{}^{\gamma}{}_{\dot{\alpha}} (DH^{\dagger}_4)^{j_4}{}_{\gamma}{}^{\dot{\alpha}},\\
\mathcal{O}^{(y)}_{W_{\rm L}^2 H H^{\dagger} D^2,2} &= \mc{T}^{(y)}_{112}
&=2 \delta^{I_1I_2} \delta^{i_3}_{j_4} \vev{12}\vev{13}\vev{24}[34] \nn \\
&&= 2 \delta^{I_1I_2} \delta^{i_3}_{j_4} W_{\rm L}{}^{I_1}_1{}^{\alpha\beta} W_{\rm L}{}^{I_2}_2{}_{\alpha}{}^{\gamma} (DH_3)_{i_3}{}_{\beta\dot{\alpha}} (DH^{\dagger}_4)^{j_4}{}_{\gamma}{}^{\dot{\alpha}},\\
\mathcal{O}^{(y)}_{W_{\rm L}^2 H H^{\dagger} D^2,3} &= \mc{T}^{(y)}_{121}
&=\left[\delta^{I_1I_2} \delta^{i_3}_{j_4} - i \epsilon^{I_1I_2J} (\tau^{J})^{i_3}_{j_4}\right] \vev{12}^2\vev{34}[34] \nn \\
&&= \left[\delta^{I_1I_2} \delta^{i_3}_{j_4} - i \epsilon^{I_1I_2J} (\tau^{J})^{i_3}_{j_4}\right] W_{\rm L}{}^{I_1}_1{}^{\alpha\beta} W_{\rm L}{}^{I_2}_2{}_{\alpha\beta} (DH_3)_{i_3}{}^{\gamma}{}_{\dot{\alpha}} (DH^{\dagger}_4)^{j_4}{}_{\gamma}{}^{\dot{\alpha}},\\
\mathcal{O}^{(y)}_{W_{\rm L}^2 H H^{\dagger} D^2,4} &= \mc{T}^{(y)}_{122}
&=\left[\delta^{I_1I_2} \delta^{i_3}_{j_4} - i \epsilon^{I_1I_2J} (\tau^{J})^{i_3}_{j_4}\right] \vev{12}\vev{13}\vev{24}[34] \nn \\
&&= \left[\delta^{I_1I_2} \delta^{i_3}_{j_4} - i \epsilon^{I_1I_2J} (\tau^{J})^{i_3}_{j_4}\right] W_{\rm L}{}^{I_1}_1{}^{\alpha\beta} W_{\rm L}{}^{I_2}_2{}_{\alpha}{}^{\gamma} (DH_3)_{i_3}{}_{\beta\dot{\alpha}} (DH^{\dagger}_4)^{j_4}{}_{\gamma}{}^{\dot{\alpha}},\label{eq:y-basisEGl}
\end{eqnarray}
with the dimension $\mc{N}=\mc{N}_{\mc{B}} \times \prod_G \mc{N}_{G} = 2 \times 2 \times 1 = 4$.

\subsection{M-basis}

For phenomenological studies, the community using the Feynman diagram approach to compute the observables indicates a preferred monomial operator basis. 
However, as one can see from eq.~(\ref{eq:y-basisEGf}) to eq.~(\ref{eq:y-basisEGl}), the result presented in y-basis may become polynomials when using the notation of Lorentz indices and non-fundamental gauge indices. 
In this sense, the y-basis expressions provide an over-complete monomial basis (m-basis) candidates by selecting superficially different monomials from those polynomials. In principle, these monomials can be expressed as linear combinations of complete and independent basis operators, thus providing a way to discern their independence. 
This procedure can be used to reduce any over-complete operator basis proposed for different phenomenological purposes as long as a general conversion algorithm is found. 
In the following subsections, we will first illustrate that the Lorentz structure of an operator can be matched onto an on-shell amplitude which can be further reduced to the y-basis amplitude following the reduction algorithm provided in our previous work~\cite{Li:2020gnx,Li:2020xlh}, then we will provide a new efficient algorithm to obtain the independent gauge m-basis, which simultaneously provided all the information needed to find the coordinates of arbitrary gauge factors on the selected m-basis as a byproduct.

%can be sometimes tedious and can be simplified to a more concise form. On the other hand, we would like to use Lorentz indices instead of spinor indices to present the Lorentz structures of operators. Based on these two considerations, we introduced the m-basis where all independent operators are simple monomials. A set of m-basis operators is naturally constructed by complete and independent monomial bases  (also called m-basis) of Lorentz structures and gauge group factors, while the y-basis Lorentz structure when expressed in terms of Lorentz indices and gauge group factors when expressed in terms of non-fundamental indices are in general polynomials, hence the problem now is converted to how to picking out independent monomial from the y-basis results.

\subsubsection{Lorentz m-basis} \label{sec:LMB}
In our program, the monomial Lorentz m-basis candidates are selected from polynomials of the y-basis operators. Any other over-complete monomial Lorentz structures proposed by readers can be reduced by the same procedure discussed in section \ref{sec:4}.
%To select an independent monomial base In the most general case, one can first construct an over-complete base of monomial operators in any method, for instance, by considering all of the Lorentz structures. 
The first step is to translate each monomial back into on-shell amplitudes expressed in angel and square bracket notations. Secondly, these on-shell amplitudes may correspond to the non-SSYT of the primary Young diagram, which can be able to convert to the linear combination of the existing complete and independent y-basis consisting of that SSYT. We provide the algorithm for this conversion in our previous work~\cite{Li:2020gnx,Li:2020xlh}, which will be discussed in detail in sec.~\ref{sec:DeYbasis}. Finally, with the coordinates of each monomial on the amplitude y-basis, one can perform the Gaussian elimination to find the independent monomials from these candidates.

%In our package, the over-complete basis is consisted of each monomial expanded from Lorentz y-basis, 
We still take the class $F_{\rm L}^2 \phi^2 D^2$ as an example. The on-shell amplitude y-basis of the class is given by eq.~(\ref{eq:y-basisEGLorf}) and eq.~(\ref{eq:y-basisEGLorl}), where the spinor indices can be translated to Lorentz indices as
\begin{align}
    \mc{B}^{(y)}_1&= -4F_{\rm{L}1\nu\mu}F_{\rm{L}2}{}^{\mu\nu}\left(D_{\lambda}\phi_3\right)\left(D^{\lambda}\phi_4\right),\\
    \mc{B}^{(y)}_2&= -4F_{\rm{L}1\mu}{}^{\nu}F_{\rm{L}2}{}^{\mu\lambda}\left(D_{\lambda}\phi_3\right)\left(D_{\nu}\phi_4\right)+4F_{\rm{L}1\mu}{}^{\nu}F_{\rm{L}2}{}^{\mu\lambda}\left(D_{\nu}\phi_3\right)\left(D_{\lambda}\phi_4\right)-2F_{\rm{L}1\nu\mu}F_{\rm{L}2}{}^{\mu\nu}\left(D_{\lambda}\phi_3\right)\left(D^{\lambda}\phi_4\right).
\end{align}
Then we choose the over-complete basis to be:
\begin{align}
    \mathcal{B}^{(m)}_1&= F_{\rm{L}1\nu\mu}F_{\rm{L}2}{}^{\mu\nu}\left(D_{\lambda}\phi_3\right)\left(D^{\lambda}\phi_4\right)=-\frac14\mathcal{B}^{(y)}_1,\\
    \mathcal{B}^{(m)}_2&=F_{\rm{L}1\mu}{}^{\nu}F_{\rm{L}2}{}^{\mu\lambda}\left(D_{\lambda}\phi_3\right)\left(D_{\nu}\phi_4\right)= \frac18\mathcal{B}^{(y)}_1-\frac18\mathcal{B}^{(y)}_2,\\
    \mathcal{B}^{(m)}_3&=F_{\rm{L}1\mu}{}^{\nu}F_{\rm{L}2}{}^{\mu\lambda}\left(D_{\nu}\phi_3\right)\left(D_{\lambda}\phi_4\right)= \frac18\mathcal{B}^{(y)}_2.
\end{align}
Therefore, we can pick out the first two basis vectors to form the Lorentz m-basis, with the conversion matrix that transform the Lorentz y-basis to Lorentz m-basis, denoted as $\mc{K}_{\mc{B}}^{(my)}$,
\begin{eqnarray}\label{eq:kmylor}
    \mc{K}_{\mc{B}}^{(my)} = \left(
\begin{array}{cc}
 -\frac{1}{4} & 0 \\
 \frac{1}{8} & -\frac{1}{8} \\
\end{array}
\right).
\end{eqnarray}

\subsubsection{Gauge m-basis}\label{sec:GaugeM}
For the gauge factors, monomial candidates like their Lorentz counterparts should also be obtained by reducing the y-basis results' polynomials. For example, from eq.~\eqref{eq:gyb}, one can directly identify two independent and complete monomial group factors as 
\begin{eqnarray}\label{eq:W2H2D2SU2m}
    T^{m}_1=\delta^{I_1I_2}\delta^{i_3}_{j_4},\quad T^{m}_2=\epsilon^{I_1I_2J}(\tau^J)^{i_3}_{j_4},
\end{eqnarray}
and the conversion matrix between the y-basis and the m-basis is manifest:
\beq
\mathcal{K}^{(my)}_{SU2}=\left(
\begin{array}{cc}
 \frac{1}{2} & 0 \\
 -\frac{i}{2} & i \\
\end{array}
\right).\label{eq:kmygauge}
\eeq

However, it is not always so lucky to have the number of superficially distinct monomial group factors equals to that of y-basis, especially when multiple adjoint representations present. For example, in the appendix.~\ref{app:1}, we find that there are ten superficially distinct monomial candidates for four gluon operators $G^4$:
\begin{eqnarray}
&& T^m_{{\rm can,}1}=d^{ABE}d^{CDE},\quad 
 T^m_{{\rm can,}2}=d^{ABE}f^{CDE},\quad 
 T^m_{{\rm can,}3}=f^{ABE}f^{CDE},\quad 
 T^m_{{\rm can,}4}=\delta^{AB}\delta^{CD},\quad \nonumber \\
&& T^m_{{\rm can,}5}=d^{CDE}f^{ABE},\quad 
 T^m_{{\rm can,}6}=\delta^{AC}\delta^{BD},\quad
 T^m_{{\rm can,}7}=d^{ACE}d^{BDE},\quad
 T^m_{{\rm can,}8}=d^{ACE}f^{BDE},\quad \nn\\
&& T^m_{{\rm can,}9}=d^{BDE}f^{ACE},\quad  
 T^m_{{\rm can,}10}=d^{CDE}f^{ABE}\quad
\end{eqnarray}
We need to select 8 independent gauge factors from the above candidates.
Unfortunately, we do not have a symbolic reduction algorithm similar to the Lorentz amplitude. To find independent ones from candidates, we propose an efficient new method to find the m-basis iteratively and also provide the way to find coordinate of arbitrary group factors on this basis, which also benefits the conversion of the base.

We start by introducing the inner product defined between two group factors of the same type:
\begin{equation}
    (T, T')=\sum_{\{a_i\}}(T^{a_1a_2\dots})^*T'{}^{a_1a_2\dots},
\end{equation}
i.e. the contraction of all the corresponding indices of $T'$ and the complex conjugate of $T$. For example, the contraction of the first candidate itself becomes:
\begin{eqnarray}
    (T_{\rm can,1}, T_{\rm can,1})&&=\sum_{\{ABCD,EF\}}(d^{ABF}d^{CDF})^*d^{ABE}d^{CDE}\nonumber\\
    &&=\frac{200}{9}.
\end{eqnarray}
With this definition of the inner product, one can iteratively determine whether a candidate gauge factor is independent of those in a candidate pool that has already been proven to be independent with each other by checking whether the metric tensor $g_{ij}=(T_i, T_j)$  is invertible formed by this candidate gauge factors and those in this pool. 
%Though in the $W_{\rm L}^4$ example the number of candidates equals the number of y-basis, in general, the number of candidates could be larger than that of y-basis, and we provide a non-trivial example in appendix.~\ref{app:1}. 
Following the algorithm, the pool is an empty set in the first step. One can pop out the first gauge factor from the candidates $T_{\rm can,1}$ and add it into the pool, which gives a trivial one-dimensional invertible metric tensor $g_{11}$, so we can retain $T_{\rm can,1}$ in the pool. The second step is to add $T_{\rm can,2}$ into the pool, and the corresponding metric tensor $g_{ij}$ becomes:
\begin{eqnarray}
    g_{ij}=\begin{pmatrix}
    g_{11} & g_{12}\\
    g_{21} & g_{22}
    \end{pmatrix}=\begin{pmatrix}
    \frac{200}{9} & 0\\
    0 & 40
    \end{pmatrix}.
\end{eqnarray}
One can verify the invertibility of this metric tensor by checking whether the determinant of the metric tensor is zero or not; if it is zero, then it means that $T_{\rm can,2}$ and $T_{\rm can,1}$ are not independent, so we can abandon the $T_{\rm can,2}$ in the pool and move on to test $T_{\rm can,3}$. Luckily, in our example, $g_{ij}$ is invertible, so we keep $T_{\rm can,2}$ in the pool as one of our gauge m-basis. 
One can continue this procedure for one candidate at a time until a number of independent monomials reach that of the y-basis group factor, which gives a set of complete and independent monomial group factors as the m-basis $\{T^{(m)}_i\}$, with the corresponding metric tensor $g_{ij}$.  In our example, one can verify that the first 8 candidates $T_{\rm can,1-8}$ are independent with each other, so finally, our m-basis group factors become 
\begin{eqnarray}
&& T^{(m)}_{1}=d^{ABE}d^{CDE},\quad 
 T^{(m)}_{2}=d^{ABE}f^{CDE},\quad 
 T^{(m)}_{3}=f^{ABE}f^{CDE},\quad 
 T^{(m)}_{4}=\delta^{AB}\delta^{CD},\quad \nonumber \\
&& T^{(m)}_{5}=d^{CDE}f^{ABE},\quad 
 T^{(m)}_{6}=\delta^{AC}\delta^{BD},\quad
 T^{(m)}_{7}=d^{ACE}d^{BDE},\quad
 T^{(m)}_{8}=d^{ACE}f^{BDE},\quad 
\end{eqnarray}
of which the metric tensor is:
\begin{eqnarray}
    g_{ij}=\begin{pmatrix}
\frac{200}{9} & 0 & 0 & 0 & 0 & \frac{40}{3} & \frac{-20}{3} & 0\\ 
0 & 40 & 0 & 0 & 0 & 0 & 0 & -20\\ 
0 & 0 & 72 & 0 & 0 & 24 & 20 & 0\\ 
0 & 0 & 0 & 64 & 0 & 8 & \frac{40}{3} & 0\\ 
0 & 0 & 0 & 0 & 40 & 0 & 0 & 20\\ 
\frac{40}{3} & 0 & 24 & 8 & 0 & 64 & 0 & 0\\ 
\frac{-20}{3} & 0 & 20 & \frac{40}{3} & 0 & 0 & \frac{200}{9} & 0\\ 
0 & -20 & 0 & 0 & 20 & 0 & 0 & 40
    \end{pmatrix}.
\end{eqnarray}

Comments are in order for the above new algorithm. First, the inner product as a contraction of all the corresponding indices of two tensors can be viewed as a dot product of two 1-dimensional vectors if one flattened all the tensor indices. This dot operation is usually very fast in a linear algebra library. Therefore one can view the construction of the metric tensor as a way to compress the information in the gauge factors onto the lower-dimensional space before checking their independence. Otherwise one needs to check the independence among vectors of very large dimension ($d=\prod_i {\rm dim}_{r_i}$, where ${\rm dim}_{r_i}$ are the dimension of the representation $r_i$).  Second, the procedure to find the independent m-basis simultaneously constructs the metric tensor, which can be used to define the dual basis to find the coordinate of an arbitrary tensor as we discuss in section~\ref{sec:DeYbasis}, where most computational complexity resides in the inversion of the metric tensor of which dimension is ${\cal N}_{G}^2$. However, if one starts with a set of flattened tensors of much larger large dimension $d\times {\cal N}_G$, one may encounter numerical accuracy issue for either solving the coordinate $c_i$ in a linear equation system on a vector space of dimension $d$ or finding the ``QR-Decomposition'' of this set of tensors, which may lead to an error in determining the independence among the group factors.
%The m-basis vectors of $SU(2)$ gauge factor of $W^4_{\rm L}$ are 
%\begin{eqnarray}
%T^{(m)}_{SU2,1} &=& \delta^{IK} \delta^{JL}, \\
%T^{(m)}_{SU2,2} &=& \delta^{IJ} \delta^{KL}, \\
%T^{(m)}_{SU2,3} &=& \delta^{IL} \delta^{JK},
%\end{eqnarray}
%With the algorithm described above, one can definitely find the corresponding transformation matrix $\mathcal{K}^{(my)}_{SU2}$ between y-basis and m-basis of $SU(2)$ gauge factor is
%\beq
%\mathcal{K}^{(my)}_{SU2}=\left(
%\begin{array}{cc}
% \frac{1}{2} & 0 \\
% -\frac{i}{2} & i \\
%\end{array}
%\right)\label{eq:kmygauge}
%\eeq
%with $\mathcal{K}^{(my)}_{SU2} T^{(y)}_{SU2} = T^{(m)}_{SU2}$, where the $T^{(y)}_{SU2}$ is formed by the two y-basis vectors given in eq.~(\ref{eq:y-basisEGGauf}-\ref{eq:y-basisEGGaul}). Since the $SU(3)$ gauge factor of $W^4_{\rm L}$ is trivial, $T^{(m)}_{SU3,1}=1$ and $\mathcal{K}^{(my)}_{SU3}=1$. 

\vspace{1cm}

Equipped with the technique for converting arbitrary Lorentz structures and group factors to a complete and independent basis, one can, in principle, verify the independence of any two operators of the same type no matter what in form the operators are written. To complete the story we combine the ${\cal K}^{(my)}$'s  in the eq.~(\ref{eq:kmylor}) and eq.~(\ref{eq:kmygauge}) in hand, and obtain the full transformation matrix $\mathcal{K}^{(my)}$as the outer product of $\mathcal{K}^{(my)}_{SU3}$, $\mathcal{K}^{(my)}_{SU2}$ and $\mathcal{K}^{(my)}_{\mathcal{B}}$, 
\beq
\mathcal{K}^{(my)} = \mathcal{K}^{(my)}_{SU3} \otimes \mathcal{K}^{(my)}_{SU2} \otimes \mathcal{K}^{(my)}_{\mathcal{B}} = \left(
\begin{array}{cccc}
 -\frac{1}{8} & 0 & 0 & 0 \\
 \frac{1}{16} & -\frac{1}{16} & 0 & 0 \\
 \frac{i}{8} & 0 & -\frac{i}{4} & 0 \\
 -\frac{i}{16} & \frac{i}{16} & \frac{i}{8} & -\frac{i}{8} \\
\end{array}
\right),
\eeq
where we have used  $\mathcal{K}^{(my)}_{SU3}=(1)$ for the ${W^2 H H^{\dagger} D^2}$.
We list the m-basis in the following:
\begin{eqnarray}\label{eq:m-basisEGf}
\mathcal{O}^{(m)}_{W_{\rm L}^2 H H^{\dagger} D^2,1} &= \mc{T}^{(m)}_{111} &= -\frac{1}{4} \delta^{I_1 I_2} \delta^{i_3}_{j_4} \vev{12}^2\vev{34}[34] \nn \\
&&= W_{\rm{L}}{}^I_{1\nu\mu} W_{\rm{L}}{}^{I\mu\nu}_2  (D_{\lambda}H_{3i})(D^{\lambda}H^{\dagger i}_4),\\
\mathcal{O}^{(m)}_{W_{\rm L}^2 H H^{\dagger} D^2,2} &= \mc{T}^{(m)}_{112} &= \frac{1}{8} \delta^{I_1 I_2} \delta^{i_3}_{j_4} (\vev{12}^2\vev{34}[34] - \vev{12}\vev{13}\vev{24} [34]) \nn \\
&&= W_{\rm{L}}{}^I_{1\mu}{}^{\nu} W_{\rm{L}}{}^I_2{}^{\mu\lambda}(D_{\lambda}H_{3i})(D_{\nu}H^{\dagger i}_4), \\
\mathcal{O}^{(m)}_{W_{\rm L}^2 H H^{\dagger} D^2,3} &= \mc{T}^{(m)}_{121} &= -\frac{1}{4} \epsilon^{I_1I_2J} (\tau^J)^{i_3}_{j_4} \vev{12}^2\vev{34}[34] \nn \\
&&= \epsilon^{IJK} (\tau^K)^{i}_{j} W_{\rm{L}}{}^I_{1\nu\mu} W_{\rm{L}}{}^{J\mu\nu}_2  (D_{\lambda}H_{3i})(D^{\lambda}H^{\dagger j}_4), \label{eq:m-basisEGm} \\
\mathcal{O}^{(m)}_{W_{\rm L}^2 H H^{\dagger} D^2,4} &= \mc{T}^{(m)}_{122} &= \frac{1}{8} \epsilon^{I_1I_2J} (\tau^J)^{i_3}_{j_4} (\vev{12}^2\vev{34}[34] - \vev{12}\vev{13}\vev{24} [34]) \nn \\
&&= \epsilon^{IJK} (\tau^K)^{i}_{j} W_{\rm{L}}{}^I_{1\mu}{}^{\nu} W_{\rm{L}}{}^J_2{}^{\mu\lambda}(D_{\lambda}H_{3i})(D_{\nu}H^{\dagger j}_4),
\label{eq:m-basisEGl}
\end{eqnarray}
where $\mc{T}^{(m)}_{ijk} \equiv T^{(m)}_{SU3,i} T^{(m)}_{SU2,j} \mc{B}^{(m)}_k$. Comparing the above m-basis with the y-basis of $W_{\rm L}^2 H H^{\dagger} D^2$ eq.~(\ref{eq:y-basisEGf}-\ref{eq:y-basisEGl}), it can be seen that the forms of the m-basis vectors are all monomials where each field is labeled with the Lorentz indices while the completeness and independence of the y-basis are kept in the transformation.

Here we want to remind the readers that the above y-basis and m-basis are obtained in the condition where all fields are distinguishable. For example, in eq.~(\ref{eq:m-basisEGm}), the two $W_{\rm L}$s are antisymmetric due to their symmetric Lorentz structure and antisymmetric gauge structure, but the W bosons in the SM must be totally symmetric. That is why we need to introduce the p-basis for repeated fields.

\subsection{P-basis}
The y-basis and m-basis are bases of flavor-blind operators where all fields are distinguishable. Suppose there are repeated fields in an operator. In that case, extra constraints require the permutation symmetry of flavor indices of the repeated fields must come from these of Lorentz and gauge factors since the operator stays the same if we exchange two repeated bosonic fields or obtains a minus sign if we exchange two repeated fermionic fields.
\begin{eqnarray}
\underbrace{\pi\circ {\cal O}^{\{f_{k},...\}}}_{\rm permute\ flavor} &=& \underbrace{\left(\pi\circ T_{{\rm G_1}}^{\{g_k,...\}}\right)\left(\pi\circ T_{{\rm G2}}^{\{h_k,...\}}\right)\cdots}_{\rm permute\ gauge}\underbrace{\left(\pi\circ{\cal B}^{\{f_k,...\}}_{\{g_{k},...\},\{h_{k},...\}}\right)}_{\rm permute\ Lorentz},
\label{eq:tperm}
\end{eqnarray}
The constraint can also be understood as the requirement of the spin statistic for amplitudes involving identical particles in an amplitude perspective,
\eq{\label{eq:exchpar}
	\mc{M}^{\rm (p)}\big( \underbrace{\phi^{a_1}(p_1),\dots,\phi^{a_m}(p_m)}_{m},\dots \big) 
	= \mc{D}_{\phi}(\pi)\mc{M}^{\rm (p)}
	\big( \underbrace{\phi^{a_{\pi(1)}}(p_1),\dots,\phi^{a_{\pi(m)}}(p_m)}_{m},\dots \big), \\
	\mc{D}_{\phi}(\pi) = \left\{\begin{array}{ll} 1 & \text{boson } \phi \\ (-1)^{{\rm sgn}(\pi)} & \text{fermion } \phi \end{array} \right. ,\qquad \pi\in S_m ,
}
where $\mc{D}_{\phi}(\pi)$ denote the representation of permutation $\pi$ for the identical particles $\phi$, and $(-1)^{\rm sgn}(\pi)$ is the signature of the permutation $\pi$.
We treat each field in operators as flavor multiplets and introduce a basis for operators with definite permutation symmetries of flavor indices, called p-basis. Furthermore, after specifying each flavor indices in a p-basis with three or more repeated fermionic fields, more redundancies appear due to flavor relations of components of operators of the p-basis. After removing these redundancies, the basis is called p'-basis. 

We will still use the type $W_{\rm L}^2 H H^{\dagger} D^2$ as an example to illustrate the method of obtaining the p-basis. Exchange of the two repeated fields $\{W_{\rm L}{}_1, W_{\rm L}{}_2\}$ form a 1-dimensional representation of the symmetric group $S_2$, as shown in eq.~(\ref{eq:exchpar}). Exchange of the Lorentz indices of the repeated fields in the type form a 2-dimensional representation of the $S_2$ because the linear space spanned by the y-basis of the Lorentz factor of the type is 2-dimensional, and any permutation of the $S_2$ acting on the basis should give a linear combination of basis vectors since the basis is complete. 

The generator of the $S_2$ group is $(1 \ 2)$, then it acting on the y-basis of the Lorentz factor of the $W_{\rm L}^2 H H^{\dagger} D^2$, eq.~(\ref{eq:W2H2D2lory}) gives the following relations,
\begin{eqnarray}
(1 \ 2) \ \vev{12}^2\vev{34}[34] &=& \vev{21}^2\vev{34}[34] = \vev{12}^2\vev{34}[34], \\
(1 \ 2) \ \vev{12}\vev{13}\vev{24}[34] &=& \vev{21}\vev{23}\vev{14}[34] = \vev{12}^2\vev{34}[34] - \vev{12}\vev{13}\vev{24}[34].
\end{eqnarray}
Using this method, we can deduce the matrix representations of $(1 \ 2)$ acting on the y-basis of the Lorentz factor of $W_{\rm L}^2 H H^{\dagger} D^2$ as
\beq
\mc{D}^{(y)}_{\mc{B}}[(1 \ 2)]=\left(
\begin{array}{cc}
 1 & 0 \\
 1 & -1
\end{array}
\right).
\eeq
The matrix representations of the $S_2$ generator acting on the m-basis are connected to that acting on the y-basis by $\mc{D}^{(m)} = \mc{K}^{(my)} \mc{D}^{(y)} (\mc{K}^{(my)})^{-1}$, so we obtain
\beq
\mc{D}^{(m)}_{\mc{B}}[(1 \ 2)]=\left(
\begin{array}{cc}
 1 & 0 \\
 -\frac{1}{2} & -1 \\
\end{array}
\right).
\eeq
Similarly, we can obtain the matrix representations of the $S_2$ generator acting on the m-basis of the $SU(2)$ gauge factor of the $W_{\rm L}^2 H H^{\dagger} D^2$, eq.~(\ref{eq:W2H2D2SU2m}), as $2 \times 2$ matrix
\beq
\mc{D}^{(m)}_{SU2}[(1 \ 2)]=\left(
\begin{array}{ccc}
 1 & 0 \\
 0 & -1
\end{array}
\right).
\eeq
The matrix representations of the generator for $SU(3)$ gauge factor is still trivial, $\mc{D}^{(m)}_{SU3}[(1 \ 2)]=\left(1\right)$, and finally the matrix representations of the $S_2$ generator acting on the full m-basis of the type should be the direct product of that of the gauge factor and that of the Lorentz factor, $\mc{D}^{(m)}(\pi) = \mc{D}^{(m)}_{SU3}(\pi) \otimes \mc{D}^{(m)}_{SU2}(\pi) \otimes \mc{D}^{(m)}_{\mc{B}}(\pi)$, so
\begin{eqnarray}
\mc{D}^{(m)}[(1 \ 2)] &=& \left(
\begin{array}{cccc}
 1 & 0 & 0 & 0 \\
 -\frac{1}{2} & -1 & 0 & 0 \\
 0 & 0 & -1 & 0 \\
 0 & 0 & \frac{1}{2} & 1 \\
\end{array}
\right).
\end{eqnarray}
In this example, the $W$ bosons in the SM have only one flavor number, which requires that the permutation symmetry of flavor indices must be totally symmetric. The generator $(1 \ 2)$ $S_2$ group can be used to generate all Young symmetrizers of $S_2$, among which the Young symmetrizer of totally symmetric representation of $S_2$ can be presented as
\beq\label{eq:p-basisEGsym}
\mc{Y}^{[2]}_1 = \frac{1}{2}\mc{Y}\left[\tiny{\young(12)}\right] = \left(
\begin{array}{cccc}
 1 & 0 & 0 & 0 \\
 -\frac{1}{4} & 0 & 0 & 0 \\
 0 & 0 & 0 & 0 \\
 0 & 0 & \frac{1}{4} & 1 \\
\end{array}
\right),
\eeq
where the superscript $\left[2\right]$ denotes the totally symmetric representation of $S_2$, and subscript labels each Young symmetrizer of the representation, which can only be $1$ in this case since $\left[2\right]$ is a one-dimensional representation. Each row of the matrix in eq.~(\ref{eq:p-basisEGsym}) gives the flavor-symmetrized m-basis in terms of the linear combinations of the original m-basis eq.~(\ref{eq:m-basisEGf}-\ref{eq:m-basisEGl}). We find that the matrix in eq.~(\ref{eq:p-basisEGsym}) is rank-2 and choose the first row and the last row to be the basis vectors of p-basis,
\beq
\mc{K}^{(pm)}=\left(
\begin{array}{cccc}
 1 & 0 & 0 & 0 \\
 0 & 0 & \frac{1}{4} & 1 \\
\end{array}
\right).
\eeq
So the two operators of the p-basis of the type $W_{\rm L}^2 H H^{\dagger} D^2$ are written explicitly as
\begin{eqnarray}\label{eq:p-basisEGf}
\mathcal{O}^{(p)}_{W_{\rm L}^2 H H^{\dagger} D^2,1} &=& -\frac{1}{8}\mc{Y}\left[\tiny{\young(12)}\right] \delta^{I_1 I_2} \delta^{i_3}_{j_4} \vev{12}^2\vev{34}[34] \nn \\
&=&  W_{\rm{L}}{}^I_{\nu\mu} W_{\rm{L}}{}^{I\mu\nu}  (D_{\lambda}H_{i})(D^{\lambda}H^{\dagger i}),\\
\mathcal{O}^{(p)}_{W_{\rm L}^2 H H^{\dagger} D^2,2} &=& \frac{1}{16}\mc{Y}\left[\tiny{\young(12)}\right] \epsilon^{I_1I_2J} (\tau^J)^{i_3}_{j_4} (\vev{12}^2\vev{34}[34] - \vev{12}\vev{13}\vev{24} [34]) \nn \\
&=& \epsilon^{IJK} (\tau^K)^{i}_{j} W_{\rm{L}}{}^I_{\mu}{}^{\nu} W_{\rm{L}}{}^J{}^{\mu\lambda}(D_{\lambda}H_{i})(D_{\nu}H^{\dagger j}),
\label{eq:p-basisEGl}
\end{eqnarray}
where the $\mc{Y}\left[\tiny{\young(12)}\right]$s in the amplitudes above act on all corresponding indices, including the $SU(2)$ indices ${I_1,I_2}$ and the spinor indices ${1,2}$ in this case. It should be noted that the amplitudes above are not exactly the amplitudes generated by the above operators, but the amplitudes corresponding to the operators by amplitude-operator correspondence.

The situation would be more complicated when an operator involves three or more repeated fields. Generally, for $m$ repeated fields, we choose the $S_m$ group generators as $(1 \ 2)$ and $(1 \ 2 \ \cdots \ m)$ and use the matrix representations of the two generators to generate matrix representations of all Young symmetrizers of the $S_m$. When $m \geq 3$ and the flavor number of the repeated fields is more than 1, we will need to deal with representations of the $S_m$ with dimension 2 or more. For example, consider type $Q^3 H e_{_\mathbb{C}}^{\dagger} D$. Labeling the fields as $Q_1 Q_2 Q_3 H_4 e_{_\mathbb{C}5}^{\dagger}$, the y-basis and m-basis of $Q^3 H e_{_\mathbb{C}}^{\dagger} D$ are as follows:
\begin{align}
    \begin{array}{c|c}
        \text{Y-basis} &  \\
        \hline
        \mc{O}^{(y)}_{Q^3 H e_{_\mathbb{C}}^{\dagger} D,1} & i\epsilon^{abc} \epsilon^{ik} \epsilon^{jl} \left(Q_{1ai} Q_{2bj}\right) \left(Q_{3ck} \sigma^{\mu} e_{_\mathbb{C}}^{\dagger}{}_5\right) D_{\mu} H_{4l} \\
        \mc{O}^{(y)}_{Q^3 H e_{_\mathbb{C}}^{\dagger} D,2} & -i\epsilon^{abc} \epsilon^{ik} \epsilon^{jl} \left(Q_{1ai} \sigma^{\mu} e_{_\mathbb{C}}^{\dagger}{}_5\right) \left(Q_{2bj} D_{\mu} Q_{3ck}\right) H_{4l} \\
        \mc{O}^{(y)}_{Q^3 H e_{_\mathbb{C}}^{\dagger} D,3} & -i\epsilon^{abc} \epsilon^{ik} \epsilon^{jl} \left(Q_{1ai} Q_{3ck}\right) \left(Q_{2bj} \sigma^{\mu} e_{_\mathbb{C}}^{\dagger}{}_5\right) D_{\mu} H_{4l} \\
        \mc{O}^{(y)}_{Q^3 H e_{_\mathbb{C}}^{\dagger} D,4} & i\epsilon^{abc} \epsilon^{ij} \epsilon^{kl} \left(Q_{1ai} Q_{2bj}\right) \left(Q_{3ck} \sigma^{\mu} e_{_\mathbb{C}}^{\dagger}{}_5\right) D_{\mu} H_{4l} \\
        \mc{O}^{(y)}_{Q^3 H e_{_\mathbb{C}}^{\dagger} D,5} & -i\epsilon^{abc} \epsilon^{ij} \epsilon^{kl} \left(Q_{1ai} \sigma^{\mu} e_{_\mathbb{C}}^{\dagger}{}_5\right) \left(Q_{2bj} D_{\mu} Q_{3ck}\right) H_{4l} \\
        \mc{O}^{(y)}_{Q^3 H e_{_\mathbb{C}}^{\dagger} D,6} & -i\epsilon^{abc} \epsilon^{ij} \epsilon^{kl} \left(Q_{1ai} Q_{3ck}\right) \left(Q_{2bj} \sigma^{\mu} e_{_\mathbb{C}}^{\dagger}{}_5\right) D_{\mu} H_{4l}
    \end{array}
\end{align}
\begin{align}
    \begin{array}{c|c}
        \text{M-basis} &  \\
        \hline
        \mc{O}^{(m)}_{Q^3 H e_{_\mathbb{C}}^{\dagger} D,1} & i\epsilon^{abc} \epsilon^{ik} \epsilon^{jl} \left(Q_{1ai} Q_{2bj}\right) \left(Q_{3ck} \sigma^{\mu} e_{_\mathbb{C}}^{\dagger}{}_5\right) D_{\mu} H_{4l} \\
        \mc{O}^{(m)}_{Q^3 H e_{_\mathbb{C}}^{\dagger} D,2} & i\epsilon^{abc} \epsilon^{ik} \epsilon^{jl} \left(Q_{1ai} \sigma^{\mu} e_{_\mathbb{C}}^{\dagger}{}_5\right) \left(Q_{2bj} D_{\mu} Q_{3ck}\right) H_{4l} \\
        \mc{O}^{(m)}_{Q^3 H e_{_\mathbb{C}}^{\dagger} D,3} & i\epsilon^{abc} \epsilon^{ik} \epsilon^{jl} \left(Q_{1ai} Q_{3ck}\right) \left(Q_{2bj} \sigma^{\mu} e_{_\mathbb{C}}^{\dagger}{}_5\right) D_{\mu} H_{4l} \\
        \mc{O}^{(m)}_{Q^3 H e_{_\mathbb{C}}^{\dagger} D,4} & i\epsilon^{abc} \epsilon^{ij} \epsilon^{kl} \left(Q_{1ai} Q_{2bj}\right) \left(Q_{3ck} \sigma^{\mu} e_{_\mathbb{C}}^{\dagger}{}_5\right) D_{\mu} H_{4l} \\
        \mc{O}^{(m)}_{Q^3 H e_{_\mathbb{C}}^{\dagger} D,5} & i\epsilon^{abc} \epsilon^{ij} \epsilon^{kl} \left(Q_{1ai} \sigma^{\mu} e_{_\mathbb{C}}^{\dagger}{}_5\right) \left(Q_{2bj} D_{\mu} Q_{3ck}\right) H_{4l} \\
        \mc{O}^{(m)}_{Q^3 H e_{_\mathbb{C}}^{\dagger} D,6} & i\epsilon^{abc} \epsilon^{ij} \epsilon^{kl} \left(Q_{1ai} Q_{3ck}\right) \left(Q_{2bj} \sigma^{\mu} e_{_\mathbb{C}}^{\dagger}{}_5\right) D_{\mu} H_{4l}
    \end{array}
\end{align}
\beq
\mc{K}^{(my)}=\left(
\begin{array}{cccccc}
	1 & 0 & 0 & 0 & 0 & 0 \\
	0 & -1 & 0 & 0 & 0 & 0 \\
	0 & 0 & -1 & 0 & 0 & 0 \\
	0 & 0 & 0 & 1 & 0 & 0 \\
	0 & 0 & 0 & 0 & -1 & 0 \\
	0 & 0 & 0 & 0 & 0 & -1 \\
\end{array}
\right)
\eeq
The m-basis of this type is 6-dimensional and the Young symmetrizers of representation $[3]$, $[2,1]$ and $[1,1,1]$ are presented in the m-basis as
\begin{eqnarray}\label{eq:youngsymofS3}
	&&\mc{Y}^{[3]}_1=\left(
	\begin{array}{cccccc}
		-\frac{1}{6} & 0 & -\frac{1}{3} & \frac{1}{3} & 0 & \frac{1}{6} \\
		\frac{1}{6} & 0 & \frac{1}{3} & -\frac{1}{3} & 0 & -\frac{1}{6} \\
		\frac{1}{3} & 0 & \frac{2}{3} & -\frac{2}{3} & 0 & -\frac{1}{3} \\
		-\frac{1}{3} & 0 & -\frac{2}{3} & \frac{2}{3} & 0 & \frac{1}{3} \\
		0 & 0 & 0 & 0 & 0 & 0 \\
		\frac{1}{6} & 0 & \frac{1}{3} & -\frac{1}{3} & 0 & -\frac{1}{6} \\
	\end{array}
	\right), \quad
	\mc{Y}^{[2,1]}_1=\left(
	\begin{array}{cccccc}
		\frac{1}{3} & 0 & \frac{2}{3} & \frac{1}{3} & 0 & -\frac{1}{3} \\
		-\frac{1}{3} & \frac{4}{3} & -\frac{2}{3} & \frac{1}{3} & -\frac{2}{3} & \frac{1}{3} \\
		0 & 0 & 0 & 0 & 0 & 0 \\
		\frac{1}{3} & 0 & \frac{2}{3} & \frac{1}{3} & 0 & -\frac{1}{3} \\
		-\frac{1}{3} & \frac{2}{3} & -\frac{2}{3} & 0 & -\frac{1}{3} & \frac{1}{3} \\
		-\frac{1}{3} & 0 & -\frac{2}{3} & -\frac{1}{3} & 0 & \frac{1}{3} \\
	\end{array}
	\right), \nn \\
	&&\mc{Y}^{[2,1]}_2=\left(
	\begin{array}{cccccc}
		\frac{1}{3} & 0 & -\frac{1}{3} & -\frac{2}{3} & 0 & -\frac{1}{3} \\
		\frac{1}{3} & -\frac{2}{3} & \frac{1}{3} & -\frac{2}{3} & \frac{4}{3} & -1 \\
		0 & 0 & 0 & 0 & 0 & 0 \\
		\frac{1}{3} & 0 & -\frac{1}{3} & -\frac{2}{3} & 0 & -\frac{1}{3} \\
		0 & -\frac{1}{3} & \frac{1}{3} & 0 & \frac{2}{3} & -\frac{1}{3} \\
		-\frac{1}{3} & 0 & \frac{1}{3} & \frac{2}{3} & 0 & \frac{1}{3} \\
	\end{array}
	\right), \quad
	\mc{Y}^{[1,1,1]}_1=\left(
	\begin{array}{cccccc}
		\frac{1}{2} & 0 & 0 & 0 & 0 & \frac{1}{2} \\
		\frac{1}{6} & 0 & 0 & 0 & 0 & \frac{1}{6} \\
		0 & 0 & 0 & 0 & 0 & 0 \\
		0 & 0 & 0 & 0 & 0 & 0 \\
		\frac{1}{3} & 0 & 0 & 0 & 0 & \frac{1}{3} \\
		\frac{1}{2} & 0 & 0 & 0 & 0 & \frac{1}{2} \\
	\end{array}
	\right).
\end{eqnarray}
From the above matrix representations of Young symmetrizers, we find that there are 6 independent basis vectors in the p-basis,
\begin{align}
    \begin{array}{c|c}
        \text{P-basis} &  \\
        \hline
        \mc{O}^{(p)}_{Q^3 H e_{_\mathbb{C}}^{\dagger} D,1} & \dfrac{1}{6}\mc{Y}\left[\tiny{\young(prs)}\right] i\epsilon^{abc} \epsilon^{ik} \epsilon^{jl} \left(Q_{pai} Q_{rbj}\right) \left(Q_{sck} \sigma^{\mu} e_{_\mathbb{C}}^{\dagger}{}_t\right) D_{\mu} H_{l} \\
        \mc{O}^{(p)}_{Q^3 H e_{_\mathbb{C}}^{\dagger} D,2} & \dfrac{1}{3}\mc{Y}\left[\tiny{\young(pr,s)}\right] i\epsilon^{abc} \epsilon^{ik} \epsilon^{jl} \left(Q_{pai} Q_{rbj}\right) \left(Q_{sck} \sigma^{\mu} e_{_\mathbb{C}}^{\dagger}{}_t\right) D_{\mu} H_{l} \\
        \mc{O}^{(p)}_{Q^3 H e_{_\mathbb{C}}^{\dagger} D,3} & \dfrac{1}{3} (r \ s)\mc{Y}\left[\tiny{\young(pr,s)}\right] i\epsilon^{abc} \epsilon^{ik} \epsilon^{jl} \left(Q_{pai} Q_{rbj}\right) \left(Q_{sck} \sigma^{\mu} e_{_\mathbb{C}}^{\dagger}{}_t\right) D_{\mu} H_{l} \\
        \mc{O}^{(p)}_{Q^3 H e_{_\mathbb{C}}^{\dagger} D,4} & \dfrac{1}{3}\mc{Y}\left[\tiny{\young(pr,s)}\right] i\epsilon^{abc} \epsilon^{ik} \epsilon^{jl} \left(Q_{pai} \sigma^{\mu} e_{_\mathbb{C}}^{\dagger}{}_t\right) \left(Q_{rbj} D_{\mu} Q_{sck}\right) H_{l} \\
        \mc{O}^{(p)}_{Q^3 H e_{_\mathbb{C}}^{\dagger} D,5} & \dfrac{1}{3}(r \ s)\mc{Y}\left[\tiny{\young(pr,s)}\right] i\epsilon^{abc} \epsilon^{ik} \epsilon^{jl} \left(Q_{pai} \sigma^{\mu} e_{_\mathbb{C}}^{\dagger}{}_t\right) \left(Q_{rbj} D_{\mu} Q_{sck}\right) H_{l} \\
        \mc{O}^{(p)}_{Q^3 H e_{_\mathbb{C}}^{\dagger} D,6} & \dfrac{1}{6}\mc{Y}\left[\tiny{\young(1,2,3)}\right] i\epsilon^{abc} \epsilon^{ik} \epsilon^{jl} \left(Q_{pai} Q_{rbj}\right) \left(Q_{sck} \sigma^{\mu} e_{_\mathbb{C}}^{\dagger}{}_t\right) D_{\mu} H_{l}
    \end{array}
\end{align}
\beq
\mc{K}^{(pm)}=\left(
\begin{array}{cccccc}
	-\frac{1}{6} & 0 & -\frac{1}{3} & \frac{1}{3} & 0 & \frac{1}{6} \\
	\frac{1}{3} & 0 & \frac{2}{3} & \frac{1}{3} & 0 & -\frac{1}{3} \\
	\frac{1}{3} & 0 & -\frac{1}{3} & -\frac{2}{3} & 0 & -\frac{1}{3} \\
	-\frac{1}{3} & \frac{4}{3} & -\frac{2}{3} & \frac{1}{3} & -\frac{2}{3} & \frac{1}{3} \\
	\frac{1}{3} & -\frac{2}{3} & \frac{1}{3} & -\frac{2}{3} & \frac{4}{3} & -1 \\
	\frac{1}{2} & 0 & 0 & 0 & 0 & \frac{1}{2} \\
\end{array}
\right).
\eeq
However, since we treat each fermion in the operators as a flavor multiplet with flavor $n_f$, the operators in the p-basis become flavor tensors of representations of $SU(n_f)$ group. The two young symmetrizers of $[2,1]$, $\mc{Y}^{[2,1]}_1$ and $\mc{Y}^{[2,1]}_2$, acting on the m-basis vectors actually give a different set of the p-basis vectors that span the same space. To illustrate this point, let us take
\begin{eqnarray}
	T^{(1)}_{prs} &=& \mc{Y}^{[2,1]}_1 M_{prs} = \frac{1}{3} \mc{Y}\left[\tiny{\young(12,3)}\right] M_{prs} = \frac{1}{3} \left(M_{prs} + M_{rps} - M_{srp} - M_{spr}\right), \\
	T^{(2)}_{prs} &=& \mc{Y}^{[2,1]}_2 M_{prs} = \frac{1}{3} \, (2 \ 3) \, \mc{Y}\left[\tiny{\young(12,3)}\right] M_{prs} = \frac{1}{3} \left(M_{psr} + M_{spr} - M_{rsp} - M_{rps}\right),
\end{eqnarray}
then $T^{(1)}_{prs}$ and $T^{(2)}_{prs}$ span a 2-dimensional representation of $S_3$ where each element of $S_3$ can be presented as
\begin{eqnarray}
	&D_{[2,1]} \left(E\right)= \left(
	\begin{array}{cc}
		1 & 0 \\
		0 & 1
	\end{array}
    \right), \quad
    D_{[2,1]} \left(1 \ 2\right)= \left(
    \begin{array}{cc}
    	1 & 0 \\
    	-1 & -1
    \end{array}
    \right), \quad
    D_{[2,1]} \left(1 \ 3\right)= \left(
    \begin{array}{cc}
    	-1 & -1 \\
    	0 & 1
    \end{array}
    \right), \\
    &D_{[2,1]} \left(2 \ 3\right)= \left(
    \begin{array}{cc}
    	0 & 1 \\
    	1 & 0
    \end{array}
    \right), \quad
    D_{[2,1]} \left(1 \ 3 \ 2\right)= \left(
    \begin{array}{cc}
    	-1 & -1 \\
    	1 & 0
    \end{array}
    \right), \quad
    D_{[2,1]} \left(1 \ 2 \ 3\right)= \left(
    \begin{array}{cc}
    	0 & 1 \\
    	-1 & -1
    \end{array}
    \right).
\end{eqnarray}
The following relations allow us to rewrite each component of $T^{(2)}_{prs}$ as linear combinations of components of $T^{(1)}_{prs}$,
\beq\label{eq:tensorcoe}
T_{p_{1}, \cdots, p_{m}}^{(i)}=\frac{d(\mu)}{m !} \sum_{\pi \in S_{m}} \left[D_{\mu}\left(\pi\right)\right]_{i 1} T_{\pi\left(p_{1}, \cdots, p_{m}\right)}^{(1)}, \quad i=1,\cdots,d(\mu).
\eeq
For example,
\begin{eqnarray}
	\frac{1}{3} \sum_{\pi \in S_{3}} \left[D_{\mu}\left(\pi\right)\right]_{2 1} T^{(1)}_{\pi(prs)}
	&=& \frac{1}{3} \left(-T^{(1)}_{rps} + T^{(1)}_{psr} + T^{(1)}_{spr} - T^{(1)}_{rsp}\right) \nn \\
	&=& (2 \ 3) \ \mc{Y}\left[\tiny{\young(12,3)}\right] T^{(1)}_{prs} \nn \\
	&=& \frac{1}{3} \, (2 \ 3) \, \mc{Y}\left[\tiny{\young(12,3)}\right] \mc{Y}\left[\tiny{\young(12,3)}\right] M_{prs} \nn \\
	&=& \frac{1}{3} \, (2 \ 3) \, \mc{Y}\left[\tiny{\young(12,3)}\right] M_{prs} \nn \\
	&=& T^{(2)}_{prs}.
\end{eqnarray}
So we should keep only one Young symmetrizers of $[2,1]$ in eq.~{(\ref{eq:youngsymofS3})}, and the independent operators, presented as flavor tensors, are called the p'-basis.
\begin{align}
    \begin{array}{c|c}
        \text{P'-basis} &  \\
        \hline
        \mc{O}^{(p')}_{Q^3 H e_{_\mathbb{C}}^{\dagger} D,1} & \dfrac{1}{6}\mc{Y}\left[\tiny{\young(prs)}\right] i\epsilon^{abc} \epsilon^{ik} \epsilon^{jl} \left(Q_{pai} Q_{rbj}\right) \left(Q_{sck} \sigma^{\mu} e_{_\mathbb{C}}^{\dagger}{}_t\right) D_{\mu} H_{l} \\
        \mc{O}^{(p')}_{Q^3 H e_{_\mathbb{C}}^{\dagger} D,2} & \dfrac{1}{3}\mc{Y}\left[\tiny{\young(pr,s)}\right] i\epsilon^{abc} \epsilon^{ik} \epsilon^{jl} \left(Q_{pai} Q_{rbj}\right) \left(Q_{sck} \sigma^{\mu} e_{_\mathbb{C}}^{\dagger}{}_t\right) D_{\mu} H_{l} \\
        \mc{O}^{(p')}_{Q^3 H e_{_\mathbb{C}}^{\dagger} D,3} & \dfrac{1}{3}\mc{Y}\left[\tiny{\young(pr,s)}\right] i\epsilon^{abc} \epsilon^{ik} \epsilon^{jl} \left(Q_{pai} \sigma^{\mu} e_{_\mathbb{C}}^{\dagger}{}_t\right) \left(Q_{rbj} D_{\mu} Q_{sck}\right) H_{l} \\
        \mc{O}^{(p')}_{Q^3 H e_{_\mathbb{C}}^{\dagger} D,4} & \dfrac{1}{6}\mc{Y}\left[\tiny{\young(p,r,s)}\right] i\epsilon^{abc} \epsilon^{ik} \epsilon^{jl} \left(Q_{pai} Q_{rbj}\right) \left(Q_{sck} \sigma^{\mu} e_{_\mathbb{C}}^{\dagger}{}_t\right) D_{\mu} H_{l}
    \end{array}
\end{align}
\beq
\mc{K}^{(p'm)}=\left(
\begin{array}{cccccc}
	-\frac{1}{6} & 0 & -\frac{1}{3} & \frac{1}{3} & 0 & \frac{1}{6} \\
	\frac{1}{3} & 0 & \frac{2}{3} & \frac{1}{3} & 0 & -\frac{1}{3} \\
	-\frac{1}{3} & \frac{4}{3} & -\frac{2}{3} & \frac{1}{3} & -\frac{2}{3} & \frac{1}{3} \\
	\frac{1}{2} & 0 & 0 & 0 & 0 & \frac{1}{2} \\
\end{array}
\right).
\eeq
The relation eq.~(\ref{eq:tensorcoe}) is valid for representations of the symmetric group with any dimension, and we always choose to keep the first Young symmetrizer for representations with dimension more than 1 in the package.

\subsection{J-basis}

 The above amplitude basis can be re-organized to be the eigenbasis of the conserved angular momentum and gauge quantum numbers for subsets of the external particles. In this case, the spin and quantum numbers for partitions of the external operator correspond to certain ultraviolet resonances. This eigenbasis can be obtained from the partial wave expansion via the Casimir action, more specifically, the Pauli-Lubanski operator~\cite{Jiang:2020rwz,Li:2020zfq}. In terms of the spinor helicity notation, the Pauli-Lubanski operator, $W_{\mu}=\frac12\epsilon_{\mu\nu\rho\sigma}P^{\nu}M^{\rho\sigma}$, can be rewritten. 
It forms a Casimir invariant $W^2$ for the Poincar\'{e} group, which has the eigenvalue $-P^2J(J+1)$. 
Apply to multi-particle states $\mathcal{I}=\{i,j,\dots\}$, $P_{\mathcal{I}}^2=(\sum_{i\in \mathcal{I}} p_i)^2$ is the total momentum square, and $J$ is the total angular momentum. 
These operators are represented by the helicity spinors as follows
\begin{align}
    P_{\mathcal{I}\alpha\dot\alpha}=\sum_{i\in\mathcal{I}}\lambda_{i\alpha}\tilde\lambda_{i\dot\alpha},\quad M_{\mathcal{I},\alpha\beta}=&i\sum_{i\in\mathcal{I}}\left( \lambda_{i\alpha}\frac{\partial}{\partial\lambda_i^{\beta}}+\lambda_{i\beta}\frac{\partial}{\partial\lambda_i^{\alpha}} \right),\quad \tilde{M}_{\mathcal{I},\dot\alpha\dot\beta}=i\sum_{i\in\mathcal{I}}\left( \tilde\lambda_{i\dot\alpha}\frac{\partial}{\partial\tilde\lambda_i^{\dot\beta}}+\tilde\lambda_{i\dot\beta}\frac{\partial}{\partial\tilde\lambda_i^{\dot\alpha}} \right).\\
    W^2_{\mathcal{I}}=&\frac18\sum_{ij\in\mathcal{I}}\left( \langle ij\rangle[j|\tilde{M}^2|i]+[ij]\langle j|M^2|i\rangle -2\langle i|M|j\rangle[i|\tilde{M}|j] \right).
\end{align}
It's straightforward to confirm that the Poincar\'{e} algebra is satisfied with
\begin{align}
    [P_{\rho},M_{\mu\nu}]=i(g_{\rho\mu}P_{\nu}-g_{\rho\nu}P_{\mu}).
\end{align}
For example, the scattering amplitude $\psi_1\phi_2 \to \psi_3\phi_4$ generated by a dim-5 operator $\mathcal{O}=(\psi_1\psi_3)\phi_2\phi_4$ is $\mathcal{A}=\langle 13\rangle$, which is exactly the eigenstate of $W^2$,
\begin{align}
    W^2_{\{1,2\}}\langle 13\rangle=-\frac34 s_{12}\langle 13\rangle.\label{eq:ex1}
\end{align}
The eigenvalue $-P^2_{\mathcal{I}}J(J+1)=\frac34s_{12}$ confirms that $J=1/2$. 
Angular momentum conservation ensures that $W^2_{\mathcal{I}}\mathcal{A}=W^2_{\bar{\mathcal{I}}}\mathcal{A}$, when $\bar{\mathcal{I}}$ is the complement of channel $\mathcal{I}$. The eigenstate of $W^2_{\mathcal{I}}$ is called a partial-wave amplitude with angular momentum $J$, among that, the independent one is called the Lorentz j-basis. 

To obtain the j-basis in given channel $\mathcal{I}$, we apply $W^2_{\mathcal{I}}$ to the y-basis in the same dimension $d$. The result must be the dim-$(d+2)$ amplitudes. Note that a Mandelstam variable $s_{\mathcal{I}}$ times the dim-$d$ y-basis is also a dim-$(d+2)$ amplitude. We can then reduce them to our dim-$(d+2)$ y-basis and solve the coefficient matrix $\mathcal{W}_{ij}$, which is defined as $W^2_{\mathcal{I}}\mathcal{B}^{y,d}_i=\sum_j\mathcal{W}_{ij}s_{\mathcal{I}}\mathcal{B}^{y,d}_j$.
\begin{align}
    W^2_{\mathcal{I}}\mathcal{B}^{y,d}_i=\sum_j\mathcal{K}^{wy}_{ij}\mathcal{B}^{y,d+2}_j,\quad s_{\mathcal{I}}\mathcal{B}^{y,d}_i=\sum_j\mathcal{K}^{sy}_{ij}\mathcal{B}^{y,d+2}_j\;\Rightarrow
    \sum_j\mathcal{W}_{ij}\mathcal{K}^{sy}_{jk}=\mathcal{K}_{ik}^{wy}\, .
\end{align}
After that, we diagonalize $\mathcal{W}_{ij}$ and obtain the j-basis. For example, to obtain the j-basis in channel $\{1,2\}$ of state $\{1/2,1/2,1/2,1/2\}$ with 2 derivatives, we first list the dim-8 y-basis as follows
\begin{align}
    \mathcal{B}^{y,8}_i=\big\{[12][34]^2\vev{34},&\; -[13][24]^2\vev{24},\; [13][24][34]\vev{34}\big\}\,.
\end{align}
Acting the $W^2$ in the channel $\{1,2\}$ gives
\begin{align}
    W^2_{\{1,2\}}\left(\begin{array}{c}
        [12][34]^2\vev{34}\\
        -[13][24]^2\vev{24}\\
        [13][24][34]\vev{34}
    \end{array}\right)=&
    \left(\begin{array}{ccc}
        0 & 0 & 0\\
        0 & -6& 4\\
        1 & 0 &-2
    \end{array}\right)\left(\begin{array}{c}
        [12][34]^2\vev{34}\\
        -[13][24]^2\vev{24}\\
        [13][24][34]\vev{34}
    \end{array}\right)\,.
\end{align}
After diagonalization, we obtain
\begin{align}
    \left(\begin{array}{ccc}
        -1& -6& 6\\
        -1& 0 & 2\\
        1 & 0 & 0
    \end{array}\right)\left(\begin{array}{ccc}
        0 & 0 & 0\\
        0 & -6& 4\\
        1 & 0 &-2
    \end{array}\right)\left(\begin{array}{ccc}
        -1& -6& 6\\
        -1& 0 & 2\\
        1 & 0 & 0
    \end{array}\right)^{-1}=\left(\begin{array}{ccc}
        -6& 0 & 0\\
         0& -2& 0\\
        0 & 0 & 0
    \end{array}\right)\,.
\end{align}
Therefore, the complete set of the J-basis in the dim-8 should be
\begin{align}
    \left(\begin{array}{c}
        \mathcal{B}^{J=2}\\
        \mathcal{B}^{J=1}\\
        \mathcal{B}^{J=0}
    \end{array}\right)=
    \left(\begin{array}{c}
        [12][34]s_{34}-6[13][24](s_{24}+s_{34})\\
        ([12][34]-2[13][24])s_{34}\\
        -[12][34]s_{34}
    \end{array}\right)=
    \left(\begin{array}{ccc}
        -1& -6& 6\\
        -1& 0 & 2\\
        1 & 0 & 0
    \end{array}\right)\left(\begin{array}{c}
        [12][34]^2\vev{34}\\
        -[13][24]^2\vev{24}\\
        [13][24][34]\vev{34}
    \end{array}\right)\,.
\end{align}
The total angular momentum for the particles $\{1,2\}$ is 2, 1, and 0 respectively. 
In our package, the function \mmaInlineCell[defined=W2Diagonalize]{Code}{W2Diagonalize} incorporates the above steps.%, the function \mmaInlineCell[defined=GetJBasisForType]{Code}{GetJBasisForType} can obtain the J-basis directly  for a given type. Their usage will be described in sec.~\ref{sec:functions}. 

A similar $J$ basis can be defined on the gauge group factor space. The meaning of $J$ becomes a series value of Casimir of the corresponding gauge group for each group of particles in the channel. The definition of the Casimirs is based on the following definition of the action of generators on a group factor:
\begin{eqnarray}
T'_{I_1I_2\dots I_N}\equiv \underset{{\cal I}}{{\mathbbm{T}}^A}\circ T_{I_1I_2\dots I_N}=
\sum_{i\in {\cal I}}^N ({\rm T}^A_{r_i})_{I_{i}}^{Z}T_{I_1\dots I_{i-1}Z I_{i+1} I_N},\label{eq:partialT}
\end{eqnarray}
where ${\cal I}$ again represents a channel of particles that the generator acts on e.g., ${\cal I}=\{1,2\}$, $\mathbbm{T}$ represents a formal operator on the gauge m-basis, ${\rm T}^A_{r_i}$ represents the generator matrices for irreducible representation $r_i$. 
%For example, it can be  $\{1, 2\}$ meaning the generators only act on the first two indices. 
For $SU(M)$ group, one needs the value of $M-1$ Casimirs constructed from generators to characterize an irreducible representation. For $SU(2)$ and $SU(3)$ we have the following Casimirs:
\begin{eqnarray}
\mathbbm{C}_2&=& \mathbbm{T}^a\mathbbm{T}^a,\ \text{for both $SU(2)$ and $SU(3)$,}\label{eq:CasimirC2}\\
\mathbbm{C}_3 &=& d^{abc}\mathbbm{T}^a\mathbbm{T}^b\mathbbm{T}^c,\ \text{for $SU(3)$ only},\label{eq:CasimirC3}
\end{eqnarray}
For a general definition of Casimirs, one can consult the textbook~\cite{ma2007group}.
We take the operator type $W^4$ as an example; the corresponding gauge m-basis is:
\begin{eqnarray}
T^{(m)}_1=\delta_{I_1I_3}\delta_{I_2I_4},\ T^{(m)}_2=\delta_{I_1I_2}\delta_{I_3I_4},\ T^{(m)}_3=\delta_{I_1I_4}\delta_{I_2I_3}.
\label{eq:Typipi}
\end{eqnarray}
\begin{eqnarray}
\underset{\{1,2\}}{\mathbbm{C}_2}\circ T^{(m)}_i=\underset{\{1,2\}}{\mathbbm{T}_a}\circ\underset{\{1,2\}}{T_a}\circ T^{(m)}_i=\underset{\{1,2\}}{({C}_2)_{ij}}{\cal T}^{(m)}_j=
\begin{pmatrix}
4 & 0 & -2\\
-2 & 0 & -2\\
2 & 0 & 4
\end{pmatrix}
\begin{pmatrix}
{  T}^{(m)}_1\\
{  T}^{(m)}_2\\
{  T}^{(m)}_3
\end{pmatrix},
\end{eqnarray}
where $\underset{\{1,2\}}{({C}_2)_{ij}}$ represents the representation matrix of $\underset{\{1,2\}}{\mathbbm{C}_2}$ on the gauge m-basis. Diagonalizing the $\underset{\{1,2\}}{({C}_2)_{ij}}$ gives you the eigenvalues as $\underset{\{1,2\}}{({C}_2)}=2,1,0$, and the corresponding eigenvectors:
\begin{eqnarray}
&&{ T}^{(J=2)}_{I_1I_2I_3I_4}=-3{ T}^{ (m)}_1+2{  T}^{ (m)}_2-3{  T}^{ (m)}_3=-3\delta_{I_1I_3}\delta_{I_2I_4}+2\delta_{I_1I_2}\delta_{I_3I_4}-3\delta_{I_1I_4}\delta_{I_2I_3}\nonumber \\
&&{ T}^{(J=1)}_{I_1I_2I_3I_4}=1{ T}^{ (m)}_1-{  T}^{ (m)}_3=\delta_{I_1I_3}\delta_{I_2I_4}-\delta_{I_1I_4}\delta_{I_2I_3}\nonumber \\
&&{ T}^{(J=0)}_{I_1I_2I_3I_4}=1\times{  T}^{ (m)}_2=\delta_{I_1I_2}\delta_{I_3I_4}.
\end{eqnarray}
Representing the $W_1$ and $W_2$ are the group to intermediate irreducible representations of quartet, doublet, and singlet, respectively. The above gauge j-basis can be obtained by the function \mmaInlineCell[defined={GaugeJBasis}]{Input}{GaugeJBasis}.

Combining the coordinates of the Lorentz and gauge j-basis, one can obtain the operator j-basis,
here we give an example of the j-basis of type $Q^3 H e_{_\mathbb{C}}^{\dagger} D$. To define the j-basis with channels, one first need to label the fields, by convention, we label the fields in increasing order of their helicities, i.e., in our example, the fields are labeled as $Q_1 Q_2 Q_3 H_4 e_{_\mathbb{C}5}^{\dagger}$.
With the function \mmaInlineCell[defined={GetJBasisForType}]{Input}{GetJBasisForType}, we can obtain the results listed below:
\begin{equation}
	\begin{array}{c|c}
		\hline
		\multicolumn{2}{c}{\mc{O}^{(p')}_{1} = \frac{1}{6}\mc{Y}\left[\tiny{\young(prs)}\right] i\epsilon^{abc} \epsilon^{ik} \epsilon^{jl} \left(Q_{pai} Q_{rbj}\right) \left(Q_{sck} \sigma^{\mu} e_{_\mathbb{C}}^{\dagger}{}_t\right) D_{\mu} H_{l}} \\
		\multicolumn{2}{c}{\mc{O}^{(p')}_{2} = \frac{1}{3}\mc{Y}\left[\tiny{\young(pr,s)}\right]i\epsilon^{abc} \epsilon^{ik} \epsilon^{jl} \left(Q_{pai} Q_{rbj}\right) \left(Q_{sck} \sigma^{\mu} e_{_\mathbb{C}}^{\dagger}{}_t\right) D_{\mu} H_{l}} \\
		\multicolumn{2}{c}{\mc{O}^{(p')}_{3} = \frac{1}{3}\mc{Y}\left[\tiny{\young(pr,s)}\right]i\epsilon^{abc} \epsilon^{ik} \epsilon^{jl} \left(Q_{pai} \sigma^{\mu} e_{_\mathbb{C}}^{\dagger}{}_t\right) \left(Q_{rbj} D_{\mu} Q_{sck}\right) H_{l}} \\
		\multicolumn{2}{c}{\mc{O}^{(p')}_{4} = \frac{1}{6}\mc{Y}\left[\tiny{\young(p,r,s)}\right] i\epsilon^{abc} \epsilon^{ik} \epsilon^{jl} \left(Q_{pai} Q_{rbj}\right) \left(Q_{sck} \sigma^{\mu} e_{_\mathbb{C}}^{\dagger}{}_t\right) D_{\mu} H_{l}} \\
		\hline
		\multicolumn{2}{c}{\text{group:} \left(\operatorname{Spin}, SU(3)_{c}, SU(2)_{w}, U(1)_{y}\right)}  \\
		\hline
		\{Q_1, Q_{2}\},\{Q_{3}, H_4, e_{_\mathbb{C}5}^{\dagger}\} & \mathcal{O}_{j} \\
		\hline
		(1,3,3,\frac{1}{3}) & -2\mc{O}^{(p')}_{1} + \frac{1}{2}\mc{O}^{(p')}_{2} + \frac{3}{2}\mc{O}^{(p')}_{3} \\
		& -6\mc{O}^{(p')}_{1} + 3\mc{O}^{(p')}_{2} \\
		\hline
		(0,3,3,\frac{1}{3}) & -[1+ 2 (r \ s)]\mc{O}^{(p')}_{2} - 2\mc{O}^{(p')}_{4} \\
		\hline
		(1,3,1,\frac{1}{3}) & [\frac{1}{2} + (r \ s)]\mc{O}^{(p')}_{2} - [\frac{1}{2} + (r \ s)]\mc{O}^{(p')}_{3} - \frac{2}{3}\mc{O}^{(p')}_{4} \\
		& [1 + 2(r \ s)]\mc{O}^{(p')}_{Q^3 H e_{_\mathbb{C}}^{\dagger} D,2} - 2\mc{O}^{(p')}_{4} \\
		\hline
		(0,3,1,\frac{1}{3}) & 2\mc{O}^{(p')}_{1} + \mc{O}^{(p')}_{2} \\
		\hline
		\{Q_1, H_4\},\{Q_{2}, Q_{3}, e_{_\mathbb{C}5}^{\dagger}\} & \mathcal{O}_{j} \\
		\hline
		(\frac{3}{2},3,3,\frac{2}{3}) & 8\mc{O}^{(p')}_{1} + \frac{7}{2}[1 + (r \ s)]\mc{O}^{(p')}_{2} + \frac{3}{2}[1 + (r \ s)]\mc{O}^{(p')}_{3} \\
		\hline
		(\frac{1}{2},3,3,\frac{1}{3}) & \mc{O}^{(p')}_{1} + \frac{1}{2}[1 + 3 (r \ s)]\mc{O}^{(p')}_{2} - \frac{3}{2}[1 + (r \ s)]\mc{O}^{(p')}_{3} - \mc{O}^{(p')}_{4} \\
		& -[1 - (r \ s)]\mc{O}^{(p')}_{2} - 2\mc{O}^{(p')}_{4} \\
		\hline
		(\frac{3}{2},3,1,\frac{2}{3}) & \frac{3}{2}[1 - (r \ s)]\mc{O}^{(p')}_{2} - \frac{1}{2}[1 - (r \ s)]\mc{O}^{(p')}_{3} - \frac{8}{2}\mc{O}^{(p')}_{4} \\
		\hline
		(\frac{1}{2},3,1,\frac{2}{3}) & -\mc{O}^{(p')}_{1} + \frac{1}{2}[1 +  (r \ s)]\mc{O}^{(p')}_{2} + \frac{1}{2}[1 - (r \ s)]\mc{O}^{(p')}_{3} - \frac{1}{3}\mc{O}^{(p')}_{4} \\
		& -2\mc{O}^{(p')}_{1} + [1 +  (r \ s)]\mc{O}^{(p')}_{2} \\
		\hline
		\{e_{_\mathbb{C}5}^{\dagger}, Q_1\},\{Q_{2}, Q_{3}, H_4\} & \mathcal{O}_{j} \\
		\hline
		(1,3,2,-\frac{5}{6}) & \mc{O}^{(p')}_{1} + [1 + (r \ s)]\mc{O}^{(p')}_{2} + \mc{O}^{(p')}_{4} \\
		& \mc{O}^{(p')}_{1} + \frac{1}{2}(r \ s)\mc{O}^{(p')}_{2} - \frac{1}{2}[2 + (r \ s)]\mc{O}^{(p')}_{3} - \frac{1}{3}\mc{O}^{(p')}_{4} \\
		& 2\mc{O}^{(p')}_{1} + (r \ s)\mc{O}^{(p')}_{2} \\
		& 2\mc{O}^{(p')}_{1} + \mc{O}^{(p')}_{2} \\
		& \frac{1}{2}[1 + 2(r \ s)]\mc{O}^{(p')}_{2} - \frac{1}{2}[1 + 2(r \ s)]\mc{O}^{(p')}_{3} - \frac{2}{3}\mc{O}^{(p')}_{4} \\
		& \mc{O}^{(p')}_{1} + [1 + (r \ s)]\mc{O}^{(p')}_{2} - \mc{O}^{(p')}_{4} \\
		\hline
		\{H_4, e_{_\mathbb{C}5}^{\dagger}\},\{Q_1, Q_{2}, Q_{3}\} & \mathcal{O}_{j} \\
		\hline
		(\frac{1}{2},1,2,-\frac{1}{2}) & \mc{O}^{(p')}_{1} + [1 + (r \ s)]\mc{O}^{(p')}_{2} + \mc{O}^{(p')}_{4} \\
		& \mc{O}^{(p')}_{1} + \frac{1}{2}(r \ s)\mc{O}^{(p')}_{2} - \frac{1}{2}[2 + (r \ s)]\mc{O}^{(p')}_{3} - \frac{1}{3}\mc{O}^{(p')}_{4} \\
		& 2\mc{O}^{(p')}_{1} + (r \ s)\mc{O}^{(p')}_{2} \\
		& 2\mc{O}^{(p')}_{1} + \mc{O}^{(p')}_{2} \\
		& \frac{1}{2}[1 + 2(r \ s)]\mc{O}^{(p')}_{2} - \frac{1}{2}[1 + 2(r \ s)]\mc{O}^{(p')}_{3} - \frac{2}{3}\mc{O}^{(p')}_{4} \\
		& \mc{O}^{(p')}_{1} + [1 + (r \ s)]\mc{O}^{(p')}_{2} - \mc{O}^{(p')}_{4} \\
		\hline
	\end{array}
\end{equation}

Let us take a closer look at the j-basis of $\{Q_1, Q_{2}\},\{Q_{3}, H_4, e_{_\mathbb{C}5}^{\dagger}\}$. Permuting the label of the three repeated $Q$s gives the following three different channels:
	\begin{equation}
		\begin{array}{cc|c|c|c}
			\hline 
			\multicolumn{5}{c}{\text{group:} \left(\operatorname{Spin}, SU(3)_{c}, SU(2)_{w}, U(1)_{y}\right)}  \\
			\hline
			& \{Q_1, Q_{2}\},\{Q_{3}, H_4, e_{_\mathbb{C}5}^{\dagger}\} & \mathcal{O}^{(y)}_{j} & \mathcal{O}^{(m)}_{j} & \mathcal{O}^{(p)}_{j} \\
			\hline
			& (1,3,3,\frac{1}{3}) & (0,-2,0,0,1,0) & (0,2,0,0,-1,0) & (-2,\frac{1}{2},0,\frac{3}{2},0,0) \\
			& & (2,0,-4,-1,0,2) & (2,0,4,-1,0,-2) & (-6,3,0,0,0,0) \\
			\hline
			& (0,3,3,\frac{1}{3}) & (-2,0,0,1,0,0) & (-2,0,0,1,0,0) & (0,-1,-2,0,0,-2) \\
			\hline
			& (1,3,1,\frac{1}{3}) & (0,0,0,0,1,0) & (0,0,0,0,-1,0) & (0,\frac{1}{2},1,-\frac{1}{2},-1,-\frac{2}{3}) \\
			& & (0,0,0,-1,0,2) & (0,0,0,-1,0,-2) & (0,1,2,0,0,-2) \\
			\hline
			& (0,3,1,\frac{1}{3}) & (0,0,0,1,0,0) & (0,0,0,1,0,0) & (2,1,0,0,0,0) \\
			\hline
			& \{Q_3, Q_1\},\{Q_{2}, H_4, e_{_\mathbb{C}5}^{\dagger}\} & \mathcal{O}^{(y)}_{j} & \mathcal{O}^{(m)}_{j} & \mathcal{O}^{(p)}_{j} \\
			\hline
			& (1,3,3,\frac{1}{3}) & (1,-1,0,-2,2,0) & (1,1,0,-2,-2,0) & (-4,0,\frac{5}{2},0,-\frac{3}{2},0) \\
			& & (2,0,-1,-4,0,2) & (2,0,1,-4,0,-2) & (-6,0,3,0,0,0) \\
			\hline
			& (0,3,3,\frac{1}{3}) & (0,0,-1,0,0,2) & (0,0,1,0,0,-2) & (0,2,1,0,0,-2) \\
			\hline
			& (1,3,1,\frac{1}{3}) & (-1,1,0,0,0,0) & (-1,-1,0,0,0,0) & (0,-1,-\frac{1}{2},-1,-\frac{1}{2},-\frac{4}{3}) \\
			& & (-2,0,1,0,0,0) & (-2,0,-1,0,0,0) & (0,-2,-1,0,0,-2) \\
			\hline
			& (0,3,1,\frac{1}{3}) & (0,0,1,0,0,0) & (0,0,-1,0,0,0) & (2,0,1,0,0,0) \\
			\hline
			& \{Q_2, Q_3\},\{Q_{1}, H_4, e_{_\mathbb{C}5}^{\dagger}\} & \mathcal{O}^{(y)}_{j} & \mathcal{O}^{(m)}_{j} & \mathcal{O}^{(p)}_{j} \\
			\hline
			& (1,3,3,\frac{1}{3}) & (1,1,0,1,1,0) & (1,-1,0,1,-1,0) & (4,\frac{5}{2},\frac{5}{2},-\frac{3}{2},-\frac{3}{2},0) \\
			& & (1,0,1,1,0,1) & (1,0,-1,1,0,-1) & (6,3,3,0,0,0) \\
			\hline
			& (0,3,3,\frac{1}{3}) & (-1,0,1,-1,0,1) & (-1,0,-1,-1,0,-1) & (0,-1,1,0,0,-2) \\
			\hline
			& (1,3,1,\frac{1}{3}) & (-1,-1,0,1,1,0) & (-1,1,0,1,-1,0) & (0,\frac{1}{2},-\frac{1}{2},\frac{1}{2},-\frac{1}{2},-\frac{4}{3}) \\
			& & (-1,0,-1,1,0,1) & (-1,0,1,1,0,-1) & (0,1,-1,0,0,-2) \\
			\hline
			& (0,3,1,\frac{1}{3}) & (1,0,-1,-1,0,1) & (1,0,1,-1,0,-1) & (-2,1,1,0,0,0) \\
			\hline
		\end{array}
	\end{equation}
	The j-bases of two channels differed by permuting labels of repeated fields may look different on flavor-blind bases, but indeed they are the same up to a permutation and an overall constant. For example, the j-basis operator corresponding to the state $(0,3,3,\frac{1}{3})$ of $\{Q_1, Q_{2}\},\{Q_{3}, H_4, e_{_\mathbb{C}5}^{\dagger}\}$ expanded on the p'-basis is given as
	\begin{eqnarray}\label{eq:jop1}
	    &&-[1+ 2 (r \ s)]\mc{O}^{(p')}_{2} - 2\mc{O}^{(p')}_{4} \nn \\ 
	    =&&\frac{1}{3} \left\{ -[1+ 2 (r \ s)]\mc{Y}\left[\tiny{\young(pr,s)}\right] - \mc{Y}\left[\tiny{\young(p,r,s)}\right] \right\} i\epsilon^{abc} \epsilon^{ik} \epsilon^{jl} \left(Q_{pai} Q_{rbj}\right) \left(Q_{sck} \sigma^{\mu} e_{_\mathbb{C}}^{\dagger}{}_t\right) D_{\mu} H_{l}
	\end{eqnarray}
	while the j-basis operator corresponding to the state $(0,3,3,\frac{1}{3})$ of $\{Q_3, Q_{1}\},\{Q_{2}, H_4, e_{_\mathbb{C}5}^{\dagger}\}$ is
	\begin{eqnarray}\label{eq:jop2}
	    &&[2+ (r \ s)]\mc{O}^{(p')}_{2} - 2\mc{O}^{(p')}_{4} \nn \\
	    =&&\frac{1}{3} \left\{ [2 + (r \ s)]\mc{Y}\left[\tiny{\young(pr,s)}\right] - \mc{Y}\left[\tiny{\young(p,r,s)}\right] \right\} i\epsilon^{abc} \epsilon^{ik} \epsilon^{jl} \left(Q_{pai} Q_{rbj}\right) \left(Q_{sck} \sigma^{\mu} e_{_\mathbb{C}}^{\dagger}{}_t\right) D_{\mu} H_{l}
	\end{eqnarray}
	One can check the operator eq.~(\ref{eq:jop2}) is connected with eq.~(\ref{eq:jop1}) by a permutation that transform $\{Q_1, Q_{2}\},\{Q_{3}, H_4, e_{_\mathbb{C}5}^{\dagger}\}$ to $\{Q_3, Q_{1}\},\{Q_{2}, H_4, e_{_\mathbb{C}5}^{\dagger}\}$, which is $(1 \ 3 \ 2)$, or $(p \ s \ r)$ in this case, with
	\begin{eqnarray}
	    (p \ s \ r) \left\{ -[1+ 2 (r \ s)]\mc{Y}\left[\tiny{\young(pr,s)}\right] - \mc{Y}\left[\tiny{\young(p,r,s)}\right] \right\} = \left\{ [2 + (r \ s)]\mc{Y}\left[\tiny{\young(pr,s)}\right] - \mc{Y}\left[\tiny{\young(p,r,s)}\right] \right\}
	\end{eqnarray}

\section{Conversion Among Different Bases}\label{sec:4}

It has always been a confusing issue to have multiple choices of operator bases in the EFT, which keep active in the literature for different purposes. For example, at dimension 6, the Warsaw basis~\cite{Grzadkowski:2010es}, the HISZ basis~\cite{Hagiwara:1993ck}, and the SILH basis~\cite{Giudice:2007fh} were introduced to address different aspects in the studies of the SMEFT operator. Therefore, it is important to systematically relate the bases to compare results among works in different areas of study. However, due to the many redundancy relations, the conversions between operator bases are usually tedious. In the package, we provide a solution by deriving the coordinate of any given operator under the y-basis, so that the relations among the operator bases could be solved easily by linear algebra. 
% The technique is already used in deriving the p-basis and j-basis, and we are going to elaborate on it in this section.

Let us take dimension-6 type $D^2 H^2 H^{\dagger2}$ as an example. In Ref.~\cite{Buchmuller:1985jz}, there are 3 operators in this type
\begin{eqnarray}\label{eq:BuchmullerD2H4}
    \mc{O}_{\partial H} = \frac{1}{2} D_{\mu} (H^{\dagger} H) D^{\mu} (H^{\dagger} H), \quad \mc{O}^{(1)}_{H} =(H^{\dagger} H) (D_{\mu} H^{\dagger} D^{\mu} H), \quad \mc{O}^{(2)}_{H} =(H^{\dagger} D^{\mu} H) (D_{\mu} H^{\dagger} H).
\end{eqnarray}
However, there are only two independent operators in this type if operators involving the EOM are not considered in this type, as in the Warsaw basis Ref.~\cite{Grzadkowski:2010es}
\begin{eqnarray}\label{eq:WarsawD2H4}
		\mathcal{O}_{H \square} &=\left(H^{\dagger} H\right) \square\left(H^{\dagger} H\right), \quad
		\mathcal{O}_{H D} &=\left(H^{\dagger} D_{\mu} H\right)^{\dagger}\left(H^{\dagger} D^{\mu} H\right).
\end{eqnarray}
It is necessary to reduce the over-complete basis eq.~(\ref{eq:BuchmullerD2H4}) to a complete basis, such as eq.~(\ref{eq:WarsawD2H4}), which can be easily done by the function \mmaInlineCell[defined=FindMCoord]{Code}{FindMCoord}. The result gives
\begin{align}\label{eq:BuchmullerToWarsaw}
    \left(\begin{array}{c}
        \mc{O}_{\partial H} \\
        \mc{O}^{(1)}_{H} \\
        \mc{O}^{(2)}_{H}
    \end{array}\right)=
    \left(\begin{array}{cc}
        -\frac{1}{2} & 0 \\
        \frac{1}{2} & 0 \\
        0 & 1
    \end{array}\right)
    \left(\begin{array}{c}
        \mathcal{O}_{H \square} \\
        \mathcal{O}_{H D}
    \end{array}\right).
\end{align}
So we conclude the over-complete basis eq.~(\ref{eq:BuchmullerD2H4}) can be reduced to either $\{\mc{O}_{\partial H}, \mc{O}^{(2)}_{H}\}$ or $\{\mc{O}^{(1)}_{H}, \mc{O}^{(2)}_{H}\}$, and its conversion relations to eq.~(\ref{eq:WarsawD2H4}) are given in eq.~(\ref{eq:BuchmullerToWarsaw}).

Furthermore, in the SILH basis Ref.~\cite{Giudice:2007fh}, they list two independent operators different from those in Warsaw basis as following
\begin{eqnarray}\label{eq:SILHD2H4}
    \mc{O}^{(1)}_{SILH}= D^{\mu} (H^{\dagger} H) D_{\mu} (H^{\dagger} H), \quad \mc{O}^{(2)}_{SILH}=(H^{\dagger} \overleftrightarrow{D}^{\mu} H)(H^{\dagger} \overleftrightarrow{D}_{\mu} H),
\end{eqnarray}
where $H^{\dagger} \overleftrightarrow{D}^{\mu} H = H^{\dagger} D^{\mu} H - D^{\mu} H^{\dagger} H$. The conversion relation between the two complete bases eq.(\ref{eq:WarsawD2H4}) and eq.(\ref{eq:SILHD2H4}) is worth known for comparison of works using the two bases. The result gives
\begin{align}\label{eq:SILHToWarsaw}
    \left(\begin{array}{c}
        \mc{O}^{(1)}_{SILH} \\
        \mc{O}^{(2)}_{SILH}
    \end{array}\right) =
    \left(\begin{array}{cc}
        -1 & 0 \\
        1 & -2
    \end{array}\right)
    \left(\begin{array}{c}
        \mathcal{O}_{H \square} \\
        \mathcal{O}_{H D}
    \end{array}\right).
\end{align}

Note that the above linear relations hold up to EOM, which means that the two sides of an equation can differ by an operator that has vanishing contributions to the on-shell amplitude in the given type of external state. The difference may contribute to other types of amplitude due to either tree-level diagram or EOM, and is supposed to have been counted in those types. When we are not focused on a particular type, extra terms should be added to the above relations, which will be solved in the near future.
In this paper, we mainly introduce the linear relation restricted to a given type. Thus all the equations in this section should be understood as modulo EOM.
% The above discussion includes redundancies induced by the IBP and some equivalence relations like the Schouten identity, but does not include redundancy induced by the EOM. In our on-shell construction of EFT operators, the EOM of a field corresponds to the square of momentum of a external particle, and is always neglected automatically. 
% To construct the transformation between different bases, We first need to convert any operator to a standard basis, which is exactly the Y-basis expressed in sec.~\ref{sec:Ybasis}. After that, The expansion coefficients on Y-basis for each basis are obtained, as well as the conversion between bases. 
We will show how to decompose a given flavor-blind operator(amplitude) into the Y-basis in sec.~\ref{sec:DeYbasis}, and the conversion between flavor-specified bases is inferred in sec.~\ref{sec:DePbasis}.

%=============================================================================

\subsection{Unique Coordinate of Operators}\label{sec:DeYbasis}

For flavor-blind operators, where each field is distinguishable, the Lorentz structure and gauge factors can be treated separately. Suppose we have a monomial operator $\mc{O} = T^{a_1,\dots,a_N}\mc{B}_{a_1,\dots,a_N}$, we can decompose it in the following steps
\eq{\label{eq:operator_decompose}
    & \mc{B}_{a_1,\dots,a_N} = \sum_i \mc{K}_{\mc{B},i} (\mc{B}^m_i)_{a_1,\dots,a_N}, \qquad T^{a_1,\dots,a_N} = \sum_j \mc{K}_{T,j} (T_j^m)^{a_1,\dots,a_N}, \\
    & \mc{O} = \sum_{i,j} \mc{K}_{\mc{B},i}\mc{K}_{T,j} \mc{T}_{ij}^m, \qquad  \mc{T}^m_{ij} \equiv (T_j^m)^{a_1,\dots,a_N}(\mc{B}^m_i)_{a_1,\dots,a_N},
}
where the final m-basis $\mc{T}^m_{ij}$ is simply the direct product of the m-basis for Lorentz structures and gauge factors. 

To get $\mc{K}_{\mc{B}}$, we first decompose $\mc{B}$ into the y-basis, for which we have a standard routine \cite{Li:2020tsi}:
%
% Since the Y-basis is build up in amplitude level naturally and the amplitude-operator correspondence, we tend to translate the operator which we are going to decompose, to helicity amplitude. They can then be further decomposed into Y-basis by following rules.
\begin{enumerate}
    \item Calculate the corresponding helicity amplitude via the follow replacements
    \begin{align}
        \psi_{i\alpha}=|i\rangle_{\alpha},\quad \psi_i^{\alpha}=\langle i|^{\alpha},\quad 
        \psi^{\dagger \dot\alpha}_i=|i]^{\dot\alpha},\quad \psi^{\dagger}_{i\dot\alpha}=[i|_{\dot\alpha},\quad iD_i^{\mu}=\frac12\langle i|^{\alpha}\sigma^{\mu}_{\alpha\dot\alpha}|i]^{\dot\alpha},\\
        F_{i\rm{L}}^{\mu\nu}=\frac12\langle i|^{\alpha} \sigma^{\mu}_{\alpha\dot\alpha}\bar\sigma^{\nu\dot{\alpha}\beta}|i\rangle_{\beta},\quad F_{i\rm{R}}^{\mu\nu}=-\frac12[i|_{\dot\alpha}\bar\sigma^{\mu\dot\alpha\alpha}\sigma^{\nu}_{\alpha\dot\beta}|i]^{\dot\beta},\quad \sigma^{\mu}_{\alpha\dot\alpha}\sigma_{\mu\beta\dot\beta}=2\epsilon_{\alpha\beta}\epsilon_{\dot\alpha\dot\beta}\, .
    \end{align}
    Other forms of building blocks can first be transformed into the above cases, for example
    \eq{
        \bar\Psi_i\gamma^\mu\Psi_j &= \psi_{iL}^\dagger \bar\sigma^\mu \psi_{jL} + \psi_{iR}^\dagger \sigma^\mu \psi_{jR} = \bra{j}\sigma^\mu|i] + \bra{i}\sigma^\mu|j], \\
        F^{\mu\nu} &= F_L^{\mu\nu} + F_R^{\mu\nu} = \frac{1}{2}\left( \bra{i}\sigma^{\mu}\bar\sigma^{\nu}\ket{i} - [i|\bar\sigma^\mu\sigma^\nu|j] \right).
    }
    Note that some conventional building blocks are usually a mixture of helicity amplitudes with different helicity assignment, which is not preferable in our construction. In some cases, it would also obscure the features of some computations, like the non-renormalization theorem \cite{Cheung:2015aba} is actually trivial from the helicity amplitude point of view. That is why we are not choosing these building blocks to construct our operators, but they can still be considered if one insists, by combining types of operators for a mixture of helicity assignment, while a given operator consisting of these building blocks would have a coordinate for a combined operator basis. We will show an example shortly.
    
    \item For any amplitudes that correspond to non-SSYT, use momentum conservation and Schouten identity to convert it towards SSYT amplitudes, \ie the y-basis. It requires a certain order of the constituting fields, and different orders would lead to different y-basis. We usually adopt the helicity-non-decreasing order proposed in \cite{Henning:2019enq}, but it is not mandatory. 
    
    The reduction can be implemented in two steps: First, use momentum conservation to replace as many momenta with lower labels in the order by those with higher labels, which include the following situation:
    \begin{itemize}
        \item Replace all the momenta of first particle by momentum conservation
            \begin{align}
                \langle i1\rangle[1j]=-\sum^N_{k=2}\langle ik\rangle[kj].\label{eq:rulep1}
            \end{align}
            It amounts to remove all the derivatives on the first field via the IBP relation. 
            \begin{align}
                \left(D^n\Psi_1\right)\cdots \simeq (-)^n\Psi_1 D^n\left(\cdots\right).
            \end{align}
            % In the sum, the terms with $k=i$ or $k=j$ would vanish, which in the corresponding operator amounts to a self-contracting building block that should be converted to other types of operators.
    
        \item Replace all the momenta of particle 2 or 3 in the following cases such that no lower label momenta would be generated,
            \eq{\label{eq:rulep3}
                & [1|p_2|i\rangle=-\sum_{k=3}^N[1|k|i\rangle,\quad \langle 1|p_2|i]=-\sum_{k=3}^N\langle 1|p_k|i],   \\
                & [1|p_3|2\rangle=-\sum_{k=4}^N[1|k|2\rangle,\quad \langle 1|p_3|2]=-\sum_{k=4}^N\langle 1|p_k|2], \\
                %            
                % p_1^2=&2\sum_{i,j\neq 1}p_i\cdot p_j=0, \\
                & p_2\cdot p_3=\sum_{\substack{i,j\neq 1\\ \{i,j\}\neq\{2,3\} }} -p_i\cdot p_j .
            }
        This is possible because on-shell conditions convert the lower label momenta generated. On the operator side, it means that EOM is involved. One could apply IBP according to eq.~\eqref{eq:rulep3} while keeping track of how some terms are converted via EOM, so that the reduction still holds at the operator level. 
        We take the type $\bar{l} l H^\dagger H D$ operators in the Warsaw basis as an example
        \eq{\label{eq:reduce_example}
            \mathcal{O}_{Hl}^{(1)} = \left(\bar{l}_p\gamma^{\mu}l_r\right)\left(H^{\dagger}i\overleftrightarrow{D}_{\mu}H\right) \leftrightarrow \mathcal{A}^{(1)}_{Hl}(l_{ri},H_j,H^{\dagger k},\bar{l}^l_p) = \delta^i_l\delta^j_k\langle 1|2-3|4] = -2\delta^i_l\delta^j_k\langle 1|3|4]
        }
        where we used momentum conservation to replace $p_2 \to -(p_1+p_3+p_4)$, and the $p_1$ and $p_4$ terms vanish by on-shell condition. On the operator side, we can do it more carefully
        \eq{
            \left(\bar{l}_p\gamma^{\mu}l_r\right)\left(H^{\dagger}i\overleftrightarrow{D}_{\mu}H\right) 
            & \stackrel{\rm IBP}{=} -2\left(\bar{l}_p\gamma^{\mu}l_r\right)\left(iD_{\mu}H^{\dagger}H\right) - iD_{\mu}\left(\bar{l}_p\gamma^{\mu}l_r\right)\left(H^{\dagger}H\right) \\
            & \stackrel{\rm EOM}{=} -2\left(\bar{l}_p\gamma^{\mu}l_r\right)\left(iD_{\mu}H^{\dagger}H\right) + \left(\bar{l}_p \mc{J}_r + \bar{\mc{J}}_p l_r\right)\left(H^{\dagger}H\right),
        }
        where the first term is our y-basis corresponding to the amplitude in eq.~\eqref{eq:reduce_example}, and for the second term we used the EOM of lepton $iD\4 l + \mc{J}=0$ for lepton source term $\bar{l}\mc{J}+h.c.$ in the Lagrangian.
        In the code, we have not kept track of these terms because it is still a purely amplitude-based algorithm. An intact reduction of operators may be implemented in future versions.
        
        \comment{
        Corresponding to the conversion of operators
        \eq{
            \Psi_1&D^2\left(\Psi_2\Psi_3\cdots\right)=\Psi_1\left(D^2\Psi_2\right)\Psi_3\left(\cdots\right)+\Psi_1\Psi_2\left(D^2\Psi_3\right)\left(\cdots\right)\\
            &+2\Psi_1\left(D^{\mu}\Psi_2\right)\left(D_{\mu}\Psi_3\right)\left(\cdots\right)+2\Psi_1\left(D^{\mu}\Psi_2\right)\Psi_3D_{\mu}\left(\cdots\right)+2\Psi_1\Psi_2\left(D^{\mu}\Psi_3\right)D_{\mu}\left(\cdots\right) \\
            & \qquad\qquad +\Psi_1\Psi_2\Psi_3D^2\left(\cdots\right),
        }
        where the terms in the first line are all convertible to other types via the EOM.}
    \end{itemize}
    
    Second, when two same-type brackets contain 4 different particles with the order $i<j<k<l$, we use the Schouten Identity to apply the following replacement
    \begin{align}
        & \langle il\rangle\langle jk\rangle = \langle ik\rangle\langle jl\rangle - \langle ij\rangle\langle kl\rangle, \label{eq:ASch}\\
        & [il][jk] = [ik][jl] - [ij][kl].\label{eq:SSch}
    \end{align}
    % In this rule, we replace the third term in eq.~(\ref{eq:ASch}-\ref{eq:SSch}) by the other two terms.
%    \item Remove derivatives on $\Psi_2$ (or $\Psi_3$) when the two spinor indices on them contract with those in building block 1 and 2, such as
%    \begin{align}
%        \Psi_{1,\alpha\dots}\left(D^{\alpha}_{\dot\alpha}D^n\Psi_2\right)^{\dots}_{\dots}\cdots &\simeq-\Psi_{1,\alpha\dots}\left(D^n\Psi_2\right)^{\dots}_{\dots}\left(D^{\alpha}_{\dot\alpha}\cdots\right),\label{eq:rulep2}\\
%        \Psi_{1,\alpha\dots}\Psi_2^{\dot\alpha\dots}\left(D^{\alpha}_{\dot\alpha}D^n\Psi_3\right)^{\dots}_{\dots}\cdots &\simeq-\Psi_{1,\alpha\dots}\Psi^{\dot\alpha\dots}_2\left(D^n\Psi_3\right)^{\dots}_{\dots}\left(D^{\alpha}_{\dot\alpha}\cdots\right).
%    \end{align}

    \item When a monomial basis other than the y-basis is defined as the standard m-basis, such as in the case that Lorentz indices rather than spinor indices are used for the building blocks, we could find the coordinates of all the m-basis operators via the above steps
    \eq{
        \mc{B}^m_i = \sum_j\mc{K}^{my}_{ij}\mc{B}^y_j,
    }
    where $\mc{K}^{my}$ should be the convertible matrix. For a given operator, we can first decompose it into the y-basis, and then use $\mc{K}^{my}$ to get the coordinate under the m-basis
    \eq{
        \mc{B} = \sum_j \mc{K}^y_j \mc{B}^y_j = \sum_{i,j} \mc{K}^y_j (\mc{K}^{my,-1})^{ji} \mc{B}^m_i \equiv \sum_i \mc{K}_{\mc{B},i} \mc{B}^m_i.
    }
\end{enumerate}

For the gauge factors, we can make use of the metric obtained in section~\ref{sec:GaugeM} and easily find the projection of a given gauge factor onto the m-basis by the inner product
\eq{
    (T^m_i,T) = \sum_j \mc{K}_{T,j} (T^m_i,T^m_j) = \sum_j g_{ij} \mc{K}_{T,j}, \quad \Rightarrow \mc{K}_{T,i} = \sum_j g^{ij} (T^m_j,T),
}
where $g^{ij}$ is the inverse metric $g^{ij}g_{jk} = \delta^i_{\ k}$, the invertibility guaranteed by the independence of the m-basis $\{T^m_i\}$.

Now that we obtain both $\mc{K}_{\mc{B}}$ and $\mc{K}_{T}$, whose direct product give a unique coordinate of the operator $\mc{O}$ (up to EOM) according to eq.~\eqref{eq:operator_decompose}. However, repeated fields have not been considered yet. In the presence of repeated fields, the m-basis itself may not be independent.

\subsection{Flavor Specified Operators and Flavor Relations}\label{sec:DePbasis}

When the flavor is turned on, there are two effects: 1. some p-basis operators must vanish due to the lack of flavors to fill in the tall Young tableau of the flavor tensor; 2. for multi-dimensional representations of the permutation group for repeated fields, only a subset of the p-basis, namely the p'-basis, is independent, while the other operators should be expressed as flavor permutations of the p'-basis. The latter leads to the flavor relations among the flavor components of the operators. We examine the two aspects with two examples.

The first example is the dimension-8 operators $H^2 H^{\dagger 2} D^4$ in Ref.~\cite{Murphy:2020rsh}. % to the corresponding p-basis of ours. 
The operators in Ref.~\cite{Murphy:2020rsh} and those generated by our package are presented respectively as
\begin{align}\label{eq:MurphyBasis}
    \begin{array}{c|c}
        \text{Ref.~\cite{Murphy:2020rsh}} & \\
        \hline
        \mathcal{O}_{H^4}^{(1)} & (D_{\mu} H^{\dagger i} D_{\nu} H_i) (D^{\nu} H^{\dagger j} D^{\mu} H_j) \\
        \mathcal{O}_{H^4}^{(2)} & (D_{\mu} H^{\dagger i} D_{\nu} H_i) (D^{\mu} H^{\dagger j} D^{\nu} H_j) \\
        \mathcal{O}_{H^4}^{(3)} & (D^{\mu} H^{\dagger i} D_{\mu} H_i) (D^{\nu} H^{\dagger j} D_{\nu} H_j) \\
    \end{array}
    \qquad
    \begin{array}{c|c}
        \text{P-basis} &  \\ 
        \hline
        \mathcal{O}_{H^2 H^{\dagger 2} D^2,1}^{(p)} & H_{i} H_{j} (D_{\mu} D_{\nu} H^{\dagger i}) (D^{\mu} D^{\nu} H^{\dagger j}) \\
        \mathcal{O}_{H^2 H^{\dagger 2} D^2,2}^{(p)} & H_{i} H^{\dagger i} (D_{\mu} D_{\nu} H_{j}) (D^{\mu} D^{\nu} H^{\dagger j}) \\
        \mathcal{O}_{H^2 H^{\dagger 2} D^2,3}^{(p)} & H_{i} (D_{\mu} H_{j}) (D_{\nu} H^{\dagger i}) (D^{\mu} D^{\nu} H^{\dagger j})
    \end{array}
\end{align}

Here we retrieved the omitted $SU(2)$ indices in Ref.~\cite{Murphy:2020rsh}. In our package, the m-basis and p-basis of type $H^2 H^{\dagger 2} D^4$ can be obtained by the function \mmaInlineCell[defined=GetBasisForType]{Code}{GetBasisForType} as
\begin{align}\label{eq:D4H4mBasis}
    \begin{array}{c|c}
        \text{M-basis} & \text{Higgs are treated as distinguishable.} \\ 
        \hline
        \mathcal{O}_1^{(m)} & H_{1i} H_{2j} (D_{\mu} D_{\nu} H_3^{\dagger i}) (D^{\mu} D^{\nu} H_4^{\dagger j}) \\
        \mathcal{O}_2^{(m)} & H_{1i} H_3^{\dagger i} (D_{\mu} D_{\nu} H_{2j}) (D^{\mu} D^{\nu} H_4^{\dagger j}) \\
        \mathcal{O}_3^{(m)} & H_{1i} (D_{\mu} H_{2j}) (D_{\nu} H_3^{\dagger i}) (D^{\mu} D^{\nu} H_4^{\dagger j}) \\
        \mathcal{O}_4^{(m)} & H_{1i} H_{2j} (D^{\mu} D^{\nu} H_4^{\dagger i}) (D_{\mu} D_{\nu} H_3^{\dagger j}) \\
        \mathcal{O}_5^{(m)} & H_{1i} H_3^{\dagger j} (D_{\mu} D_{\nu} H_{2j}) (D^{\mu} D^{\nu} H_4^{\dagger i}) \\
        \mathcal{O}_6^{(m)} & H_{1i} (D_{\mu} H_{2j}) (D^{\mu} D^{\nu} H_4^{\dagger i}) (D_{\nu} H_3^{\dagger j}) \\
    \end{array}
    \qquad
    \begin{array}{c|c}
        \text{P-basis} & \text{Higgs are treated as flavor multiplets.} \\ 
        \hline
        \mathcal{O}_1^{(p)} & \frac{1}{4}\mc{Y}\left[\tiny{\young(pr),\young(st)}\right] H_{pi} H_{rj} (D_{\mu} D_{\nu} H_s^{\dagger i}) (D^{\mu} D^{\nu} H_t^{\dagger j}) \\
        \mathcal{O}_2^{(p)} & \frac{1}{4}\mc{Y}\left[\tiny{\young(pr),\young(st)}\right] H_{pi} H_s^{\dagger i} (D_{\mu} D_{\nu} H_{rj}) (D^{\mu} D^{\nu} H_t^{\dagger j}) \\
        \mathcal{O}_3^{(p)} & \frac{1}{4}\mc{Y}\left[\tiny{\young(pr),\young(st)}\right] H_{pi} (D_{\mu} H_{rj}) (D_{\nu} H_s^{\dagger i}) (D^{\mu} D^{\nu} H_t^{\dagger j}) \\
        \mathcal{O}_4^{(p)} & \frac{1}{4}\mc{Y}\left[\tiny{\young(p,r),\young(s,t)}\right] H_{pi} H_{rj} (D_{\mu} D_{\nu} H_s^{\dagger i}) (D^{\mu} D^{\nu} H_t^{\dagger j}) \\
        \mathcal{O}_5^{(p)} & \frac{1}{4}\mc{Y}\left[\tiny{\young(p,r),\young(s,t)}\right] H_{pi} H_s^{\dagger i} (D_{\mu} D_{\nu} H_{rj}) (D^{\mu} D^{\nu} H_t^{\dagger j}) \\
        \mathcal{O}_6^{(p)} & \frac{1}{4}\mc{Y}\left[\tiny{\young(p,r),\young(s,t)}\right] H_{pi} (D_{\mu} H_{rj}) (D_{\nu} H_s^{\dagger i}) (D^{\mu} D^{\nu} H_t^{\dagger j}) \\
    \end{array}
\end{align}
with the conversion matrix
\begin{eqnarray}
	\mc{K}^{(pm)} = \left(
	\begin{array}{cccccc}
		\frac{1}{2} & 0 & 0 & \frac{1}{2} & 0 & 0 \\
		0 & \frac{1}{2} & 0 & \frac{1}{2} & \frac{1}{2} & 1 \\
		0 & 0 & \frac{1}{2} & -\frac{1}{2} & 0 & -\frac{1}{2} \\
		\frac{1}{2} & 0 & 0 & -\frac{1}{2} & 0 & 0 \\
		0 & \frac{1}{2} & 0 & -\frac{1}{2} & -\frac{1}{2} & -1 \\
		0 & 0 & \frac{1}{2} & \frac{1}{2} & 0 & \frac{1}{2} \\
	\end{array}
	\right)
\end{eqnarray}
where all Higgs are treated as distinguishable in the m-basis and are labeled with $1,2,3,4$ while they are treated as repeated flavor multiplets in the p-basis and are labeled with $p,r,s,t$. The coefficient matrix of eq.~(\ref{eq:MurphyBasis}) expanded on the m-basis eq.~(\ref{eq:D4H4mBasis}) is given by \mmaInlineCell[defined=FindMCoord]{Code}{FindMCoord} as
\beq
C^{(m)}=\left(
	\begin{array}{cccccc}
		1 & 1 & 2 & 0 & 0 & 0 \\
		1 & 0 & 0 & 0 & 0 & 0 \\
		0 & 1 & 0 & 0 & 0 & 0
	\end{array}
	\right),
\eeq
thus
\beq
C^{(p)}= C^{(m)} \left(\mc{K}^{(pm)}\right)^{-1} = \left(
	\begin{array}{cccccc}
		1 & 1 & 2 & 1 & 1 & 2 \\
		1 & 0 & 0 & 1 & 0 & 0 \\
		0 & 1 & 0 & 0 & 1 & 0
	\end{array}
	\right).
\eeq
For the SMEFT where there is only one generation of Higgs, the last 3 terms of the p-basis in eq.~(\ref{eq:D4H4mBasis}) vanish and $C^{(p)}$ becomes a square matrix.
\beq
C^{(p)}=\left(
	\begin{array}{ccc}
		1 & 1 & 2 \\
		1 & 0 & 0 \\
		0 & 1 & 0
	\end{array}
	\right), \quad
	\left(\begin{array}{c}
        \mathcal{O}^{(1)}_{H^4} \\
        \mathcal{O}^{(2)}_{H^4} \\
        \mathcal{O}^{(3)}_{H^4}
    \end{array}\right)=
    \left(\begin{array}{ccc}
        1 & 1 & 2 \\
        1 & 0 & 0 \\
        0 & 1 & 0
    \end{array}\right)
    \left(\begin{array}{c}
        \mathcal{O}^{(p)}_{H^2 H^{\dagger 2} D^2,1} \\
        \mathcal{O}^{(p)}_{H^2 H^{\dagger 2} D^2,2} \\
        \mathcal{O}^{(p)}_{H^2 H^{\dagger 2} D^2,3}
    \end{array}\right).
\eeq

Next, let's look at a more complicated example, where mixed symmetry of flavors is involved. The operators in the type $W L Q^3$ in Ref.~\cite{Murphy:2020rsh} can be rewritten in our notation as,
\begin{align}\label{eq:MurphyBasis2}
    \begin{array}{c|c}
        \text{Ref.~\cite{Murphy:2020rsh}} & \\
        \hline
        \mathcal{O}_{W L Q^3}^{(1)} & \epsilon^{abc}\epsilon^{jl} \epsilon^{km} (\tau^{I})_m^i W_{\mu \nu}^{I} \left(L_{pi} \sigma^{\mu \nu} Q_{raj}\right) \left(Q_{sbk} Q_{tcl}\right) \\
        \mathcal{O}_{W L Q^3}^{(2)} & \epsilon^{abc} \epsilon^{il} \epsilon^{km} (\tau^{I})_m^j W_{\mu \nu}^{I} \left(L_{pi} Q_{tcl}\right) \left(Q_{raj} \sigma^{\mu \nu} Q_{sbk}\right) \\
        \mathcal{O}_{W L Q^3}^{(3)} & \epsilon^{abc} \epsilon^{il} \epsilon^{km} (\tau^{I})_m^j W_{\mu \nu}^{I} \left(L_{pi} \sigma^{\mu \nu} Q_{sbk}\right) \left(Q_{raj} Q_{tcl}\right) \\
    \end{array}
\end{align}
and the p-basis of this type is listed below.
\begin{align}
    \begin{array}{c|c}
        \text{P-basis} &  \\ 
        \hline
        \mathcal{O}_{W L Q^3,1}^{(p)} & \frac{1}{6}\mathcal{Y}\left[\tiny{\young(rst)}\right] \epsilon ^{abc} \epsilon ^{il} \epsilon ^{km} \left(\tau ^I\right){}_m^j W^{I}_{\mu\nu } \left(L_{pi} \sigma^{\mu\nu } Q_{ra j}\right) \left(Q_{sb k} Q_{tc l}\right)  \\
        \mathcal{O}_{W L Q^3,2}^{(p)} & \frac{1}{6}\mathcal{Y}\left[\tiny{\young(rst)}\right] \epsilon ^{abc} \epsilon ^{il} \epsilon ^{km} \left(\tau ^I\right){}_m^j W^{I}_{\mu\nu } \left(L_{pi} Q_{tc l}\right) \left(Q_{ra j} \sigma^{\mu\nu } Q_{sb k}\right) \\
        \mathcal{O}_{W L Q^3,3}^{(p)} & \frac{1}{3}\mathcal{Y}\left[\tiny{\young(rs,t)}\right] \epsilon ^{abc} \epsilon ^{il} \epsilon ^{km} \left(\tau ^I\right){}_m^j W^{I}_{\mu\nu } \left(L_{pi} \sigma^{\mu\nu } Q_{ra j}\right) \left(Q_{sb k} Q_{tc l}\right) \\
        \mathcal{O}_{W L Q^3,4}^{(p)} & \frac{1}{3}(s \ t)\mathcal{Y}\left[\tiny{\young(rs,t)}\right] \epsilon ^{abc} \epsilon ^{il} \epsilon ^{km} \left(\tau ^I\right){}_m^j W^{I}_{\mu\nu } \left(L_{pi} \sigma^{\mu\nu } Q_{ra j}\right) \left(Q_{sb k} Q_{tc l}\right) \\
        \mathcal{O}_{W L Q^3,5}^{(p)} & \frac{1}{3}\mathcal{Y}\left[\tiny{\young(rs,t)}\right] \epsilon ^{abc} \epsilon ^{il} \epsilon ^{km} \left(\tau ^I\right){}_m^j W^{I}_{\mu\nu } \left(L_{pi} \sigma^{\mu\nu } Q_{sb k}\right) \left(Q_{ra j} Q_{tc l}\right) \\
        \mathcal{O}_{W L Q^3,6}^{(p)} & \frac{1}{3}(s \ t)\mathcal{Y}\left[\tiny{\young(rs,t)}\right] \epsilon ^{abc} \epsilon ^{il} \epsilon ^{km} \left(\tau ^I\right){}_m^j W^{I}_{\mu\nu } \left(L_{pi} \sigma^{\mu\nu } Q_{sb k}\right) \left(Q_{ra j} Q_{tc l}\right) \\
        \mathcal{O}_{W L Q^3,7}^{(p)} & \frac{1}{3}\mathcal{Y}\left[\tiny{\young(rs,t)}\right] \epsilon ^{abc} \epsilon ^{il} \epsilon ^{km} \left(\tau ^I\right){}_m^j W^{I}_{\mu\nu } \left(L_{pi} Q_{tc l}\right) \left(Q_{ra j} \sigma^{\mu\nu } Q_{sb k}\right) \\
        \mathcal{O}_{W L Q^3,8}^{(p)} & \frac{1}{3}(s \ t)\mathcal{Y}\left[\tiny{\young(rs,t)}\right] \epsilon ^{abc} \epsilon ^{il} \epsilon ^{km} \left(\tau ^I\right){}_m^j W^{I}_{\mu\nu } \left(L_{pi} Q_{tc l}\right) \left(Q_{ra j} \sigma^{\mu\nu } Q_{sb k}\right) \\
        \mathcal{O}_{W L Q^3,9}^{(p)} & \frac{1}{6}\mathcal{Y}\left[\tiny{\young(r,s,t)}\right] \epsilon ^{abc} \epsilon ^{il} \epsilon ^{km} \left(\tau ^I\right){}_m^j W^{I}_{\mu\nu } \left(L_{pi} \sigma^{\mu\nu } Q_{ra j}\right) \left(Q_{sb k} Q_{tc l}\right)
    \end{array}
\end{align}
Following the similar method above, one can find
\begin{eqnarray}
    \left(\begin{array}{c}
        \mathcal{O}^{(1)}_{W L Q^3} \\
        \mathcal{O}^{(2)}_{W L Q^3} \\
        \mathcal{O}^{(3)}_{W L Q^3}
    \end{array}\right)=
    \left(
\begin{array}{ccccccccc}
 1 & 0 & 2 & 0 & -1 & 0 & 0 & 0 & 3 \\
 0 & 1 & 0 & 0 & 0 & 0 & 1 & 0 & 0 \\
 1 & 0 & 0 & -1 & 1 & 1 & 0 & 0 & -1 \\
\end{array}
\right)
    \left(\begin{array}{c}
        \mathcal{O}^{(p)}_{W L Q^3,1} \\
        \mathcal{O}^{(p)}_{W L Q^3,2} \\
        \mathcal{O}^{(p)}_{W L Q^3,3} \\
        \mathcal{O}^{(p)}_{W L Q^3,4} \\
        \mathcal{O}^{(p)}_{W L Q^3,5} \\
        \mathcal{O}^{(p)}_{W L Q^3,6} \\
        \mathcal{O}^{(p)}_{W L Q^3,7} \\
        \mathcal{O}^{(p)}_{W L Q^3,8} \\
        \mathcal{O}^{(p)}_{W L Q^3,9}
    \end{array}\right).
\end{eqnarray}
The operators in (\ref{eq:MurphyBasis2}) do not present any information on flavor structures, although not all flavor components of the operators are independent. The flavor constraints are clear once these operators are expanded on p-basis. For each operator in (\ref{eq:MurphyBasis2}), it is straightforward to decompose the operator into different irreducible representations of $SU(n_f)$ and $S_3$ as
\begin{eqnarray}
    \mc{O}_{prst} = \frac{1}{6}\mc{Y}\left[\tiny{\young(rst)}\right] \mc{O}_{prst} + \frac{1}{3}\mc{Y}\left[\tiny{\young(rs,t)}\right] \mc{O}_{prst} + \frac{1}{3}\mc{Y}\left[\tiny{\young(rt,s)}\right] \mc{O}_{prst} + \frac{1}{6}\mc{Y}\left[\tiny{\young(r,s,t)}\right] \mc{O}_{prst}.
\end{eqnarray}
For example,
\begin{eqnarray}\label{eq:WLQ3coe}
    \left(\begin{array}{c}
        \frac{1}{6}\mc{Y}\left[\tiny{\young(rst)}\right] \mathcal{O}^{(1)}_{W L Q^3,prst} \\
        \frac{1}{3}\mc{Y}\left[\tiny{\young(rs,t)}\right] \mathcal{O}^{(1)}_{W L Q^3,prst} \\
        \frac{1}{3}\mc{Y}\left[\tiny{\young(rt,s)}\right] \mathcal{O}^{(1)}_{W L Q^3,prst} \\
        \frac{1}{6}\mc{Y}\left[\tiny{\young(r,s,t)}\right] \mathcal{O}^{(1)}_{W L Q^3,prst} \\
        \hline
        \frac{1}{6}\mc{Y}\left[\tiny{\young(rst)}\right] \mathcal{O}^{(2)}_{W L Q^3,prst} \\
        \frac{1}{3}\mc{Y}\left[\tiny{\young(rs,t)}\right] \mathcal{O}^{(2)}_{W L Q^3,prst} \\
        \frac{1}{3}\mc{Y}\left[\tiny{\young(rt,s)}\right] \mathcal{O}^{(2)}_{W L Q^3,prst} \\
        \frac{1}{6}\mc{Y}\left[\tiny{\young(r,s,t)}\right] \mathcal{O}^{(2)}_{W L Q^3,prst} \\
        \hline
        \frac{1}{6}\mc{Y}\left[\tiny{\young(rst)}\right] \mathcal{O}^{(3)}_{W L Q^3,prst} \\
        \frac{1}{3}\mc{Y}\left[\tiny{\young(rs,t)}\right] \mathcal{O}^{(3)}_{W L Q^3,prst} \\
        \frac{1}{3}\mc{Y}\left[\tiny{\young(rt,s)}\right] \mathcal{O}^{(3)}_{W L Q^3,prst} \\
        \frac{1}{6}\mc{Y}\left[\tiny{\young(r,s,t)}\right] \mathcal{O}^{(3)}_{W L Q^3,prst}
    \end{array}\right)=
    \left(
\begin{array}{ccccccccc}
\Rd1 & \Rd0 & \Rd0 & \Rd0 & \Rd0 & \Rd0 & \Rd0 & \Rd0 & \Rd0 \\
 \Rd0 & \Rd0 & \Rd2 & \Rd0 & \Rd{-1} & \Rd0 & \Rd0 & \Rd0 & \Rd0 \\
 0 & 0 & 0 & 0 & 0 & 0 & 0 & 0 & 0 \\
 \Rd0 & \Rd0 & \Rd0 & \Rd0 & \Rd0 & \Rd0 & \Rd0 & \Rd0 & \Rd3 \\
 \hline
 \Rd0 & \Rd1 & \Rd0 & \Rd0 & \Rd0 & \Rd0 & \Rd0 & \Rd0 & \Rd0 \\
 \Rd0 & \Rd0 & \Rd0 & \Rd0 & \Rd0 & \Rd0 & \Rd1 & \Rd0 & \Rd0 \\
 0 & 0 & 0 & 0 & 0 & 0 & 0 & 0 & 0 \\
 0 & 0 & 0 & 0 & 0 & 0 & 0 & 0 & 0 \\
 \hline
 1 & 0 & 0 & 0 & 0 & 0 & 0 & 0 & 0 \\
 \Rd0 & \Rd0 & \Rd0 & \Rd0 & \Rd1 & \Rd0 & \Rd0 & \Rd0 & \Rd0 \\
 \Rd0 & \Rd0 & \Rd0 & \Rd{-1} & \Rd0 & \Rd1 & \Rd0 & \Rd0 & \Rd0 \\
 0 & 0 & 0 & 0 & 0 & 0 & 0 & 0 & -1
\end{array}
\right)
    \left(\begin{array}{c}
        \mathcal{O}^{(p)}_{W L Q^3,1} \\
        \mathcal{O}^{(p)}_{W L Q^3,2} \\
        \mathcal{O}^{(p)}_{W L Q^3,3} \\
        \mathcal{O}^{(p)}_{W L Q^3,4} \\
        \mathcal{O}^{(p)}_{W L Q^3,5} \\
        \mathcal{O}^{(p)}_{W L Q^3,6} \\
        \mathcal{O}^{(p)}_{W L Q^3,7} \\
        \mathcal{O}^{(p)}_{W L Q^3,8} \\
        \mathcal{O}^{(p)}_{W L Q^3,9}
    \end{array}\right).
\end{eqnarray}
Where we used the orthogonality of different Young symmetrizers. It seems that one can choose the linear independent rows from the matrix as independent operators, for example, the red rows in eq.~(\ref{eq:WLQ3coe}). The other rows should be understood as the following flavor constraints,
\begin{eqnarray}\label{eq:flcons1}
\begin{aligned}
    \mc{Y}\left[\tiny{\young(rt,s)}\right] \mathcal{O}^{(1)}_{W L Q^3,prst} &= 0, \\
    \mc{Y}\left[\tiny{\young(rt,s)}\right] \mathcal{O}^{(2)}_{W L Q^3,prst} &= 0, \\
    \mc{Y}\left[\tiny{\young(r,s,t)}\right] \mathcal{O}^{(2)}_{W L Q^3,prst} &= 0, \\
    \mc{Y}\left[\tiny{\young(rst)}\right] \mathcal{O}^{(3)}_{W L Q^3,prst} &= \mc{Y}\left[\tiny{\young(rst)}\right] \mathcal{O}^{(1)}_{W L Q^3,prst}, \\
    \mc{Y}\left[\tiny{\young(r,s,t)}\right] \mathcal{O}^{(3)}_{W L Q^3,prst} &= -\frac{1}{3}\mc{Y}\left[\tiny{\young(r,s,t)}\right] \mathcal{O}^{(1)}_{W L Q^3,prst}.
\end{aligned}
\end{eqnarray}
However, one should remember that the following p'-basis is the truly independent and complete basis for three or more repeated fields,
\begin{align}\label{eq:WLQ3p'}
    \begin{array}{c|c}
        \text{P'-basis} &  \\ 
        \hline
        \mathcal{O}_{W L Q^3,1}^{(p')} & \frac{1}{6}\mathcal{Y}\left[\tiny{\young(rst)}\right] i\epsilon ^{abc} \epsilon ^{il} \epsilon ^{km} \left(\tau ^I\right){}_m^j W^{I}_{\mu\nu } \left(L_{pi} \sigma^{\mu\nu } Q_{ra j}\right) \left(Q_{sb k} Q_{tc l}\right)  \\
        \mathcal{O}_{W L Q^3,2}^{(p')} & \frac{1}{6}\mathcal{Y}\left[\tiny{\young(rst)}\right] i\epsilon ^{abc} \epsilon ^{il} \epsilon ^{km} \left(\tau ^I\right){}_m^j W^{I}_{\mu\nu } \left(L_{pi} Q_{tc l}\right) \left(Q_{ra j} \sigma^{\mu\nu } Q_{sb k}\right) \\
        \mathcal{O}_{W L Q^3,3}^{(p')} & \frac{1}{3}\mathcal{Y}\left[\tiny{\young(rs,t)}\right] i\epsilon ^{abc} \epsilon ^{il} \epsilon ^{km} \left(\tau ^I\right){}_m^j W^{I}_{\mu\nu } \left(L_{pi} \sigma^{\mu\nu } Q_{ra j}\right) \left(Q_{sb k} Q_{tc l}\right) \\
        \mathcal{O}_{W L Q^3,4}^{(p')} & \frac{1}{3}\mathcal{Y}\left[\tiny{\young(rs,t)}\right] i\epsilon ^{abc} \epsilon ^{il} \epsilon ^{km} \left(\tau ^I\right){}_m^j W^{I}_{\mu\nu } \left(L_{pi} \sigma^{\mu\nu } Q_{sb k}\right) \left(Q_{ra j} Q_{tc l}\right) \\
        \mathcal{O}_{W L Q^3,5}^{(p')} & \frac{1}{3}\mathcal{Y}\left[\tiny{\young(rs,t)}\right] i\epsilon ^{abc} \epsilon ^{il} \epsilon ^{km} \left(\tau ^I\right){}_m^j W^{I}_{\mu\nu } \left(L_{pi} Q_{tc l}\right) \left(Q_{ra j} \sigma^{\mu\nu } Q_{sb k}\right) \\
        \mathcal{O}_{W L Q^3,6}^{(p')} & \frac{1}{6}\mathcal{Y}\left[\tiny{\young(r,s,t)}\right] i\epsilon ^{abc} \epsilon ^{il} \epsilon ^{km} \left(\tau ^I\right){}_m^j W^{I}_{\mu\nu } \left(L_{pi} \sigma^{\mu\nu } Q_{ra j}\right) \left(Q_{sb k} Q_{tc l}\right),
    \end{array}
\end{align}
which indicates the operator $\frac{1}{3}\mc{Y}\left[\tiny{\young(rt,s)}\right] \mathcal{O}^{(3)}_{W L Q^3,prst}$ is redundant. So there is another flavor constraint:
\begin{eqnarray}\label{eq:flcons2}
    \mc{Y}\left[\tiny{\young(rt,s)}\right] \mathcal{O}^{(3)}_{W L Q^3,prst} = \frac{1}{2} \left(
    -(s \ t) \mc{Y}\left[\tiny{\young(rs,t)}\right] \mathcal{O}^{(1)}_{W L Q^3,prst} + (s \ t) \mc{Y}\left[\tiny{\young(rs,t)}\right] \mathcal{O}^{(3)}_{W L Q^3,prst} \right).
\end{eqnarray}
To conclude, after taking account of the flavor structures, the operator basis of the type $W L Q^3$ can be chosen as the p'-basis (\ref{eq:WLQ3p'}), or equivalently the operator basis (\ref{eq:MurphyBasis2}) in Ref.~\cite{Murphy:2020rsh} together with the flavor constrains eq.~(\ref{eq:flcons1}) and eq.~(\ref{eq:flcons2}).

%%%%%%%%%%%%%%%%%%%%%%%%%%%%%%%%%%%%%%%%%%%%%%%%%%%%%%%%%%%%%%
\subsection{Alternative Operator Basis}

With the technique of finding unique coordinates of any given operator, we can easily implement many tasks that were thought to be tedious. 
\begin{enumerate}
    \item We can find conversion matrices between various operator bases in the literature. Suppose there are two independent operator bases $Q^{(1)}_i$ and $Q^{(2)}_i$, both being independent and complete. We can find their coordinates under our standard m-basis
    \eq{
        Q^{(1)}_i = \sum_j \mc{K}^{(1)}_{ij}\mc{O}^m_j, \quad Q^{(2)}_i = \sum_j \mc{K}^{(2)}_{ij}\mc{O}^m_j.
    }
    The conversion matrices must be invertible due to the linear independence within each basis. Therefore we have
    \eq{
        Q^{(1)}_i = \sum_{j,k} \mc{K}^{(1)}_{ij}(\mc{K}^{(2)})^{-1}_{jk} Q^{(2)}_k
    }
    
    \item We can also find linear relations among over-complete basis. If an operator basis $Q'_I$ is over-complete, with number $n>\mathbb{d}$ the actual dimension of the operator space, the coordinates of them form a $n\times \mathbb{d}$ matrix $\mc{K}'_{Ii}$. Not only can we select an independent subset of basis from its independent rows, but we can also use the solutions to the linear equation $\sum_I c_I \mc{K}'_{Ii} = 0$ to induce the redundancy relations among the operators $\sum_I c_I Q'_I = 0$.
    
    \item We can customize our m-basis by constructing the over-complete basis with selected rules. As explained previously, in the current version of the package ABC4EFT, we obtain the m-basis operators by selecting independent ones from an over-complete basis of operators from the monomials present in the y-basis. It is also possible to build an independent basis out of particular building blocks that are more conventional or phenomenologically preferred. For example, the $F/\tilde{F}$ basis of gauge field strength is more frequently used in the literature than the chiral basis $F_{L/R}$, with relations $F = F_L + F_R$, $\tilde{F} = i(F_L - F_R)$. Note that such building blocks are in the reducible representation of the Lorentz group, which means that they generate particles with different helicities. On the other hand, our predetermined types all involve fields of only irreducible representations of the Lorentz group. Therefore it is courteous to merge several types into one bigger type,
    %It is not an operator in a single type in the sense we have been working with. However, we can build operator basis for a larger operator space, 
    which we may call ``helicity-inclusive'' type of operators, whose basis is the direct sum of operator basis for types that differ only by signs of the helicities. For instance, the type we studied in section~\ref{sec:3}, $W_L^2HH^\dagger D^2$, is in a bigger ``helicity-inclusive'' type $W^2HH^\dagger D^2$, which also includes the type $W_R^2HH^\dagger D^2$ and $W_LW_RHH^\dagger D^2$. While the basis for $W_R^2HH^\dagger D^2$ is similar to that for $W_L^2HH^\dagger D^2$ obtained in eq.~(\ref{eq:m-basisEGf}-\ref{eq:m-basisEGl}), the basis for $W_LW_RHH^\dagger D^2$ is given below
    \eq{
        & \mc{O}^{(m)}_{W_{\rm L} W_{\rm R} H H^\dagger D^2,1} = \epsilon^{IJK}(\tau^K)^i_j W_{\rm L}^I{}_{\rho}{}^{\mu} W_{\rm R}^{J\rho\nu}(D_\mu D_\nu H^{\dagger j}) H_i \\
        & \mc{O}^{(m)}_{W_{\rm L} W_{\rm R} H H^\dagger D^2,2} = W_{\rm L}^I{}_{\rho}{}^{\mu} W_{\rm R}^{I\rho\nu}(D_\mu D_\nu H^{\dagger i}) H_i
    }
    After taking into account that $W$ boson has only one flavor, we obtain the p-basis for the ``helicity-inclusive'' type $W^2H^2D^2$:
    \eq{\label{eq:W2H2D2-basis}
        \mc{O}_{W^2H^2D^2}^p{}_i = \{\mc{O}^{(m)}_{W_{\rm L}^2 H H^\dagger D^2,1},\mc{O}^{(m)}_{W_{\rm L}^2 H H^\dagger D^2,4},\mc{O}^{(m)}_{W_{\rm R}^2 H H^\dagger D^2,1},\mc{O}^{(m)}_{W_{\rm R}^2 H H^\dagger D^2,4},\mc{O}^{(m)}_{W_{\rm L} W_{\rm R} H H^\dagger D^2,1},\mc{O}^{(m)}_{W_{\rm L} W_{\rm R} H H^\dagger D^2,2} \}
    }
    Now we can construct all kinds of operators involving $W_{\mu\nu}$ or $\tilde{W}_{\mu\nu}$ and work out their coordinates under the basis eq.~\eqref{eq:W2H2D2-basis} by the expansion $W=W_L+W_R$ and $\tilde{W}=i(W_L-W_R)$, so that an independent set of 6 operators can be selected as an operator basis. It can also be used to examine the basis present in the literature, such as Ref.~\cite{Murphy:2020rsh}
    \eq{
    & Q_{W^2H^2D^2}^{(1)} = (D^\mu H^\dagger D^\nu H) W^I_{\mu\rho}W^{I\rho}_\nu = -\frac14\mc{O}_{W^2H^2D^2}^p{,}_1 -\frac14\mc{O}_{W^2H^2D^2}^p{,}_3 + \mc{O}_{W^2H^2D^2}^p{,}_6 \\
    & Q_{W^2H^2D^2}^{(2)} = (D^\mu H^\dagger D_\mu H) W^I_{\nu\rho}W^{I\nu\rho} = \mc{O}_{W^2H^2D^2}^p{,}_1 + \mc{O}_{W^2H^2D^2}^p{,}_3 \\
    & Q_{W^2H^2D^2}^{(3)} = (D^\mu H^\dagger D_\mu H) W^I_{\nu\rho}\tilde{W}^{I\nu\rho} = i(\mc{O}_{W^2H^2D^2}^p{,}_1 - \mc{O}_{W^2H^2D^2}^p{,}_3) \\
    & Q_{W^2H^2D^2}^{(4)} =i\epsilon^{IJK}\left(D^{\mu}H^{\dagger}\tau^ID^{\nu}H\right)W^J_{\mu\rho}W_{\nu}^{K\rho}= i(\mc{O}_{W^2H^2D^2}^p{,}_2 + \mc{O}_{W^2H^2D^2}^p{,}_4 + 2\mc{O}_{W^2H^2D^2}^p{,}_5) \\
    & Q_{W^2H^2D^2}^{(5)} =\epsilon^{IJK}\left(D^{\mu}H^{\dagger}\tau^ID^{\nu}H\right)\left(W^J_{\mu\rho}\tilde{W}_{\nu}^{K\rho}-\tilde{W}^J_{\mu\rho}W_{\nu}^{K\rho}\right)= -2i\mc{O}_{W^2H^2D^2}^p{,}_5 \\
    & Q_{W^2H^2D^2}^{(6)} =i\epsilon^{IJK}\left(D^{\mu}H^{\dagger}\tau^ID^{\nu}H\right)\left(W^J_{\mu\rho}\tilde{W}_{\nu}^{K\rho}+\tilde{W}^J_{\mu\rho}W_{\nu}^{K\rho}\right)= -2(\mc{O}_{W^2H^2D^2}^p{,}_2 - \mc{O}_{W^2H^2D^2}^p{,}_4) \\
    }
    It's easy to check that the conversion matrix is invertible, proving that the 6 operators are indeed a complete basis for the type $W^2H^2D^2$.
\end{enumerate}

\comment{
From  $\{T^{m}_i\}$ and $g_{ij}$
 one can define the dual m-basis $\{\tilde{T}^{m}_i\}$ by:
 \begin{eqnarray}
     \tilde{T}^{m}_i = \sum_{j}(g)^{-1}_{ij}T^{m}_j,
 \end{eqnarray}
such that $(\tilde{T}^{(m)}_i,T^{m}_j)=\delta_{ij}$.
Since m-basis is an independent and complete basis, any gauge factor  $T$ of the same type must be able to express as a linear combination of the m-basis gauge factors $T=\sum_i c_i T^{m}_i$, while the coordinates $c_j$ can be obtained by taking the inner product with the corresponding dual basis $\tilde{T}^{m}_i$:
\begin{eqnarray}\label{eq:projonm}
    c_i=(\tilde{T}^{m}_i,T).
\end{eqnarray}

For gauge factors, we first read out the group factors from the operator by replacing the contraction of the same gauge indices between fields with $\delta$ of different indices and restoring the all the indices of generator matrices whenever there is a trace structure.
For example, $H_iH^{\dagger i}$ will be replaced by $\delta^i_jH_iH^{\dagger j}$, and the corresponding group factor is $\delta^i_j$, ${\rm Tr}(WWWW)$ with $W\equiv W^I(\tau^I)_i^j$ will be replaced by $(\tau^I)_i^j(\tau^J)_j^k(\tau^K)_k^l(\tau^L)_l^iW^IW^JW^KW^L$, and the corresponding group factor is $(\tau^I)_i^j(\tau^J)_j^k(\tau^K)_k^l(\tau^L)_l^i$. The conversion to the m-basis gauge factors is obtained by taking the inner product with the dual m-basis vector as discussed in eq.~\eqref{eq:projonm}:
\begin{eqnarray}
    c_i=(\tilde{T}^{m}_i,T).\nonumber
\end{eqnarray}

We take the Warsaw's basis in type $\bar{l}lH^{\dagger}HD$ as an example,
\begin{align}
    \left\{\begin{array}{l}
        \mathcal{O}_{Hl}^{(1)} = \left(\bar{l}_p\gamma^{\mu}l_r\right)\left(H^{\dagger}i\overleftrightarrow{D}_{\mu}H\right) \\
        \mathcal{O}_{Hl}^{(3)} = \left(\bar{l}_p\tau^I\gamma^{\mu}l_r\right)\left(H^{\dagger}i\tau^I\overleftrightarrow{D}_{\mu}H\right)
    \end{array}\right. \Rightarrow
    \left\{\begin{array}{l}
        \mathcal{A}^{(1)}_{Hl}(l_{ri},H_j,H^{\dagger k},\bar{l}^l_p)=\delta^i_l\delta^j_k\langle 1|2-3|4] \\
        \mathcal{A}^{(3)}_{Hl}(l_{ri},H_j,H^{\dagger k},\bar{l}^l_p)= \left(\tau^I\right)^i_l\left(\tau^I\right)^j_k\langle 1|2-3|4] 
    \end{array}\right. ,\label{eq:warsaw}\\
    \begin{array}{c}
        \text{Operator} \\
        \text{Y-basis:}
    \end{array}\left\{\begin{array}{l}
        \mathcal{O}_{1}^{(y)} = \left(\bar{l}_p\gamma^{\mu}l_r\right)\left(iD_{\mu}H^{\dagger}H\right) \\
        \mathcal{O}_{2}^{(y)} = \left(\bar{l}^i_p\gamma^{\mu}l_{rj}\right)\left(iD_{\mu}H^{\dagger j}H_i\right)
    \end{array}\right. \Rightarrow
    \left\{\begin{array}{l}
        \mathcal{A}^{(y)}_{1}(l_{ri},H_j,H^{\dagger k},\bar{l}^l_p)=\delta^i_l\delta^j_k\langle 1|3|4] \\
        \mathcal{A}^{(y)}_{2}(l_{ri},H_j,H^{\dagger k},\bar{l}^l_p)= \delta^i_k\delta^j_l\langle 1|3|4] 
    \end{array}\right. .
\end{align}
Apply eq.~\ref{eq:rulep2} and fierz identity to eq.~\ref{eq:warsaw},
\begin{align}
    \mathcal{A}^{(1)}_{Hl}(l_{ri},H_j,H^{\dagger k},\bar{l}^l_p)=&2\delta^i_l\delta^j_k\langle 1|3|4]=2\mathcal{A}^{(y)}_{1},\\
    \mathcal{A}^{(3)}_{Hl}(l_{ri},H_j,H^{\dagger k},\bar{l}^l_p)=&\left(\delta^i_k\delta^j_l-\frac12\delta^i_l\delta^j_k\right)\langle 1|3|4]=-\frac12 \mathcal{A}^{(y)}_{1}+\mathcal{A}^{(y)}_{2}.
\end{align}

}

\comment{
%=============================================================================
\subsection{Decomposition to M-basis}\label{sec:DeMbasis}
We can use the method above to reduce any complete base $\{\mathcal{O}_i\}$, as well as M-basis $\{\mathcal{O}^{(m)}_i\}$.
\begin{align}
    \mathcal{O}_i=\sum_j\mathcal{C}^{y}_{ij}\mathcal{O}^{(y)}_j,\quad
    \mathcal{O}^{(m)}_i=\sum_j\mathcal{K}^{my}_{ij}\mathcal{O}^{(y)}_j.
\end{align}
Then the coefficient matrix of base $\{\mathcal{O}_i\}$ to M-basis is obtained as below
\begin{align}
    \mathcal{O}_i=\sum_j C^m_{ij}\mathcal{O}^m_j,\quad C^m_{ij}=\sum_k C^y_{ik}\left(\mathcal{K}^{my}\right)^{-1}_{kj}.
\end{align}
We give an example of transformation from Warsaw's basis to our M-basis in type $\bar{l}lH^{\dagger}HD$.
\begin{align}
    \begin{array}{c|c}
        \text{Warsaw} &  \\
        \hline
        \mathcal{O}_{Hl}^{(1)} & \left(\bar{l}_p\gamma^{\mu}l_r\right)\left(H^{\dagger}i\overleftrightarrow{D}_{\mu}H\right) \\
        \mathcal{O}_{Hl}^{(3)} & \left(\bar{l}_p\tau^I\gamma^{\mu}l_r\right)\left(H^{\dagger}i\tau^I\overleftrightarrow{D}_{\mu}H\right) \\
    \end{array}
    \qquad
    \begin{array}{c|c}
        \text{M-basis} &  \\
        \hline
        \mathcal{O}_1^{(m)} & \left(\bar{l}_p^i\gamma^{\mu}l_{ri}\right)\left(H^{\dagger j}iD_{\mu}H_j\right) \\
        \mathcal{O}_2^{(m)} & \left(\bar{l}^i_p\tau^I\gamma^{\mu}l_{rj}\right)\left(H^{\dagger j}i\tau^ID_{\mu}H_i\right) \\
    \end{array}
\end{align}
The solution is below.
\begin{align}
    \left(\begin{array}{c}
        \mathcal{O}^{(1)}_{Hl} \\
        \mathcal{O}^{(3)}_{Hl}
    \end{array}\right)=
    \left(\begin{array}{cc}
        2 & 0 \\
        -2 & 4
    \end{array}\right)
    \left(\begin{array}{c}
        \mathcal{O}^{(m)}_{1} \\
        \mathcal{O}^{(m)}_{2}
    \end{array}\right).
\end{align}
}

\subsection{Reduction of Over-complete Basis}\label{sec:OverCombasis}

With the complete m-basis, it is able to expand any over-complete basis of any type on the corresponding m-basis and reduce the over-complete basis to its complete subset. Taking the type $D^2 L L^{\dagger} Q Q^{\dagger}$ as an example, it is straightforward to write down the following over-completed operators based on the properties of $\gamma$ matrices,
\begin{align}\label{eq:overcompletebasis}
	\begin{array}{c|c}
		D^2 L L^{\dagger} Q Q^{\dagger} & \\
		\hline
		\mathcal{O}^{(1)} & (D_{\mu} L_{pi} D^{\mu} Q_{raj}) (L^{\dagger}_s{}^i Q^{\dagger}_t{}^{aj}) \\
		\mathcal{O}^{(2)} & (D_{\mu} L_{pi} D^{\mu} Q_{raj}) (L^{\dagger}_s{}^j Q^{\dagger}_t{}^{ai}) \\
		\mathcal{O}^{(3)} & (D_{\mu} L_{pi} Q_{raj}) (D^{\mu} L^{\dagger}_s{}^i Q^{\dagger}_t{}^{aj}) \\
		\mathcal{O}^{(4)} & (D_{\mu} L_{pi} Q_{raj}) (D^{\mu} L^{\dagger}_s{}^j Q^{\dagger}_t{}^{ai}) \\
		\mathcal{O}^{(5)} & (D_{\mu} L_{pi} D_{\nu} Q_{raj}) (L^{\dagger}_s{}^i \bar{\sigma}^{\mu\nu} Q^{\dagger}_t{}^{aj}) \\
		\mathcal{O}^{(6)} & (D_{\mu} L_{pi} D_{\nu} Q_{raj}) (L^{\dagger}_s{}^j \bar{\sigma}^{\mu\nu} Q^{\dagger}_t{}^{ai}) \\
		\mathcal{O}^{(7)} & (D^{\nu} L_{pi} \sigma_{\mu} L^{\dagger}_s{}^i) (D^{\mu} Q_{raj} \sigma_{\nu} Q^{\dagger}_t{}^{aj}) \\
		\mathcal{O}^{(8)} & (D^{\nu} L_{pi} \sigma_{\mu} Q^{\dagger}_t{}^{aj}) (D^{\mu} Q_{raj} \sigma_{\nu} L^{\dagger}_s{}^i) \\
		\mathcal{O}^{(9)} & (D^{\mu} L_{pi} \sigma_{\mu} L^{\dagger}_s{}^i) (D^{\nu} Q_{raj} \sigma_{\nu} Q^{\dagger}_t{}^{aj}). \\
	\end{array}
\end{align}
Each of eq.~(\ref{eq:overcompletebasis}) can be expanded on m-basis
\begin{align}
	\begin{array}{c|c}
		\text{M-basis} & \\ 
		\hline
		\mathcal{O}_1^{(m)} & (L_{pi} Q_{raj}) (D_{\mu} L^{\dagger}_s{}^i D^{\mu} Q^{\dagger}_t{}^{aj}) \\
		\mathcal{O}_2^{(m)} & (L_{pi} \sigma_{\mu\nu} Q_{raj}) (D^{\mu} L^{\dagger}_s{}^i D^{\nu} Q^{\dagger}_t{}^{aj}) \\
		\mathcal{O}_3^{(m)} & (L_{pi} Q_{raj}) (D_{\mu} L^{\dagger}_s{}^j D^{\mu} Q^{\dagger}_t{}^{ai}) \\
		\mathcal{O}_4^{(m)} & (L_{pi} \sigma_{\mu\nu} Q_{raj}) (D^{\mu} L^{\dagger}_s{}^j D^{\nu} Q^{\dagger}_t{}^{ai}) \\
	\end{array}
\end{align}
Using \mmaInlineCell[defined=FindMCoord]{Code}{FindMCoord}, and the coefficient matrix is obtained as
\begin{eqnarray}\label{eq:coematrix}
	\left(
	\begin{array}{cccc}
		\color{red}{1} & \color{red}{0} & \color{red}{0} & \color{red}{0} \\
		\color{red}{0} & \color{red}{0} & \color{red}{1} & \color{red}{0} \\
		\color{red}{-\frac{1}{2}} & \color{red}{\frac{1}{2}} & \color{red}{0} & \color{red}{0} \\
		\color{red}{0} & \color{red}{0} & \color{red}{-\frac{1}{2}} & \color{red}{\frac{1}{2}} \\
		0 & 1 & 0 & 0 \\
		0 & 0 & 0 & 1 \\
		-1 & -1 & 0 & 0 \\
		1 & -1 & 0 & 0 \\
		0 & 0 & 0 & 0 \\
	\end{array}
	\right),
\end{eqnarray}
where the coefficients in the last row being all 0 is the consequence of $\mathcal{O}^{(9)}$ corresponding to the EOM redundancy. From eq.~(\ref{eq:coematrix}) one can choose 4 independent rows of the matrix as a complete basis of $D^2 L L^{\dagger} Q Q^{\dagger}$, for example, the 4 rows which are marked red in eq.~(\ref{eq:coematrix}).

Here we give another example of the dim-9 type $d_{_\mathbb{C}}^{\dagger} e_{_\mathbb{C}} L^{\dagger} Q u^2_{_\mathbb{C}}$, it is straightforward to write down the following basis:
\begin{align}\label{eq:overcompletebasis2}
	\begin{array}{c|c}
		d_{_\mathbb{C}}^{\dagger} e_{_\mathbb{C}} L^{\dagger} Q u^2_{_\mathbb{C}} & \\
		\hline
		\mathcal{O}^{(1)} & ( d_{_\mathbb{C}}^{\dagger}{}_{ub} L^{\dagger}_v{}^i ) ( e_{_\mathbb{C}}{}_p Q_{rai} ) ( u_{_\mathbb{C}}{}_s^a u_{_\mathbb{C}}{}_t^b )\\
		\mathcal{O}^{(2)} & ( d_{_\mathbb{C}}^{\dagger}{}_{ub} L^{\dagger}_v{}^i ) ( e_{_\mathbb{C}}{}_p Q_{rai} ) ( u_{_\mathbb{C}}{}_s^b u_{_\mathbb{C}}{}_t^a ) \\
		\mathcal{O}^{(3)} & ( d_{_\mathbb{C}}^{\dagger}{}_{ub} L^{\dagger}_v{}^i ) ( e_{_\mathbb{C}}{}_p u_{_\mathbb{C}}{}_s^a ) ( Q_{rai} u_{_\mathbb{C}}{}_t^b ) \\
		\mathcal{O}^{(4)} & ( d_{_\mathbb{C}}^{\dagger}{}_{ub} L^{\dagger}_v{}^i ) ( e_{_\mathbb{C}}{}_p u_{_\mathbb{C}}{}_s^b ) ( Q_{rai} u_{_\mathbb{C}}{}_t^a ) \\
		\mathcal{O}^{(5)} & ( d_{_\mathbb{C}}^{\dagger}{}_{ub} L^{\dagger}_v{}^i ) ( e_{_\mathbb{C}}{}_p u_{_\mathbb{C}}{}_t^b ) ( Q_{rai} u_{_\mathbb{C}}{}_s^a ) \\
		\mathcal{O}^{(6)} & ( d_{_\mathbb{C}}^{\dagger}{}_{ub} L^{\dagger}_v{}^i ) ( e_{_\mathbb{C}}{}_p u_{_\mathbb{C}}{}_t^a ) ( Q_{rai} u_{_\mathbb{C}}{}_s^b ) \\
		\mathcal{O}^{(7)} & ( e_{_\mathbb{C}}{}_p \sigma_{\mu} d_{_\mathbb{C}}^{\dagger}{}_{ub} ) ( Q_{rai} \sigma^{\mu} L^{\dagger}_v{}^i ) ( u_{_\mathbb{C}}{}_s^a u_{_\mathbb{C}}{}_t^b ) \\
		\mathcal{O}^{(8)} & ( e_{_\mathbb{C}}{}_p \sigma_{\mu} d_{_\mathbb{C}}^{\dagger}{}_{ub} ) ( Q_{rai} \sigma^{\mu} L^{\dagger}_v{}^i ) ( u_{_\mathbb{C}}{}_s^b u_{_\mathbb{C}}{}_t^a ) \\
		\mathcal{O}^{(9)} & ( e_{_\mathbb{C}}{}_p \sigma_{\mu} d_{_\mathbb{C}}^{\dagger}{}_{ub} ) ( u_{_\mathbb{C}}{}_s^a \sigma^{\mu} L^{\dagger}_v{}^i ) ( Q_{rai} u_{_\mathbb{C}}{}_t^b ) \\
		\mathcal{O}^{(10)} & ( e_{_\mathbb{C}}{}_p \sigma_{\mu} d_{_\mathbb{C}}^{\dagger}{}_{ub} ) ( u_{_\mathbb{C}}{}_s^b \sigma^{\mu} L^{\dagger}_v{}^i ) ( Q_{rai} u_{_\mathbb{C}}{}_t^a ) \\
		\mathcal{O}^{(11)} & ( e_{_\mathbb{C}}{}_p \sigma_{\mu} d_{_\mathbb{C}}^{\dagger}{}_{ub} ) ( u_{_\mathbb{C}}{}_t^b \sigma^{\mu} L^{\dagger}_v{}^i ) ( u_{_\mathbb{C}}{}_s^a Q_{rai} ) \\
		\mathcal{O}^{(12)} & ( e_{_\mathbb{C}}{}_p \sigma_{\mu} d_{_\mathbb{C}}^{\dagger}{}_{ub} ) ( u_{_\mathbb{C}}{}_t^a \sigma^{\mu} L^{\dagger}_v{}^i ) ( u_{_\mathbb{C}}{}_s^b Q_{rai} ) \\
		\mathcal{O}^{(13)} & ( Q_{rai} \sigma_{\mu} d_{_\mathbb{C}}^{\dagger}{}_{ub} ) ( u_{_\mathbb{C}}{}_s^a \sigma^{\mu} L^{\dagger}_v{}^i ) ( e_{_\mathbb{C}}{}_p u_{_\mathbb{C}}{}_t^b ) \\
		\mathcal{O}^{(14)} & ( Q_{rai} \sigma_{\mu} d_{_\mathbb{C}}^{\dagger}{}_{ub} ) ( u_{_\mathbb{C}}{}_s^b \sigma^{\mu} L^{\dagger}_v{}^i ) ( e_{_\mathbb{C}}{}_p u_{_\mathbb{C}}{}_t^a ) \\
		\mathcal{O}^{(15)} & ( Q_{rai} \sigma_{\mu} d_{_\mathbb{C}}^{\dagger}{}_{ub} ) ( u_{_\mathbb{C}}{}_t^b \sigma^{\mu} L^{\dagger}_v{}^i ) ( e_{_\mathbb{C}}{}_p u_{_\mathbb{C}}{}_s^a ) \\
		\mathcal{O}^{(16)} & ( Q_{rai} \sigma_{\mu} d_{_\mathbb{C}}^{\dagger}{}_{ub} ) ( u_{_\mathbb{C}}{}_t^a \sigma^{\mu} L^{\dagger}_v{}^i ) ( e_{_\mathbb{C}}{}_p u_{_\mathbb{C}}{}_s^b ) \\
		\mathcal{O}^{(17)} & ( u_{_\mathbb{C}}{}_s^a \sigma_{\mu} d_{_\mathbb{C}}^{\dagger}{}_{ub} ) ( u_{_\mathbb{C}}{}_t^b \sigma^{\mu} L^{\dagger}_v{}^i ) ( e_{_\mathbb{C}}{}_p Q_{rai} ) \\
		\mathcal{O}^{(18)} & ( u_{_\mathbb{C}}{}_s^b \sigma_{\mu} d_{_\mathbb{C}}^{\dagger}{}_{ub} ) ( u_{_\mathbb{C}}{}_t^a \sigma^{\mu} L^{\dagger}_v{}^i ) ( e_{_\mathbb{C}}{}_p Q_{rai} ) \\
		\mathcal{O}^{(19)} & ( d_{_\mathbb{C}}^{\dagger}{}_{ub} L^{\dagger}_v{}^i ) ( e_{_\mathbb{C}}{}_p \sigma_{\mu\nu} Q_{rai} ) ( u_{_\mathbb{C}}{}_s^a \sigma^{\mu\nu} u_{_\mathbb{C}}{}_t^b )\\
		\mathcal{O}^{(20)} & ( d_{_\mathbb{C}}^{\dagger}{}_{ub} L^{\dagger}_v{}^i ) ( e_{_\mathbb{C}}{}_p \sigma_{\mu\nu} Q_{rai} ) ( u_{_\mathbb{C}}{}_s^b \sigma^{\mu\nu} u_{_\mathbb{C}}{}_t^a ) \\
		\mathcal{O}^{(21)} & ( d_{_\mathbb{C}}^{\dagger}{}_{ub} L^{\dagger}_v{}^i ) ( e_{_\mathbb{C}}{}_p \sigma_{\mu\nu} u_{_\mathbb{C}}{}_s^a ) ( Q_{rai} \sigma^{\mu\nu} u_{_\mathbb{C}}{}_t^b ) \\
		\mathcal{O}^{(22)} & ( d_{_\mathbb{C}}^{\dagger}{}_{ub} L^{\dagger}_v{}^i ) ( e_{_\mathbb{C}}{}_p \sigma_{\mu\nu} u_{_\mathbb{C}}{}_s^b ) ( Q_{rai} \sigma^{\mu\nu} u_{_\mathbb{C}}{}_t^a ) \\
		\mathcal{O}^{(23)} & ( d_{_\mathbb{C}}^{\dagger}{}_{ub} L^{\dagger}_v{}^i ) ( e_{_\mathbb{C}}{}_p \sigma_{\mu\nu} u_{_\mathbb{C}}{}_t^b ) ( Q_{rai} \sigma^{\mu\nu} u_{_\mathbb{C}}{}_s^a ) \\
		\mathcal{O}^{(24)} & ( d_{_\mathbb{C}}^{\dagger}{}_{ub} L^{\dagger}_v{}^i ) ( e_{_\mathbb{C}}{}_p \sigma_{\mu\nu} u_{_\mathbb{C}}{}_t^a ) ( Q_{rai} \sigma^{\mu\nu} u_{_\mathbb{C}}{}_s^b )
	\end{array}
\end{align}
The basis is obviously over-complete due to Fierz identities. The independent flavor-blind basis given in our code is the m-basis
\begin{align}
	\begin{array}{c|c}
		\text{M-basis} & \\ 
		\hline
		\mathcal{O}_1^{(m)} & ( d_{_\mathbb{C}}^{\dagger}{}_{ub} L^{\dagger}_v{}^i ) ( e_{_\mathbb{C}}{}_p Q_{rai} ) ( u_{_\mathbb{C}}{}_s^a u_{_\mathbb{C}}{}_t^b )\\
		\mathcal{O}_2^{(m)} & ( d_{_\mathbb{C}}^{\dagger}{}_{ub} L^{\dagger}_v{}^i ) ( e_{_\mathbb{C}}{}_p u_{_\mathbb{C}}{}_s^a ) ( Q_{rai} u_{_\mathbb{C}}{}_t^b )\\
		\mathcal{O}_3^{(m)} & ( d_{_\mathbb{C}}^{\dagger}{}_{ub} L^{\dagger}_v{}^i ) ( e_{_\mathbb{C}}{}_p Q_{rai} ) ( u_{_\mathbb{C}}{}_s^b u_{_\mathbb{C}}{}_t^a )\\
		\mathcal{O}_4^{(m)} & ( d_{_\mathbb{C}}^{\dagger}{}_{ub} L^{\dagger}_v{}^i ) ( e_{_\mathbb{C}}{}_p u_{_\mathbb{C}}{}_s^b ) ( Q_{rai} u_{_\mathbb{C}}{}_t^a )\\
	\end{array}
\end{align}
The function \mmaInlineCell[defined=FindMCoord]{Code}{FindMCoord} can be used to find the coefficients of each operator in (\ref{eq:overcompletebasis2}) expanded on m-basis. The result is as following
\begin{eqnarray}\label{eq:coematrix2}
	\left(
	\begin{array}{cccc}
		\color{red}{1} & \color{red}{0} & \color{red}{0} & \color{red}{0} \\
		\color{red}{0} & \color{red}{0} & \color{red}{1} & \color{red}{0} \\
		\color{red}{0} & \color{red}{1} & \color{red}{0} & \color{red}{0} \\
		\color{red}{0} & \color{red}{0} & \color{red}{0} & \color{red}{1} \\
		-1 & -1 & 0 & 0 \\
		0 & 0 & -1 & -1 \\
		2 & 0 & 0 & 0 \\
		0 & 0 & 2 & 0 \\
		0 & 2 & 0 & 0 \\
		0 & 0 & 0 & 2 \\
		-2 & -2 & 0 & 0 \\
		0 & 0 & -2 & -2 \\
		-2 & -2 & 0 & 0 \\
		0 & 0 & -2 & -2 \\
		0 & 2 & 0 & 0 \\
		0 & 0 & 0 & 2 \\
		2 & 0 & 0 & 0 \\
		0 & 0 & 2 & 0 \\
		4 & 8 & 0 & 0 \\
		0 & 0 & 4 & 8 \\
		8 & 4 & 0 & 0 \\
		0 & 0 & 8 & 4 \\
		4 & -4 & 0 & 0 \\
		0 & 0 & 4 & -4 \\
	\end{array}
	\right).
\end{eqnarray}
One can check that the above matrix has rank 4 and the independent operators of $d_{_\mathbb{C}}^{\dagger} e_{_\mathbb{C}} L^{\dagger} Q u^2_{_\mathbb{C}}$ can be chosen as 4 independent rows of matrix (\ref{eq:coematrix2}), for example, the 4 red rows in (\ref{eq:coematrix2}).

\section{Brief Introduction on ABC4EFT}\label{sec:5}

In this section, first we will show how to include a user-defined model in the Mathematica package, then we will introduce the Mathematica functions to analyze EFT operators and give the complete basis of the EFT operators in such a model. To load the package, one should first locate the directory of the package ABC4EFT, and on this directory, one starts with
\begin{mmaCell}[defined=ABC4EFT]{Input}
  << ABC4EFT\`
\end{mmaCell}

%\subsection{Model Input and Main Functions}
The package allows the users to define a model involving various fields, gauge groups, and global groups. Here is an example of writing a function to define the SMEFT.
\begin{mmaCell}[pattern={model_,nf_,model,nf},defined={SetAttributes,Module,ModelIni,AddGroup,GaugeBoson,AddField,Flavor,Dim,Hermitian}]{Code}
  (* Define SMEFT *)
  SetAttributes[DefSMEFT, HoldFirst];
  DefSMEFT[model_,nf_:3]:= Module[{},
  ModelIni[model];
  AddGroup[model,"U1b"];
  AddGroup[model,"U1l"];
  AddGroup[model,"SU3c",GaugeBoson->"G"];
  AddGroup[model,"SU2w",GaugeBoson->"W"];
  AddGroup[model,"U1y",GaugeBoson->"B"];
  AddField[model,"Q",-1/2,{"SU3c"->{1,0},"SU2w"->{1},"U1y"->1/6,"U1b"->1/3},Flavor->nf];
  AddField[model,"uc",-1/2,{"SU3c"->{0,1},"U1y"->-2/3,"U1b"->-1/3},Flavor->nf];
  AddField[model,"dc",-1/2,{"SU3c"->{0,1},"U1y"->1/3,"U1b"->-1/3},Flavor->nf];
  AddField[model,"L",-1/2,{"SU2w"->{1},"U1y"->-1/2,"U1l"->1},Flavor->nf];
  AddField[model,"ec",-1/2,{"U1y"->1,"U1l"->-1},Flavor->nf];
  AddField[model,"H",0,{"SU2w"->{1},"U1y"->1/2}]
  ]
  DefSMEFT[SMEFT];
\end{mmaCell}
Where the inputs of this function are the name of the model and the number of generations of fermions that would be taken into account in the model. The user-defined function \mmaInlineCell{Code}{DefSMEFT} invokes the functions \mmaInlineCell[defined=ModelIni]{Code}{ModelIni}, \mmaInlineCell[defined=AddGroup]{Code}{AddGroup} and \mmaInlineCell[defined=AddField]{Code}{AddField} in the package, where \mmaInlineCell[defined=ModelIni]{Code}{ModelIni} just initializes the model, and \mmaInlineCell[defined=AddGroup]{Code}{AddGroup}, \mmaInlineCell[defined=AddField]{Code}{AddField} will be introduced in the following content.  

The function \mmaInlineCell[defined=AddGroup]{Code}{AddGroup} adds a global group or a gauge group to the model, along with options that can name the corresponding gauge boson if the added group is gauge and assign the certain list of indices to a group.
\begin{mmaCell}[pattern={model_,groupname_String},defined={Options,AddGroup,GaugeBoson,Index,OptionsPattern}]{Code}
  Options[AddGroup]={GaugeBoson->None,Index->"default"};
  AddGroup[model_,groupname_String,OptionsPattern[]]
\end{mmaCell}
It should be noted that the input \mmaInlineCell[pattern=groupname_String]{Code}{groupname_String} must be a string that formed by the commonly used name of the group, for example, \mmaInlineCell{Code}{"SU3"}, \mmaInlineCell{Code}{"SU2"} and \mmaInlineCell{Code}{"U1"}, and a extra char to label a certain group, for example, \mmaInlineCell{Code}{"b"}, \mmaInlineCell{Code}{"l"}, \mmaInlineCell{Code}{"c"}, \mmaInlineCell{Code}{"w"} and \mmaInlineCell{Code}{"y"}, in order to make the code recognize and load the corresponding group. For now, the package only contains the group profile of $SU(3)$, $SU(2)$ and $U(1)$, and we will give a instruction of how to add more groups in appendix.

The function \mmaInlineCell[defined=AddField]{Code}{AddField} adds a field to the model. The inputs of \mmaInlineCell[defined=AddField]{Code}{AddField} are the name of the model, name of the added field in string, helicity of the field, list of representations of the field under each group in the model and some options, which include number of flavor generations, mass dimension, Hermiticity and chirality of the field.
\begin{mmaCell}[pattern={model_,field_String,hel_,Greps_List},defined={Flavor,Dim,Hermitian,Options,Chirality,AddField,OptionsPattern}]{Code}
  Options[AddField]={Flavor->1,Dim->"default",Hermitian->False,Chirality->{}};
  AddField[model_,field_String,hel_,Greps_List,OptionsPattern[]]
\end{mmaCell}
There are a few things about the options that are worth discussing. As mentioned before, a massless chiral field in most models satisfies the relation $|h|=d-1$,
where $h$ and $d$ are helicity and mass dimension of the field, while graviton, with $|h|=2$ and $d=2$, does not satisfy the relation. So it is necessary to include the option that specifies the mass dimension of a field \mmaInlineCell[defined=Dim]{Code}{Dim}. Another thing is that the Hermitian conjugate of a field will be automatically added to the model if the option  \mmaInlineCell[defined=Hermitian]{Code}{Hermitian} is set to \mmaInlineCell{Code}{False}.

We also introduce the most important functions in this section, and leave some auxiliary functions that one may be of interest in Appendix C. 

The function \mmaInlineCell[pattern={model_,dim_},defined=StatResult]{Code}{StatResult[model_,dim_]} presents the statistic result of a model at certain mass dimension. For example,
\begin{mmaCell}[defined={SMEFT,StatResult}]{Code}
  StatResult[SMEFT,8];
\end{mmaCell}
\begin{mmaCell}{Print}
  Done! time used: 0.2870018
  number of real types->541
  number of real terms->1266
  number of real operators->44807
\end{mmaCell}

The function \mmaInlineCell[pattern={model_,dim_},defined=AllTypesC]{Input}{AllTypesC[model_,dim_]} presents all complex types in a model at certain mass dimension. For example, all complex types in SMEFT at mass dimension 6 can be obtained by
\begin{mmaCell}[defined={AllTypesC,SMEFT}]{Input}
  AllTypesC[SMEFT,6]
\end{mmaCell}
\begin{mmaCell}[]{Output}
  <|\mmaSup{FL}{3}->\big\{\mmaSup{BL}{3},BL\mmaSup{WL}{2},\mmaSup{WL}{3},BL\mmaSup{GL}{2},\mmaSup{GL}{3}\big\},\mmaSup{\(\pmb{\psi}\)}{4}->\big\{dcec\mmaSup{uc}{2},ecLQuc,dc\mmaSup{Q}{2}uc,L\mmaSup{Q}{3}\big\},
  FL\(\pmb{\phi}\)\mmaSup{\(\pmb{\psi}\)}{2}->\big\{BLecH\(\pmb{\dagger}\)L,BLdcH\(\pmb{\dagger}\)Q,BLHQuc,ecH\(\pmb{\dagger}\)LWL,dcH\(\pmb{\dagger}\)QWL,HQucWL,dcGLH\(\pmb{\dagger}\)Q,GLHQuc\big\},
  \mmaSup{FL}{2}\mmaSup{\(\pmb{\phi}\)}{2}->\big\{\mmaSup{BL}{2}HH\(\pmb{\dagger}\),BLHH\(\pmb{\dagger}\)WL,HH\(\pmb{\dagger}\)\mmaSup{WL}{2},\mmaSup{GL}{2}HH\(\pmb{\dagger}\)\big\},\mmaSup{\(\pmb{\psi}\)}{2}\mmaSup{\(\pmb{\psi}\)\(\pmb{\dagger}\)}{2}->\big\{\mmaSup{ec}{2}\mmaSup{ec\(\pmb{\dagger}\)}{2},ecec\(\pmb{\dagger}\)LL\(\pmb{\dagger}\),dc\(\pmb{\dagger}\)ecLQ\(\pmb{\dagger}\),dcdc\(\pmb{\dagger}\)LL\(\pmb{\dagger}\),
  LL\(\pmb{\dagger}\)ucuc\(\pmb{\dagger}\),ecec\(\pmb{\dagger}\)QQ\(\pmb{\dagger}\),dcec\(\pmb{\dagger}\)L\(\pmb{\dagger}\)Q,\mmaSup{L}{2}\mmaSup{L\(\pmb{\dagger}\)}{2},dcdc\(\pmb{\dagger}\)ecec\(\pmb{\dagger}\),ecec\(\pmb{\dagger}\)ucuc\(\pmb{\dagger}\),\mmaSup{dc}{2}\mmaSup{dc\(\pmb{\dagger}\)}{2},dcdc\(\pmb{\dagger}\)ucuc\(\pmb{\dagger}\),\mmaSup{uc}{2}\mmaSup{uc\(\pmb{\dagger}\)}{2},
  ec\mmaSup{Q\(\pmb{\dagger}\)}{2}uc,dcL\(\pmb{\dagger}\)Q\(\pmb{\dagger}\)uc,dc\(\pmb{\dagger}\)LQuc\(\pmb{\dagger}\),ec\(\pmb{\dagger}\)\mmaSup{Q}{2}uc\(\pmb{\dagger}\),LL\(\pmb{\dagger}\)QQ\(\pmb{\dagger}\),dcdc\(\pmb{\dagger}\)QQ\(\pmb{\dagger}\),QQ\(\pmb{\dagger}\)ucuc\(\pmb{\dagger}\),\mmaSup{Q}{2}\mmaSup{Q\(\pmb{\dagger}\)}{2}\big\},
  D\mmaSup{\(\pmb{\phi}\)}{2}\(\pmb{\psi}\)\(\pmb{\psi}\)\(\pmb{\dagger}\)->\big\{Decec\(\pmb{\dagger}\)HH\(\pmb{\dagger}\),DHH\(\pmb{\dagger}\)LL\(\pmb{\dagger}\),Ddc\mmaSup{H\(\pmb{\dagger}\)}{2}uc\(\pmb{\dagger}\),Ddcdc\(\pmb{\dagger}\)HH\(\pmb{\dagger}\),DHH\(\pmb{\dagger}\)ucuc\(\pmb{\dagger}\),Ddc\(\pmb{\dagger}\)\mmaSup{H}{2}uc,DHH\(\pmb{\dagger}\)QQ\(\pmb{\dagger}\)\big\},
  \mmaSup{D}{2}\mmaSup{\(\pmb{\phi}\)}{4}->\big\{\mmaSup{D}{2}\mmaSup{H}{2}\mmaSup{H\(\pmb{\dagger}\)}{2}\big\},\mmaSup{\(\pmb{\phi}\)}{3}\mmaSup{\(\pmb{\psi}\)}{2}->\big\{ecH\mmaSup{H\(\pmb{\dagger}\)}{2}L,dcH\mmaSup{H\(\pmb{\dagger}\)}{2}Q,\mmaSup{H}{2}H\(\pmb{\dagger}\)Quc\big\},\mmaSup{\(\pmb{\phi}\)}{6}->\big\{\mmaSup{H}{3}\mmaSup{H\(\pmb{\dagger}\)}{3}\big\}|>
\end{mmaCell}

The function \mmaInlineCell[pattern={model_,dim_},defined={GenerateOperatorList}]{Code}{GenerateOperatorList[model_,dim_]} presents all independent operators in a model at certain mass dimension as monomial p-basis. For example,
\begin{mmaCell}[defined={SMEFT,GenerateOperatorList}]{Input}
  GenerateOperatorList[SMEFT,5]
\end{mmaCell}
\begin{mmaCell}{Print}
  Generating types of operators ...
  Time spent: 0.1098517
\end{mmaCell}
\begin{mmaCell}{Output}
  <|\mmaSup{\(\pmb{\phi}\)}{2}\mmaSup{\(\pmb{\psi}\)}{2}-><|\mmaSup{H}{2}\mmaSup{L}{2}-><|\{L->\{2\},H->\{2\}\}->\big\{\mmaSup{\(\pmb{\epsilon}\)}{ik}\mmaSup{\(\pmb{\epsilon}\)}{jl}\mmaSub{H}{k}\mmaSub{H}{l}\big(\mmaSub{L}{pi}\mmaSub{L}{rj}\big)\big\}|>|>|>
\end{mmaCell}
The output presents the only EFT operator $LLHH$ in the SMEFT at mass dimension 5, with the flavor structure of the two $L$s and the two $H$s symmetric. 

The function \mmaInlineCell[pattern={model_,type_},defined={GetBasisForType,OptionsPattern}]{Code}{GetBasisForType[model_,type_,OptionsPattern[]]} gives the m-basis and p-basis of a type in a model. The inputs are a user-defined model and a type formed by derivatives, fields and their Hermitian conjugates in the model. For example, the model could be the SMEFT and the type could be \mmaInlineCell{Input}{\mmaSup{"D"}{4}\mmaSup{"H"}{2}\mmaSup{"H†"}{2}}.
\begin{mmaCell}[defined={SMEFT,GetBasisForType,OptionsPattern}]{Input}
  GetBasisForType[SMEFT,\mmaSup{"D"}{4}\mmaSup{"H"}{2}\mmaSup{"H†"}{2}]
\end{mmaCell}
\begin{mmaCell}[defined={SMEFT,GetBasisForType,OptionsPattern}]{Output}
  <|"basis"->\big\{\mmaSub{H}{i}\mmaSub{H}{j}(\mmaSub{D}{\(\pmb{\mu}\)}\mmaSub{D}{\(\pmb{\nu}\)}\mmaSup{H\(\pmb{\dagger}\)}{i})(\mmaSup{D}{\(\pmb{\mu}\)}\mmaSup{D}{\(\pmb{\nu}\)}\mmaSup{H\(\pmb{\dagger}\)}{j}),\mmaSub{H}{i}\mmaSup{H\(\pmb{\dagger}\)}{i}(\mmaSub{D}{\(\pmb{\mu}\)}\mmaSub{D}{\(\pmb{\nu}\)}\mmaSub{H}{j})(\mmaSup{D}{\(\pmb{\mu}\)}\mmaSup{D}{\(\pmb{\nu}\)}\mmaSup{H\(\pmb{\dagger}\)}{j}),
  \mmaSub{H}{i}(\mmaSub{D}{\(\pmb{\mu}\)}\mmaSub{H}{j})(\mmaSub{D}{\(\pmb{\nu}\)}\mmaSup{H\(\pmb{\dagger}\)}{i})(\mmaSup{D}{\(\pmb{\mu}\)}\mmaSup{D}{\(\pmb{\nu}\)}\mmaSup{H\(\pmb{\dagger}\)}{j}),\mmaSub{H}{i}\mmaSub{H}{j}(\mmaSub{D}{\(\pmb{\mu}\)}\mmaSub{D}{\(\pmb{\nu}\)}\mmaSup{H\(\pmb{\dagger}\)}{j})(\mmaSup{D}{\(\pmb{\mu}\)}\mmaSup{D}{\(\pmb{\nu}\)}\mmaSup{H\(\pmb{\dagger}\)}{i}),\mmaSub{H}{i}\mmaSup{H\(\pmb{\dagger}\)}{j}(\mmaSub{D}{\(\pmb{\mu}\)}\mmaSub{D}{\(\pmb{\nu}\)}\mmaSub{H}{j})(\mmaSup{D}{\(\pmb{\mu}\)}\mmaSup{D}{\(\pmb{\nu}\)}\mmaSup{H\(\pmb{\dagger}\)}{i}),
  \mmaSub{H}{i}(\mmaSub{D}{\(\pmb{\mu}\)}\mmaSub{H}{j})(\mmaSub{D}{\(\pmb{\nu}\)}\mmaSup{H\(\pmb{\dagger}\)}{j})(\mmaSup{D}{\(\pmb{\mu}\)}\mmaSup{D}{\(\pmb{\nu}\)}\mmaSup{H\(\pmb{\dagger}\)}{i})\big\},"p-basis"-><|\{H->\{2\},H\(\pmb{\dagger}\)->\{2\}\}->\big\{\big\{\mmaFrac{1}{2},0,0,\mmaFrac{1}{2},0,0\big\},
  \big\{0,\mmaFrac{1}{2},0,\mmaFrac{1}{2},\mmaFrac{1}{2},1\big\},\big\{0,0,\mmaFrac{1}{2},-\mmaFrac{1}{2},0,-\mmaFrac{1}{2}\big\}\big\}|>|>
\end{mmaCell}
Here the p-basis operators are flavor-specified operators presented by linear combinations of flavor-blind operators in m-basis \mmaInlineCell[]{Output}{"basis"}. The result in \mmaInlineCell[]{Output}{"p-basis"} presents the symmetry of flavor indices of the repeated fields in this type, which in this case are the Higgs bosons, and the coefficients of each p-basis operator expanded on the flavor-blind m-basis. The function \mmaInlineCell[defined={GetBasisForType}]{Input}{GetBasisForType} can also give the monomial p-basis after de-symmetrization by adding a option \mmaInlineCell[defined={DeSym}]{Input}{DeSym->True}.
\begin{mmaCell}[defined={SMEFT,GetBasisForType,OptionsPattern,DeSym}]{Input}
  GetBasisForType[SMEFT,\mmaSup{"D"}{4}\mmaSup{"H"}{2}\mmaSup{"H†"}{2},DeSym->True]
\end{mmaCell}
\begin{mmaCell}[defined={SMEFT,GetBasisForType,OptionsPattern}]{Output}
  <|\{H->\{2\},H\(\pmb{\dagger}\)->\{2\}\}->\big\{\mmaSub{H}{i}\mmaSub{H}{j}(\mmaSub{D}{\(\pmb{\mu}\)}\mmaSub{D}{\(\pmb{\nu}\)}\mmaSup{H\(\pmb{\dagger}\)}{i})(\mmaSup{D}{\(\pmb{\mu}\)}\mmaSup{D}{\(\pmb{\nu}\)}\mmaSup{H\(\pmb{\dagger}\)}{j}),\mmaSub{H}{i}\mmaSup{H\(\pmb{\dagger}\)}{i}(\mmaSub{D}{\(\pmb{\mu}\)}\mmaSub{D}{\(\pmb{\nu}\)}\mmaSub{H}{j})(\mmaSup{D}{\(\pmb{\mu}\)}\mmaSup{D}{\(\pmb{\nu}\)}\mmaSup{H\(\pmb{\dagger}\)}{j}),
  \mmaSub{H}{i}(\mmaSub{D}{\(\pmb{\mu}\)}\mmaSub{H}{j})(\mmaSub{D}{\(\pmb{\nu}\)}\mmaSup{H\(\pmb{\dagger}\)}{i})(\mmaSup{D}{\(\pmb{\mu}\)}\mmaSup{D}{\(\pmb{\nu}\)}\mmaSup{H\(\pmb{\dagger}\)}{j})\big\}|>
\end{mmaCell}

The function \mmaInlineCell[pattern={model_,operator_},defined={FindYCoord,OptionsPattern}]{Code}{FindYCoord[model_,operator_,OptionsPattern[]]} will reduce any operator to our y-basis and obtain the coordinate. The input format of the operator is similar to that of the \textbf{FeynRule} for more acceptably. Except that fields and indices are entered in string format. We show how to input fields and tensor structures in Tab.~\ref{tab:inputSM}. We list some of examples in Tab.~\ref{tab:inputcases}. 
\begin{table}
\centering
    \begin{tabular}{|c|p{4cm}||c|l|}
    \hline
        fields & input format & group tensors & input format \\
        \hline
        $e_{_\mathbb{C}}{}_p$ & \mmaInlineCell[]{Input}{"ec"["p"]} & $\delta^i_j$ & \mmaInlineCell[]{Input}{del2["i","j"]}\\
        $e^{\dagger}_{_\mathbb{C} p}$ & \mmaInlineCell[]{Input}{"ec†"["p"]} & $\delta^a_b$ & \mmaInlineCell[]{Input}{del3["a","b"]}\\
        $L_{pi}$ & \mmaInlineCell[]{Input}{"L"["i","p"]}  & $\delta^{IJ}$ & \mmaInlineCell[]{Input}{del3n["I","J"]} \\
        $L^{\dagger i}_p$ & \mmaInlineCell[]{Input}{"L†"["i","p"]} & $\delta^{AB}$ & \mmaInlineCell[]{Input}{del8n["A","B"]} \\
        $Q_{pai}$ & \mmaInlineCell[]{Input}{"Q"["i","p","a"]} & $\left(\tau^I\right)^i_j$ & \mmaInlineCell[]{Input}{\(\pmb{\tau}\)["I","j","i"]}\\
        $Q^{\dagger ai}_p$ & \mmaInlineCell[]{Input}{"Q†"["i","p","a"]} & $\left(\lambda^A\right)^a_b$ & \mmaInlineCell[]{Input}{\(\pmb{\lambda}\)["A","b","a"]}\\
        $u_{_\mathbb{C}}{}_p^a$ & \mmaInlineCell[]{Input}{"uc"["p","a"]} & $\epsilon^{ij}$ & \mmaInlineCell[]{Input}{eps2a["i","j"]}\\
        $u^{\dagger}_{_\mathbb{C} pa}$ & \mmaInlineCell[]{Input}{"uc†"["p","a"]} & $\epsilon_{ij}$ & \mmaInlineCell[]{Input}{eps2f["i","j"]}\\
        $d_{_\mathbb{C}}{}_p^a$ & \mmaInlineCell[]{Input}{"dc"["p","a"]} & $\epsilon^{IJK}$ & \mmaInlineCell[]{Input}{eps3n["I","J","K"]}\\
        $d^{\dagger}_{_\mathbb{C} pa}$ & \mmaInlineCell[]{Input}{"dc†"["p","a"]} & $\epsilon^{abc}$ & \mmaInlineCell[]{Input}{eps3a["a","b","c"]}\\
        $B_{L\mu\nu}$ or $B_{L}^{\mu\nu}$ & \mmaInlineCell[]{Input}{FS["BL","\(\pmb{\mu}\)","\(\pmb{\nu}\)"]} & $\epsilon_{abc}$ & \mmaInlineCell[]{Input}{eps3f["a","b","c"]}\\
        $B_{R\mu\nu}$ or $B_{R}^{\mu\nu}$ & \mmaInlineCell[]{Input}{FS["BR","\(\pmb{\mu}\)","\(\pmb{\nu}\)"]} & $f^{ABC}$ & \mmaInlineCell[]{Input}{fabc["A","B","C"]}\\
        $W^I_{L\mu\nu}$ or $W_{L}^{I\mu\nu}$ & \mmaInlineCell[]{Input}{FS["WL","\(\pmb{\mu}\)","\(\pmb{\nu}\)","I"]} & $d^{ABC}$ & \mmaInlineCell[]{Input}{dabc["A","B","C"]}\\
        $W^I_{R\mu\nu}$ or $W_{R}^{I\mu\nu}$ & \mmaInlineCell[]{Input}{FS["WR","\(\pmb{\mu}\)","\(\pmb{\nu}\)","I"]} & $\sigma^{\mu}$ or $\sigma_{\mu}$ & \mmaInlineCell[]{Input}{sigma["\(\pmb{\mu}\)"]}\\
        $G^A_{L\mu\nu}$ or $G_{L}^{A\mu\nu}$ & \mmaInlineCell[]{Input}{FS["GL","\(\pmb{\mu}\)","\(\pmb{\nu}\)","A"]} & $\bar\sigma^{\mu}$ or $\bar\sigma_{\mu}$ & \mmaInlineCell[]{Input}{sigmab["\(\pmb{\mu}\)"]}\\
        $G^A_{R\mu\nu}$ or $G_{R}^{A\mu\nu}$ & \mmaInlineCell[]{Input}{FS["GR","\(\pmb{\mu}\)","\(\pmb{\nu}\)","A"]} & $\gamma^{\mu}$ or $\gamma_{\mu}$ & \mmaInlineCell[]{Input}{Ga["\(\pmb{\mu}\)"]}\\
        $H_i$ & \mmaInlineCell[]{Input}{"H"["i"]} & $\sigma^{\mu\nu}$ or $\sigma_{\mu\nu}$  & \mmaInlineCell[]{Input}{sigmaT["\(\pmb{\mu}\)","\(\pmb{\nu}\)"]}\\
        $H^{\dagger i}$ & \mmaInlineCell[]{Input}{"H†"["i"]} & $\bar\sigma^{\mu\nu}$ or $\bar\sigma_{\mu\nu}$ & \mmaInlineCell[]{Input}{sigmabT["\(\pmb{\mu}\)","\(\pmb{\nu}\)"]} \\
        \hline
    \end{tabular}
    \caption{In the function \textbf{FindYCoord}, the input format for the SM fields and tensor structures}
    \label{tab:inputSM}
\end{table}

\begin{table}
\centering
    \begin{tabular}{c|c}
        building blocks & input format \\
        \hline
        $D_{\mu}e_{_\mathbb{C} p}$ & \mmaInlineCell[]{Input}{DC["ec"["p"],"mu"]} \\
        $L^{\dagger i}_p\bar\sigma^{\mu}L_{ri}$ & \mmaInlineCell[]{Input}{Dot["L†"["p","i"],sigmab["mu"],"L"["r","j"]] del2["j","i"]}\\
        \hline
        \multirow{2}*{$G^A_{R\mu\nu} Q^{\dagger ai}_p\bar\sigma^{\mu\nu}\left(\lambda^A\right)_a^b d^{\dagger}_{_\mathbb{C} pb}H_i$} & \mmaInlineCell[]{Input}{FS["GR","mu","nu","A"] "H"["i"] \(\pmb{\lambda}\)["A","a","b"] del2["i","j"]} \\ 
        & \mmaInlineCell[]{Input}{*Dot["Q†"["j","p","a"],sigmabT["mu","nu"],"dc†"["r","b"]]}
    \end{tabular}
    \caption{Some examples for the input format.}
    \label{tab:inputcases}
\end{table}

\begin{mmaCell}[defined={SMEFT,FindYCoord,OptionsPattern}]{Input}
  FindYCoord[SMEFT,del2["i","k"] del2["j","l"] DC[DC["H†"["k"],"mu"],"nu"]\\ DC[DC["H†"["l"],"mu"],"nu"] "H"["i"] "H"["j"]]
\end{mmaCell}
\begin{mmaCell}[defined={SU2,SU3,Lor,TProduct,type}]{Output}
  <|SU2->\{1, 0\}, SU3->\{1\}, Lor->\Big\{\mmaFrac{1}{4}, 0, 0\Big\}, TProduct->\Big\{\mmaFrac{1}{4}, 0, 0, 0, 0, 0\Big\}, type->\mmaSup{H}{2}\mmaSup{H†}{2}\mmaSup{D}{4} |>
\end{mmaCell}
Where the solutions associated to \mmaInlineCell[]{Input}{"SU2"}, \mmaInlineCell[]{Input}{"SU3"}, and  \mmaInlineCell[]{Input}{"Lor"} are the coordinates in gauge and Lorentz Y-basis respectively, and the result associated to \mmaInlineCell[]{Input}{"TProduct"} is the tensor product of the first three terms. We also provide the function \mmaInlineCell[pattern={model_,operator_},defined={FindMCoord,OptionsPattern}]{Code}{FindMCoord[model_,operator_,OptionsPattern[]]} to reduce any operator to our M-basis and P-basis. The input format is the same as \mmaInlineCell[pattern={model_,operator_},defined={FindYCoord,OptionsPattern}]{Code}{FindYCoord[model_,operator_,OptionsPattern[]]}.

The function \mmaInlineCell[pattern={model_,type_,partition_},defined={GetJBasisForType}]{Code}{GetJBasisForType[model_,type_,partition_]} presents all possible eigenvalues of $W^2_{\mathcal{I}}$ and gauge representations of a type in a model for a channel $\mathcal{I}$, along with the corresponding eigenstates as J-basis operators. The inputs are the model, the type and certain channel. For example, consider the type $H H^{\dagger} W_{\rm R}^2 D^2$ in SMEFT. Labeling the fields as $H_1, H^{\dagger}_2, W_{\rm R}{}_3, W_{\rm R}{}_4$, the J-basis of channel $\{H_1,W_{\rm R}{}_3\} \rightarrow \{H^{\dagger}_2,W_{\rm R}{}_4\}$ can be obtained as follows.
\begin{mmaCell}[defined={GetJBasisForType,SMEFT}]{Input}
  GetJBasisForType[SMEFT,"H""H\(\pmb{\dagger}\)"\mmaSup{"WR"}{2}\mmaSup{"D"}{2},\{\{1,3\},\{2,4\}\}]
\end{mmaCell}
\begin{mmaCell}[]{Output}
  <|"basis"->\big\{\mmaSub{H}{i}\mmaSup{H\(\pmb{\dagger}\)}{i}\big(\mmaSup{D}{\(\pmb{\mu}\)}\mmaSup{WR}{I\(\pmb{\lambda}\)\(\pmb{\nu}\)}\big)\big(\mmaSub{D}{\(\pmb{\mu}\)}\mmaSubSup{WR}{ \(\pmb{\nu}\)\(\pmb{\lambda}\)}{I}\big),\mmaSub{H}{i}\mmaSubSup{WR}{ \(\pmb{\nu}\)\(\pmb{\lambda}\)}{I}\big(\mmaSub{D}{\(\pmb{\mu}\)}\mmaSup{H\(\pmb{\dagger}\)}{i}\big)\big(\mmaSup{D}{\(\pmb{\mu}\)}\mmaSup{WR}{I\(\pmb{\lambda}\)\(\pmb{\nu}\)}\big),\mmaSubSup{\(\pmb{\tau}\)}{ j}{Ki}\mmaSup{\(\pmb{\epsilon}\)}{IJK}\mmaSub{H}{i}\mmaSup{H\(\pmb{\dagger}\)}{j}\big(\mmaSub{D}{\(\pmb{\mu}\)}\mmaSubSup{WR}{ \(\pmb{\nu}\)\(\pmb{\lambda}\)}{I}\big)\big(\mmaSup{D}{\(\pmb{\mu}\)}\mmaSup{WR}{J\(\pmb{\lambda}\)\(\pmb{\nu}\)}\big),
  \mmaSubSup{\(\pmb{\tau}\)}{ j}{Ki}\mmaSup{\(\pmb{\epsilon}\)}{IJK}\mmaSub{H}{i}\mmaSubSup{WR}{ \(\pmb{\nu}\)\(\pmb{\lambda}\)}{I}\big(\mmaSub{D}{\(\pmb{\mu}\)}\mmaSup{H\(\pmb{\dagger}\)}{j}\big)\big(\mmaSup{D}{\(\pmb{\mu}\)}\mmaSup{WR}{J\(\pmb{\lambda}\)\(\pmb{\nu}\)}\big)\big\},"groups"->\{SU3c,SU2w,Spin\},
  "j-basis"->\{<|\{\mmaSub{H}{1},\mmaSub{WR}{3}\}->\{\{0,0\},\{3\},2\},\{\mmaSub{H\(\pmb{\dagger}\)}{2},\mmaSub{WR}{4}\}->\{\{0,0\},\{3\},2\}|>->\{\{8i,-6i,-4,3\}\},
  <|\{\mmaSub{H}{1},\mmaSub{WR}{3}\}->\{\{0,0\},\{3\},1\},\{\mmaSub{H\(\pmb{\dagger}\)}{2},\mmaSub{WR}{4}\}->\{\{0,0\},\{3\},1\}|>->\{\{0,-2i,0,1\}\},
  <|\{\mmaSub{H}{1},\mmaSub{WR}{3}\}->\{\{0,0\},\{1\},2\},\{\mmaSub{H\(\pmb{\dagger}\)}{2},\mmaSub{WR}{4}\}->\{\{0,0\},\{1\},2\}|>->\{\{-4i,3i,-4,3\}\},
  <|\{\mmaSub{H}{1},\mmaSub{WR}{3}\}->\{\{0,0\},\{1\},1\},\{\mmaSub{H\(\pmb{\dagger}\)}{2},\mmaSub{WR}{4}\}->\{\{0,0\},\{1\},1\}|>->\{\{0,i,0,1\}\}\}|>
\end{mmaCell}
%expand on y/m-basis?

\section{Conclusion}\label{sec:6}

In this work, we present a general procedure to construct the independent and complete operator bases for generic Lorentz invariant EFTs, which is implemented into a publicly available Mathematica package: ABC4EFT (Amplitude Basis Construction for Effective Field Theories). This package provides the unified construction of Lorentz structure, gauge structure, and flavor structure by the Young tensor method, based on the on-shell amplitude operator correspondence. Our procedure can be applied to any EFT with the Lorentz symmetry and any gauge symmetry and any field content, and generate the complete and independent basis up to any mass dimension. Compared to our previous work, we improved the algorithm for finding the m-basis gauge factors with the metric tensor method; we accelerated the method for finding the p-basis coordinate and performing the de-symmetrization by directly constructing the representation matrix for Young symmetrizers, we also added the new routine for finding the j-basis operators.

This Young tensor basis provides a modern view of the EFT operators from the on-shell perspective. It has many advantages: 1) As the mass dimension goes higher, the Feynman rules of effective operators become very complicated, especially for the operators with repeated fields, while the corresponding amplitude basis can be obtained much more easily and the tree-level bootstrap of on-shell amplitudes has the potential to massively simplify the calculations. 2) The contributions of operators to helicity amplitudes are manifest, providing insights for analysis of phase space and amplitude interference. 3) The on-shell basis will also benefit the renormalization group calculation of the higher dimensional operators. 

In principle, one can write down many different kinds of bases in an EFT framework. These different bases are related by relations such as the equation of motion, the Fierz identify, etc. Our on-shell amplitude construction provides routine for reducing any operator into our standard basis. This provides automatic conversions among different bases, making it possible to compare results from various literature.   

%many different kinds of relations,  due to redun Any operator could convert to this on-shell basis. 

Although in this paper we only demonstrate this package using the SMEFT operators as an example, it could be easily applied to other EFTs including, but not limited to, the LEFT, the $\nu$SMEFT, the two Higgs doublet model EFT, the left-right symmetric EFT, and the dark matter EFT, etc. Future developments will also include EFT with soft particles like the non-linear sigma models.
One needs to modify the field contents and gauge structure in the package to generate the on-shell amplitude basis for these EFTs. The package ABC4EFT can be downloaded from the HEPForge website~\footnote{https://abc4eft.hepforge.org}.

\section*{Acknowledgments} 

J.H.Y. is supported by the National Science Foundation of China under Grants No. 12022514, No. 11875003 and No. 12047503, and National Key Research and Development Program of China Grant No. 2020YFC2201501, No. 2021YFA0718304, and CAS Project for Young Scientists in Basic Research YSBR-006, the Key Research Program of the CAS Grant No. XDPB15. 
M.-L.X. is supported in part by the U.S. Department of Energy under contracts No. DE-AC02-06CH11357 at Argonne and No.DE-SC0010143 at Northwestern.
H.-L.L. is supported by F.R.S.-FNRS through the IISN convention "Theory of Fundamental Interactions” (N : 4.4517.08).

\appendix

\section{A Non-trivial Example for Finding Gauge M-basis}\label{app:1}
Here we present a non-trivial example for finding the gauge m-basis from candidates, where the number of m-basis candidates is larger than that of independent ones. The type of operator we will analyze is $G_L^AG_L^BG_L^CG_L^D$. 
We can convert each field to the one with fundamental indices only:
\begin{eqnarray}
   && (G)_{a_1a_2a_3}=(\lambda^A)_{a_2}^a\epsilon_{aa_1a_3}G_L^A\sim \young({{a_1}}{{a_2}},{{a_3}})\\
    && (G)_{b_1b_2b_3}=(\lambda^B)_{b_2}^b\epsilon_{bb_1b_3}G_L^B\sim \young({{b_1}}{{b_2}},{{b_3}})\\
    && (G)_{c_1c_2c_3}=(\lambda^C)_{c_2}^c\epsilon_{cc_1c_3}G_L^C\sim \young({{c_1}}{{c_2}},{{c_3}})\\
    && (G)_{d_1d_2d_3}=(\lambda^D)_{d_2}^d\epsilon_{dd_1d_3}G_L^D\sim \young({{d_1}}{{d_2}},{{d_3}})
\end{eqnarray}
The overall conversion factor is:
\begin{eqnarray}
    CF=(\lambda^A)_{a_2}^a\epsilon_{aa_1a_3}(\lambda^B)_{b_2}^a\epsilon_{bb_1b_3}(\lambda^C)_{c_2}^c\epsilon_{cc_1c_3}(\lambda^D)_{d_2}^d\epsilon_{dd_1d_3}
\end{eqnarray}
The corresponding singlet Young tableaux and y-basis becomes is: 
\begin{eqnarray}
   T^y_1=\epsilon^{a_1a_3c_2}\epsilon^{a_2b_3c_3}\epsilon^{b_1c_1d_2}\epsilon^{b_2d_1d_3}\sim\young({{a_1}}{{a_2}}{{b_1}}{{b_2}},{{a_3}}{{b_3}}{{c_1}}{{d_1}},{{c_2}}{{c_3}}{{d_2}}{{d_3}})\quad 
T^y_2=\epsilon^{a_1a_3c_2}\epsilon^{a_2b_2c_3}\epsilon^{b_1b_3d_2}\epsilon^{c_1d_1d_3}\sim\young({{a_1}}{{a_2}}{{b_1}}{{c_1}},{{a_3}}{{b_2}}{{b_3}}{{d_1}},{{c_2}}{{c_3}}{{d_2}}{{d_3}})\\ 
T^y_3=\epsilon^{a_1a_3b_3}\epsilon^{a_2c_1c_3}\epsilon^{b_1c_2d_2}\epsilon^{b_2d_1d_3}\sim\young({{a_1}}{{a_2}}{{b_1}}{{b_2}},{{a_3}}{{c_1}}{{c_2}}{{d_1}},{{b_3}}{{c_3}}{{d_2}}{{d_3}})\quad 
T^y_4=\epsilon^{a_1a_3b_2}\epsilon^{a_2b_1b_3}\epsilon^{c_1c_3d_2}\epsilon^{c_2d_1d_3}\sim\young({{a_1}}{{a_2}}{{c_1}}{{c_2}},{{a_3}}{{b_1}}{{c_3}}{{d_1}},{{b_2}}{{b_3}}{{d_2}}{{d_3}})\\ 
T^y_5=\epsilon^{a_1a_3b_3}\epsilon^{a_2b_2c_3}\epsilon^{b_1c_2d_2}\epsilon^{c_1d_1d_3}\sim\young({{a_1}}{{a_2}}{{b_1}}{{c_1}},{{a_3}}{{b_2}}{{c_2}}{{d_1}},{{b_3}}{{c_3}}{{d_2}}{{d_3}})\quad 
T^y_6=\epsilon^{a_1a_3b_3}\epsilon^{a_2b_2c_2}\epsilon^{b_1c_3d_2}\epsilon^{c_1d_1d_3}\sim\young({{a_1}}{{a_2}}{{b_1}}{{c_1}},{{a_3}}{{b_2}}{{c_3}}{{d_1}},{{b_3}}{{c_2}}{{d_2}}{{d_3}})\\ 
T^y_7=\epsilon^{a_1a_3b_2}\epsilon^{a_2b_3c_3}\epsilon^{b_1c_2d_2}\epsilon^{c_1d_1d_3}\sim\young({{a_1}}{{a_2}}{{b_1}}{{c_1}},{{a_3}}{{b_3}}{{c_2}}{{d_1}},{{b_2}}{{c_3}}{{d_2}}{{d_3}})\quad
T^y_8=\epsilon^{a_1a_3b_2}\epsilon^{a_2b_3c_2}\epsilon^{b_1c_3d_2}\epsilon^{c_1d_1d_3}\sim\young({{a_1}}{{a_2}}{{b_1}}{{c_1}},{{a_3}}{{b_3}}{{c_3}}{{d_1}},{{b_2}}{{c_2}}{{d_2}}{{d_3}})
\end{eqnarray}
After contraction with conversion factor, the y-basis gauge factor expressed with adjoint and anti-fundamental indices becomes polynomials of invariant tensors:
\begin{eqnarray}
&&    CF\cdot T^y_1=8d^{ABE}d^{CDE}+16\delta^{AC}\delta^{BD}+\frac{16}{3}\delta^{AB}\delta^{CD}-8id^{CDE}f^{ABE}+8id^{ABE}f^{CDE}+8f^{ABE}f^{CDE},\nn\\ 
&&CF\cdot T^y_2=-16d^{ACE}d^{BDE}+\frac{64}{3}\delta^{AC}\delta^{BD}-16id^{BDE}f^{ACE}-16id^{ACE}f^{BDE}+16f^{ACE}f^{BDE},\nn\\ 
&&CF\cdot T^y_3=-16d^{ACE}d^{BDE}+\frac{64}{3}\delta^{AC}\delta^{BD}+16id^{BDE}f^{ACE}+16id^{ACE}f^{BDE}+16f^{ACE}f^{BDE},\nn\\ 
&&CF\cdot T^y_4=64\delta^{AB}\delta^{CD},\nn\\ 
&&CF\cdot T^y_5=32d^{ABE}d^{CDE}+\frac{16}{3}\delta^{AB}\delta^{CD},\nn\\ 
&&CF\cdot T^y_6=16d^{ABE}d^{CDE}+\frac{-16}{3}\delta^{AB}\delta^{CD}+16id^{ABE}f^{CDE},\nn\\ 
&&CF\cdot T^y_7=16d^{ABE}d^{CDE}+\frac{32}{3}\delta^{AB}\delta^{CD}+16id^{CDE}f^{ABE},\nn\\ 
&&CF\cdot T^y_8=8d^{ABE}d^{CDE}+\frac{-32}{3}\delta^{AB}\delta^{CD}+8id^{CDE}f^{ABE}+8id^{ABE}f^{CDE}-8f^{ABE}f^{CDE},
\end{eqnarray}
from the above polynomials one can find 10 superficially different monomial invarant tensors as our M-basis candidates:
\begin{eqnarray}
&& T^m_{{\rm can,}1}=d^{ABE}d^{CDE},\quad 
 T^m_{{\rm can,}2}=d^{ABE}f^{CDE},\quad 
 T^m_{{\rm can,}3}=f^{ABE}f^{CDE},\quad 
 T^m_{{\rm can,}4}=\delta^{AB}\delta^{CD},\quad \nonumber \\
&& T^m_{{\rm can,}5}=d^{CDE}f^{ABE},\quad 
 T^m_{{\rm can,}6}=\delta^{AC}\delta^{BD},\quad
 T^m_{{\rm can,}7}=d^{ACE}d^{BDE},\quad
 T^m_{{\rm can,}8}=d^{ACE}f^{BDE},\quad \nn\\
&& T^m_{{\rm can,}9}=d^{BDE}f^{ACE},\quad  
 T^m_{{\rm can,}10}=d^{CDE}f^{ABE}\quad
\end{eqnarray}
Our algorithm can select 8 out of 10 from the above candidate by iteratively constructing the metric tensor $g_{ij}$, finally our M-basis is:
\begin{eqnarray}
&&    T^m_{1}=d^{ABE}d^{CDE},\quad 
 T^m_{2}=d^{ABE}f^{CDE},\quad 
 T^m_{3}=f^{ABE}f^{CDE},\quad 
 T^m_{4}=\delta^{AB}\delta^{CD},\quad\nonumber \\
&& T^m_{5}=d^{CDE}f^{ABE},\quad 
 T^m_{6}=\delta^{AC}\delta^{BD}\quad
 T^m_{7}=d^{ACE}d^{BDE},\quad
 T^m_{8}=d^{ACE}f^{BDE}.
\end{eqnarray}

% \mmaInlineCell[]{Code}{tList[SU2] = {del2, eps2a, eps2f, \[Tau], del3n, eps3n};
\section{Group Profile for $SU(N)$ Gauge Symmetry}\label{app:groupprofile}
For the purposes of the symbolic and numerical manipulation of $SU(N)$ gauge group invariant tensors, we need a group profile file for each gauge group, it contains  information for: 
\begin{itemize}
    \item The list names of basic invariant tensors stored in  \mmaInlineCell[]{Code}{tList[group]}. 
    
    For example, for the $SU(2)$ group we have:
    
%    \begin{equation}
%        \text{\mmaInlineCell[]{Input}{tList[SU2] = List[del2, eps2a, eps2f, \(\pmb{\tau}\), del3n, eps3n];}}\nonumber
%    \end{equation}
    {\centering\begin{tabular}{c}
         \mmaInlineCell[]{Input}{tList[SU2] = \big\{del2, eps2a, eps2f, \(\pmb{\tau}\), del3n, eps3n\big\};}
         %\mmaInlineCell[]{Input}{\(\pmb{\lambda}\)["A","b","a"]}
    \end{tabular}
    \par
    }
    where  \mmaInlineCell[]{Code}{del2} represents the Kronecker delta for the fundamental and anti-fundamental indices, \mmaInlineCell[]{Code}{eps2a} and \mmaInlineCell[]{Code}{eps2f} represents 2nd-rank $\epsilon$ tensors with anti-fundamental and fundamental indices,  \mmaInlineCell[]{Input}{\(\pmb{\tau}\)} represents $SU(2)$ generator, \mmaInlineCell[]{Code}{del3n} and \mmaInlineCell[]{Code}{eps3n} represent Kronecker delta and 3n-rand $\epsilon$ tensor with adjoint indices.
    
    \item The name list of totally anti-symmetric tensor stored in \mmaInlineCell[]{Code}{tasList[group]}. 
    
    For example, for $SU(2)$ we have:
    
     {\centering\begin{tabular}{c}
         \mmaInlineCell[]{Input}{tasList[SU2] = \big\{eps2a, eps2f, eps3n\big\};}
         %\mmaInlineCell[]{Input}{\(\pmb{\lambda}\)["A","b","a"]}
    \end{tabular}
    \par
    }
    
    \item The tensor rank and tensor dimension of basic invariant tensors are stored in a global variable \mmaInlineCell[]{Code}{tAssumptions}. 
    
    For example, for the $SU(2)$ generator $(\tau^I)_i^j$ as a rank-3 Mathematica tensor:
     \begin{mmaCell}[]{Input}
     \mmaUnd{τ}[I,i,j]
    \end{mmaCell}
    is registered in \mmaInlineCell[]{Code}{tAssumptions} as:
    \begin{mmaCell}[defined={Arrays, Reals, AppendTo}]{Input}
     AppendTo[tAssumptions, \mmaUnd{τ} \mmaUnd{∈} Arrays[\{3, 2, 2\}, Reals]];
     \end{mmaCell}

    \item The ordered indices type of basic invariant tensors stored in \mmaInlineCell[]{Code}{tRep[group]}. 
    
    As the example above, for the $SU(2)$ generator $\tau$ as a rank-3 Mathematica tensor, we register their indices type orders as:
    \begin{mmaCell}[defined={Arrays, Reals, AppendTo}]{Input}
     tRep[\mmaUnd{τ}] = \{\{2\}, \{1\}, \{-1\}\};
     \end{mmaCell}
    meaning the first index transform as adjoint representation, the second index transforms as fundamental representation, the last index transforms as anti-fundamental representation.
    
    \item The relations between invariant tensors and their complex conjugate stored in  \mmaInlineCell[]{Code}{TensorConj[]}. For example the $SU(2)$ generators:
    \begin{mmaCell}[defined={Arrays, Reals, AppendTo}]{Input}
     TensorConj[\mmaUnd{τ}[I_, a_, b_]] := \mmaUnd{τ}[I, b, a];
     \end{mmaCell}
    
    \item Replacement rules stored in  \mmaInlineCell[]{Code}{tSimp[group]} for the contraction of basic invariant group tensors such as Levi-Civita $\epsilon$, Kronecker delta $\delta$, group structure constant $f^{abc}$ and generators $(T^A)^{b}_{a}$. For example, for $SU(2)$ group we have:
    \begin{mmaCell}[pattern={i_,j_,k_,a_,b_,c_,J_,k_,l_,m_,n_x_,y_,z_,w_}]{Input}
    tSimp[SU2] = Hold[Block[\{\},
    del2[i_, j_] del2[j_, k_] := del2[i, k];
    del2[i_, i_] := 2;
    del3n[i_, i_] := 3;
    del3n[a_, c_] del3n[a_, b_] := del3n[c, b];
    del3n[a_, b_] del3n[b_, c_] := del3n[a, c];
    del3n[a_, c_] del3n[b_, c_] := del3n[a, b];
    del3n[b_, c_] del3n[a_, b_] := del3n[a, c];
    del3n[a_, b_]^2 := 3;   
    del2[a_, c_] \mmaUnd{τ}[J_, a_, b_] := \mmaUnd{τ}[J, c, b];
    del2[c_, a_] \mmaUnd{τ}[J_, b_, a_] := \mmaUnd{τ}[J, b, c];
    \mmaUnd{τ}[i_, j_, j_] := 0;
    \mmaUnd{τ}[a_, i_, j_] \mmaUnd{τ}[a_, k_, l_] := 
     2 del2[l, i] del2[j, k] - del2[l, k] del2[j, i];
    eps2a[x_, y_] eps2f[w_, z_] := 
     del2[x, w] del2[y, z] - del2[x, z] del2[y, w];
    eps3n[i_, j_, k_] eps3n[l_, m_, n_] := 
     Det@Outer[del3n, {i, j, k}, {l, m, n}];
    del3n[a_, d_] eps3n[a_, b_, c_] := eps3n[d, b, c];
    del3n[a_, d_] eps3n[b_, a_, c_] := eps3n[b, d, c];
    del3n[a_, d_] eps3n[c_, b_, a_] := eps3n[c, b, d];
    eps2f[i_, j_] del2[i_, k_] := eps2f[k, j];
    eps2f[i_, j_] del2[j_, k_] := eps2f[i, k];
    eps2a[i_, j_] del2[k_, i_] := eps2a[k, j];
    eps2a[i_, j_] del2[k_, j_] := eps2a[i, k]; ]]
    \end{mmaCell}
    they include contractions between and within Kronecker delta, contractions between delta and generators, contractions between $\epsilon$ and Kronecker delta, Fierz identity for generators, conversion of $\epsilon$ tensors to the $\delta$ products. In addition, there is one more replacement rule need to be stored independently in the global variable \mmaInlineCell[]{Code}{tY2M}, which is used to reduce of the number of generators when the presence of the matrix product of two generators:
    \begin{mmaCell}[pattern={i_,j_,k_,a_,b_,c_,J_,m_},defined={AssociateTo}]{Input}
    AssociateTo[tY2M, \{\mmaUnd{τ}[a_, j_, k_]\mmaUnd{τ}[b_, k_, m_] :> 
    Module[\{dummy = Unique[]\}, 
     I eps3n[a, b, dummy] \mmaUnd{τ}[dummy, j, m] + del3n[a, b] del2[m, j]]
    \}];
    \end{mmaCell}
    Additionaly, the conversion from the structure tensors into trace of generators are also useful in the simplification, therefore we individually store them in the global variable \mmaInlineCell[]{Code}{tM2Y}, for example, for $SU(3)$ group we have:
    \begin{mmaCell}[defined={Module,Unique,AssociateTo},pattern={a_,b_,c_}]{Input}
    AssociateTo[tM2Y, \{fabc[a_, b_, c_] :> 
      Module[\{d1 = Unique[], d2 = Unique[], d3 = Unique[]\}, 
        -(I/4) \mmaUnd{λ}[a, d1,d2] (\mmaUnd{λ}[b, d2, d3] \mmaUnd{λ}[c, d3, d1] -\mmaUnd{λ}[c, d2, d3] \mmaUnd{λ}[b, d3, d1])],
        dabc[a_, b_, c_] :> 
      Module[\{d1 = Unique[], d2 = Unique[], d3 = Unique[]\}, 
        1/4 \mmaUnd{λ}[a, d1,d2] (\mmaUnd{λ}[b, d2, d3] \mmaUnd{λ}[c, d3, d1] + \mmaUnd{λ}[c,d2, d3] \mmaUnd{λ}[b, d3, d1])]
   \}];
    \end{mmaCell}

    \item The numerical values of these basic invariant tensors stored in \mmaInlineCell[]{Code}{tVal[group]}.
    
    For example, for $SU(2)$ group we have:
    \begin{mmaCell}[defined={LeviCivitaTensor,IdentityMatrix}]{Input}
  tVal[SU2] = \big\{del2 -> IdentityMatrix[2], eps2f -> LeviCivitaTensor[2],  
  eps2a -> LeviCivitaTensor[2], \(\pmb{\tau}\) -> GellMann[2],
  del3n -> IdentityMatrix[3], eps3n -> LeviCivitaTensor[3]\big\}.
    \end{mmaCell}

    \item The name of the Levi-Civita tensor used for translating colums of singlet Young tableaux columns \mmaInlineCell[]{Code}{tYDcol[group]}.
    
    For example, for $SU(2)$ group we have:
    
        {\centering\begin{tabular}{c}
         \mmaInlineCell[]{Input}{tYDcol[SU2] = \big\{eps2a\big\};}
         %\mmaInlineCell[]{Input}{\(\pmb{\lambda}\)["A","b","a"]}
    \end{tabular}
    \par
    }
    as the indices in the Young tableaux are fundamental ones, we need use $\epsilon$ with anti-fundamental indices to contract them.
    
    \item The additional tensors need to convert fundamental indices to non-fundamental ones stored in \mmaInlineCell[pattern={ind_, num_,rep_},defined=CF]{Code}{CF[rep_, num_, ind_]}. 
    
    As we have shwon in 
    eq.~\eqref{eq:example_T_gluon}, the conversion prefactors such as $\epsilon_{jk}\left(\tau^I\right){}_i^k$ and $\epsilon_{acd}\left(\lambda^A\right){}_b^d$ to contract the products of $\epsilon$'s we obtained from the translation of the singlet Young tableaux. The \mmaInlineCell[pattern={ind_, num_,rep_},defined=CF]{Code}{rep_} are the representation of the corresponding gauge group expressed in Dynkin coefficients, which is needed to specified by user for each representation (including fundamental and singlet representation for consistency).  For example, for the $SU(2)$ group we have:
    
     \begin{mmaCell}[defined={TensorContract, Subscript},pattern={num_,ind_}]{Input}
    CF[\{0\}, num_, ind_] := 1
    CF[\{1\}, num_, ind_] := del2[ind, Subscript[num, 1]]
    CF[\{-1\}, num_, ind_] := eps2f[Subscript[num, 1], ind]
    CF[\{2\}, num_, ind_] := 
    TensorContract[eps2f\mmaUnd{⊗}\mmaUnd{τ}, \{\{1, 5\}\}][Subscript[num, 1], ind, Subscript[num, 2]]
     \end{mmaCell}
    where we have specified the conversion factors for singlet, fundamental, anti-fundamental and adjoint representations.
    On the right hand are of the form tensors appended with free indices in order. \mmaInlineCell[pattern={ind_, num_,rep_},defined=CF]{Code}{num} represents the position of the representation in a representation list ready to be analyzed, \mmaInlineCell[pattern={ind_, num_,rep_},defined=CF]{Code}{ind} is the non-fundamental indices. Take the last line, the convert factor for the adjoint representation as an example, the right hand side corresponds to $\left(\tau^I\right){}_i^k\epsilon_{jk}$, with correspondence $I\to$\mmaInlineCell[]{Code}{ind}, $i\to$\mmaInlineCell[defined={Subscript}]{Code}{Subscript[num, 1]} and $j\to$\mmaInlineCell[defined={Subscript}]{Code}{Subscript[num, 2]}, the dummy index $k$ is formally suppressed in the function \mmaInlineCell[pattern={ind_, num_,rep_},defined=TensorContract]{Code}{TensorContract}.

    \item The setting of output symbols for basic invariant tensors stored in  \mmaInlineCell[]{Code}{tOut[group]}.  
    
    For example, again for the $SU(2)$ generator $\tau$ we have:
    \begin{mmaCell}[]{Input}
     tOut[\mmaUnd{τ}] = 
        PrintTensor[<|"tensor" -> PrintTensor[<|"tensor" -> "\mmaUnd{τ}", "upind" -> \{#1\}|>], 
        "upind" -> \{#3\}, "downind" -> \{#2\}|>]&;
    \end{mmaCell}
    This tells us that the printed format of $\tau$ should have a adjoint index in the superscript of the $\tau$ with whatever label presented by  \mmaInlineCell[]{Code}{#1}, and additional lower and upper indices represented by  \mmaInlineCell[]{Code}{#2}  and \mmaInlineCell[]{Code}{#3} for the grouped object $\tau$ and adjoint index.
    
    \item The generators for different irreducible representations stored in  \mmaInlineCell[pattern={rep_}]{Code}{TGen[rep_]}.  
    
    These are used for getting j-basis operators.  For example, for $SU(2)$ group we have:
    \begin{mmaCell}[defined={TensorContract, Subscript},pattern={num_,ind_}]{Input}
    TGen[\{1\}] := 1/2 \mmaUnd{τ}[#1, #2, #3] &
    TGen[\{-1\}] := -1/2 \mmaUnd{τ}[#1, #3, #2] &
    TGen[\{2\}] := -I eps3n[#1, #2, #3] &
     \end{mmaCell}
    where \mmaInlineCell[pattern={rep_}]{Code}{{1}}, \mmaInlineCell[pattern={rep_}]{Code}{{-1}}, \mmaInlineCell[pattern={rep_}]{Code}{{2}} represents fundamental, anti-fundamental and adjoint representation respectively. 
    The returned generators must be a function of invariant tensor with three arguments \mmaInlineCell[pattern={rep_}]{Code}{#1}, \mmaInlineCell[pattern={rep_}]{Code}{#2} and \mmaInlineCell[pattern={rep_}]{Code}{#3}, where \mmaInlineCell[pattern={rep_}]{Code}{#1} is the label for the generator, i.e. an adjoint index, \mmaInlineCell[pattern={rep_}]{Code}{#3} is the index that contract with the index of the fields in the transformation, i.e. an index of type of conjugate representation of \mmaInlineCell[pattern={rep_}]{Code}{rep}, \mmaInlineCell[pattern={rep_}]{Code}{#2} is the free index of the transformed fields, i.e. an index of type of  \mmaInlineCell[pattern={rep_}]{Code}{rep}. 
    As an illustration, a generator acting on a field of fundamental representation yields:
    \begin{eqnarray}
        \tau^{I}\circ u_i\to u'_i = (\tau^{I})_i^j u_j,
    \end{eqnarray}
    thus  \mmaInlineCell[pattern={rep_}]{Code}{#1}, \mmaInlineCell[pattern={rep_}]{Code}{#2}, \mmaInlineCell[pattern={rep_}]{Code}{#3} corresponds to $I$,  $i$ and $j$ respectively.  The order of the position should be consistent with the order of indices of those invariant tensors claimed in  \mmaInlineCell[pattern={rep_}]{Code}{tRep} and  \mmaInlineCell[pattern={rep_}]{Code}{tAssumptions}.
\end{itemize}

Lastly, we comment on the addition of higher representation fields in the Model. If we use only fundamenal indices to represents the field and leave the symmetry among indices implicit, then the only piece of information that needs to be added to the group profile file is a identity conversion factor in \mmaInlineCell[pattern={ind_, num_,rep_},defined=CF]{Code}{CF[rep_, num_, ind_]}. For example, if we use $\Delta_{abc}$ to represents Quartet, with all three fundamental indices totally symmetrized, we need to add:
   \begin{mmaCell}[defined={TensorContract, Subscript},pattern={num_,ind_}]{Input}
    CF[\{3\}, num_, ind_] := 1
    \end{mmaCell}
This is enough for the use of the function \mmaInlineCell[defined=GetBasisForType]{Code}{GetBasisForType}, however the function \mmaInlineCell[defined=GetJBasisForType]{Code}{GetJBasisForType} will not work properly in this setup. Alternatively, one may insist on expressing the field of higher dimensional representation with a single index of that representation, for example, one may use  $\Delta_I$ to express quartet. In this case, to ensure our Young tablueax method work, we need to find the conversion factor as invariant tensor $\Gamma^I_{abc}$ such that $\Gamma^I_{abc}\Delta_I=\Delta_{abc}$, then one need to register this new invariant tenor in \mmaInlineCell[]{Code}{tList} for its name, in \mmaInlineCell[]{Code}{tAssumptions} for its shape, in \mmaInlineCell[]{Code}{tRep} for its indices type, in \mmaInlineCell[]{Code}{TensorConj} for its conjugate tensor, in \mmaInlineCell[]{Code}{tVal} its numerical value, in \mmaInlineCell[]{Code}{tOut} its output string form. After these registeration, one can set \mmaInlineCell[pattern={ind_, num_,rep_}]{Code}{CF[rep_, num_, ind_]} in the following:
   \begin{mmaCell}[defined={TensorContract, Subscript},pattern={num_,ind_}]{Input}
    CF[\{3\}, num_, ind_] := \mmaUnd{Γ}[ind, Subscript[num,1], Subscript[num,2], Subscript[num,3]]
    \end{mmaCell}
To enable the function \mmaInlineCell[defined=GetJBasisForType]{Code}{GetJBasisForType}, one needed to register the generator of quartet (named for example as \mmaInlineCell[]{Code}{Tq}) as an another new invariant tensor with the same procedures above, after the registration, one can add it to the \mmaInlineCell[]{Code}{TGen}, to specify the action of generators on the quartet field:
    \begin{mmaCell}[defined={TensorContract, Subscript},pattern={num_,ind_}]{Input}
    TGen[\{3\}] :=  Tq[#1, #2, #3] &
    TGen[\{-3\}] :=  Tq[#1, #3, #2] &
    \end{mmaCell}

%the list of basic invariant tensors stored in  \mmaInlineCell[]{Code}{tList[group]}; Replacement rules stored in  \mmaInlineCell[]{Code}{tSimp[group]} for the contraction of basic invariant group tensors such as Levi-Civita $\epsilon$, Kronecker delta $\delta$, group structure constant $f^{abc}$ and generators $(T^A)^{b}_{a}$; the numerical value of these basic invariant tensors stored in \mmaInlineCell[]{Code}{tVal[group]}; The name of the Levi-Civita tensor used for translating the singlet Young tableaux \mmaInlineCell[]{Code}{tYDcol[group]}; The name list of totally anti-symmetric tensor stored in \mmaInlineCell[]{Code}{tasList[group]}; The additional tensors need to convert fundamental indices to non-fundamental ones stored in \mmaInlineCell[pattern={ind_, num_,rep_},defined=CF]{Code}{CF[rep_, num_, ind_]}; The assumption of the tensor dimension and symmetric properties of the basic invariant tensors stroed in the global variable \mmaInlineCell[]{Code}{tAssumptions}; The indices type of basic invariant tensors stored in \mmaInlineCell[]{Code}{tRep[group]}; The output symbols for basic invariant tensors stored in  \mmaInlineCell[]{Code}{tOut[group]}; The generators acting on the specific type of indices stored in  \mmaInlineCell[]{Code}{TGen[rep]}.

\section{Auxiliary Mathematica Functions}\label{sec:functions}

In this section, we present some auxiliary functions which might be useful for readers who are interested in the internal structure of the code. 

The function \mmaInlineCell[pattern={dim_},defined=LorentzList]{Code}{LorentzList[dim_]} enumerates all possible states at a given dimension. These states by default contain particles with helicitis $\{0,\pm\frac12,\pm 1,\pm 2\}$. People who need to take more particles of different spins into account, may add the option \mmaInlineCell[defined=HelicityInclude]{Code}{HelicityInclude}. 
\begin{mmaCell}[defined={LorentzList,OptionsPattern}]{Code}
  LorentzList[6]
\end{mmaCell}
\begin{mmaCell}[]{Output}
  \big\{\{\{-1, -1, -1\}, 0\},\big\{\big\{-1/2,-1/2,-1/2,-1/2\big\},0\big\},\big\{\big\{-1,-1/2,-1/2,0\big\},0\big\},\\
    \{\{-1,-1,0,0\},0\},\big\{\big\{-1/2,-1/2,1/2,1/2\big\},0\big\},\big\{\big\{-1/2,0,0,1/2\big\},1\big\},\\
    \{\{0,0,0,0\},2\},\big\{\big\{-1/2,-1/2,0,0,0\big\},0\big\},\{\{0,0,0,0,0,0\},0\}\big\}
\end{mmaCell}

The function \mmaInlineCell[pattern={state_,k_},defined={SSYT,OptionsPattern}]{Input}{SSYT[state_,k_,OptionsPattern[]]} enumerates the independent helicity amplitudes for a special state and derivatives.
\begin{mmaCell}[defined={SSYT,OutMode}]{Input}
  SSYT[\{1,1,0,0\}, 2, OutMode->"amplitude"]
\end{mmaCell}
\begin{mmaCell}[]{Output}
  \big\{ab[3,4] \mmaSup{sb[1,2]}{2} sb[3,4], ab[3,4] sb[1,2] sb[1,3] sb[2,4]\big\}
\end{mmaCell}
\begin{mmaCell}[defined={SSYT,OutMode}]{Input}
  SSYT[\{1,1,0,0\}, 2, OutMode->"amplitude output"]
\end{mmaCell}
\begin{mmaCell}[]{Output}
  \big\{ -\mmaSup{[12]}{2}\mmaSub{s}{34}, <34>[12][13][24]\big\}
\end{mmaCell}
Options \mmaInlineCell[defined=OutMode]{Input}{OutMode->} could control the output format. Note that \mmaInlineCell[]{Code}{ab[i,j]} and  \mmaInlineCell[]{Code}{sb[i,j]} represent $\vev{ij}$ and $[ij]$ respectively. 

The function \mmaInlineCell[pattern={state_,k_,posRepeat_},defined={LorentzBasisAux,OptionsPattern}]{Code}{LorentzBasisAux[state_, k_, posRepeat_, OptionsPattern[]]} obtain the independent Lorentz m-basis in a given state and repeat fields. For instance, a complete set of operators including 4 fermions with helicities $h=-1/2$ without derivatives is given by
\begin{mmaCell}[defined={LorentzBasisAux}]{Input}
  LorentzBasisAux[\{-1/2,-1/2,-1/2,-1/2\},0,\{\{1,2\},\{3,4\}\}]
\end{mmaCell}
\begin{mmaCell}[]{Output}
  <|"basis"->\{ch\(\pmb{\psi}\)[\mmaSub{\(\pmb{\psi}\)}{1},1,\mmaSub{\(\pmb{\psi}\)}{2}] ch\(\pmb{\psi}\)[\mmaSub{\(\pmb{\psi}\)}{3},1,\mmaSub{\(\pmb{\psi}\)}{4}], ch\(\pmb{\psi}\)[\mmaSub{\(\pmb{\psi}\)}{1},1,\mmaSub{\(\pmb{\psi}\)}{3}] ch\(\pmb{\psi}\)[\mmaSub{\(\pmb{\psi}\)}{2},1,\mmaSub{\(\pmb{\psi}\)}{4}]\}, \\
  "Trans"->\{\{1,0\},\{0,1\}\}, "Ybasis"->\{ab[1,2] ab[3,4], ab[1,3] ab[2,4]\},\\
  "generators"-><|\{1,2\}->\{\{\{1,0\},\{1,-1\}\},\{\{1,0\},\{1,-1\}\}\},\\
    \{3,4\}->\{\{\{1,0\},\{1,-1\}\},\{\{1,0\},\{1,-1\}\}\}|>|>
\end{mmaCell}
Where \mmaInlineCell[pattern={state_,k_,posRepeat_},defined=LorentzBasisAux]{Code}{LorentzBasisAux[state_,k_,posRepeat_]["Trans"]} is the conversion matrix between Lorentz m-basis and y-basis, \mmaInlineCell[pattern={state_,k_,posRepeat_},defined=LorentzBasisAux]{Code}{LorentzBasisAux[state_,k_,posRepeat_]["generators"]} show the conversion matrix after the permutations. One may add a option \mmaInlineCell[defined=AlderZero]{Code}{AlderZero->{__}} to require that some particles are soft.\footnote{When we say particle $i$ is soft, which means that amplitude vanishs when $p_i\to 0$, such as Goldstones.}

The function \mmaInlineCell[pattern={group_,replist_,posRepeat_},defined={GaugeBasisAux}]{Code}{GaugeBasisAux[group_,replist_,posRepeat_]} gives the gauge m-basis of a list of representations for certain group, and the matrix representations of generators of $S_m$ group on the gauge m-basis, where $m$ is the number of repeated fields. For example, consider the $SU(2)$ gauge structure of $H^2 H^{\dagger 2} D^4$ and label the field as $H_1,H_2,H^{\dagger}_3,H^{\dagger}_4$. The $SU(2)$ representations of fields are labeled by the Dynkin coefficients, but for the $SU(2)$, both fundamental representation and anti-fundamental representation are of the Dynkin coefficient \mmaInlineCell[]{Code}{{1}}, so we label the anti-fundamental representation as \mmaInlineCell[]{Code}{{-1}} instead. The repeated fields are $H_1,H_2$ and $H^{\dagger}_3,H^{\dagger}_4$.
\begin{mmaCell}[defined={GaugeBasisAux,SU2}]{Input}
  GaugeBasisAux[SU2,\{\{1\},\{1\},\{-1\},\{-1\}\},\{\{1,2\},\{3,4\}\}]
\end{mmaCell}
\begin{mmaCell}[]{Output}
  <|"basis"->\{del2[i,k] del2[j,l],del2[i,l] del2[j,k]\},
  "generators"-><|\{1,2\}->\{\{\{0,1\},\{1,0\}\},\{\{0,1\},\{1,0\}\}\},
    \{3,4\}->\{\{\{0,1\},\{1,0\}\},\{\{0,1\},\{1,0\}\}\}|>|>
\end{mmaCell}
The outputs are the gauge m-basis of $H^2 H^{\dagger 2} D^4$, $\delta^i_k \delta^j_l$ and $\delta^i_l \delta^j_k$, and the matrix representations of generators of the two $S_2$ groups on the m-basis.

The function \mmaInlineCell[pattern={amp_,ch_},defined=W2]{Code}{W2[amp_,ch_]} evaluates the result of the $W^2_{\mathcal{I}}$ applied to an amplitude
\begin{mmaCell}[defined={W2,ab,sb}]{Code}
  W2[ab[1,2] sb[2,3], {1,2}]
\end{mmaCell}
\begin{mmaCell}[defined={ab,sb}]{Output}
  \mmaFrac{3}{4} \mmaSup{ab[1,2]}{2} sb[1,2] sb[2,3]
\end{mmaCell}

The function \mmaInlineCell[pattern={state_,k_,ch_},defined=W2Diagonalize]{Code}{W2Diagonalize[state_,k_,ch_]} enumerates the independent j-basis in a given state.
\begin{mmaCell}[defined={W2Diagonalize,Ampform}]{Code}
  W2Diagonalize[{1, 0, 1, 0}, 2, {1, 2}] //Ampform
\end{mmaCell}
\begin{mmaCell}[defined={ab,sb}]{Output}
  <|"basis"->\big\{\mmaSup{[13]}{2}\mmaSub{s}{24}, -\mmaSup{[13]}{2}\mmaSub{s}{34}\big\}, "j"->\{2, 1\}, "transfer"->\{\{-4, 3\}, \{0, 1\}\},\\ 
  "j-basis"->\big\{-4 \mmaSup{[13]}{2}\mmaSub{s}{24} -3 \mmaSup{[13]}{2}\mmaSub{s}{34}, - \mmaSup{[13]}{2}\mmaSub{s}{34}\big\}|>
\end{mmaCell}
Where \mmaInlineCell[pattern={state_,k_,ch_},defined=W2Diagonalize]{Code}{W2Diagonalize[state_,k_,ch_]["basis"]} exhibits the Lorentz y-basis,\\ \mmaInlineCell[pattern={state_,k_,ch_},defined=W2Diagonalize]{Code}{W2Diagonalize[state_,k_,ch_]["j-basis"]} are the result for J-basis with angular momentum showed in \mmaInlineCell[pattern={state_,k_,ch_},defined=W2Diagonalize]{Code}{W2Diagonalize[state_,k_,ch_]["j"]} respectively. \mmaInlineCell[pattern={state_,k_,ch_},defined=W2Diagonalize]{Code}{W2Diagonalize[state_,k_,ch_]["transfer"]} is the convertion matrix from y-basis to j-basis.

The function \mmaInlineCell[pattern={amp_,num_,ch_},defined={PWExpand}]{Code}{PWExpand[amp_,num_,ch_]} calculates the partial wave expansion of local amplitudes, with the input of particle number \mmaInlineCell[pattern={amp_,num_,ch_},defined={PWExpand}]{Code}{num_} and channel \mmaInlineCell[pattern={amp_,num_,ch_},defined={PWExpand}]{Code}{ch_}.
\begin{mmaCell}[defined={PWExpand}]{Input}
  PWExpand[ab[1,3]sb[1,3], 4, \{1,2\}]
\end{mmaCell}
\begin{mmaCell}[]{Output}
  <|"j"-> \{1,0\}, "j-basis"-> \{-2 ab[2,4] sb[2,4] - ab[3,4] sb[3,4], ab[3,4] sb[3,4]\},\\
  "coeff"-> \{-1/2, -1/2\}|>
\end{mmaCell}
Where \mmaInlineCell[pattern={amp_,num_,ch_},defined={PWExpand}]{Code}{PWExpand[amp_,num_,ch_]["j-basis"]} are the Lorentz J-basis in the same dimension, along with angular momentum showed in \mmaInlineCell[pattern={amp_,num_,ch_},defined={PWExpand}]{Code}{PWExpand[amp_,num_,ch_]["j"]}. \mmaInlineCell[pattern={amp_,num_,ch_},defined={PWExpand}]{Code}{PWExpand[amp_,num_,ch_]["coeff"]} is the expansion coefficient.

\section{On-shell Amplitude Bases in the SMEFT}

Here we list the on-shell amplitude bases in the SMEFT at the mass dimensions 6, 7, and 8. The dimension-6 and dimension-7 on-shell amplitude bases are full, while the dimension-8 on-shell amplitude bases only contain the Bosonic part due to limited space. The on-shell amplitude bases are not in one-to-one correspondence with the operator bases we listed before, for example, in Ref.~\cite{Li:2020gnx}, but these are both complete and independent bases connected by linear transformations.

\subsection{Dimension-6 On-shell Amplitude Basis}

$F_{\rm{L}}^3$

\begin{align}\begin{array}{c|ll}
		
		\multirow{1}*{$B_{\rm{L}}^3   $}
		
		&-  \langle 12\rangle  \langle 13\rangle  \langle 23\rangle   & \\
		
\end{array}\end{align}

\begin{align}\begin{array}{c|ll}
		
		\multirow{1}*{$B_{\rm{L}} W_{\rm{L}}^2   $}
		
		&-  \delta ^{I_2I_3} \langle 12\rangle  \langle 13\rangle  \langle 23\rangle   & \\
		
\end{array}\end{align}

\begin{align}\begin{array}{c|ll}
		
		\multirow{1}*{$W_{\rm{L}}^3   $}
		
		&-  \epsilon ^{I_1I_2I_3} \langle 12\rangle  \langle 13\rangle  \langle 23\rangle   & \\
		
\end{array}\end{align}

\begin{align}\begin{array}{c|ll}
		
		\multirow{1}*{$B_{\rm{L}} G_{\rm{L}}^2   $}
		
		&-  \delta ^{A_2A_3} \langle 12\rangle  \langle 13\rangle  \langle 23\rangle   & \\
		
\end{array}\end{align}

\begin{align}\begin{array}{c|ll}
		
		\multirow{1}*{$G_{\rm{L}}^3   $}
		
		&-  f^{A_1A_2A_3} \langle 12\rangle  \langle 13\rangle  \langle 23\rangle   & \\
		
\end{array}\end{align}

$\psi ^4$

\begin{align}\begin{array}{c|ll}
		
		\multirow{2}*{$d_{_\mathbb{C}} e_{_\mathbb{C}} u_{_\mathbb{C}}^2   $}
		
		& C_{f_1,f_2,f_3f_4}^{[2]} \epsilon _{a_1a_3a_4} \langle 13\rangle  \langle 24\rangle   & \\
		
		& C_{f_1,f_2,f_3f_4}^{[1,1]} \epsilon _{a_1a_3a_4} \langle 12\rangle  \langle 34\rangle   & \\
		
\end{array}\end{align}

\begin{align}\begin{array}{c|ll}
		
		\multirow{2}*{$e_{_\mathbb{C}} L Q u_{_\mathbb{C}}   $}
		
		& C_{f_1,f_2,f_3,f_4} \delta _{a_4}^{a_3} \epsilon ^{i_2i_3} \langle 12\rangle  \langle 34\rangle   & \\
		
		& C_{f_1,f_2,f_3,f_4} \delta _{a_4}^{a_3} \epsilon ^{i_2i_3} \langle 13\rangle  \langle 24\rangle   & \\
		
\end{array}\end{align}

\begin{align}\begin{array}{c|ll}
		
		\multirow{4}*{$d_{_\mathbb{C}} Q^2 u_{_\mathbb{C}}   $}
		
		& C_{f_1,f_2f_3,f_4}^{[2]} \delta _{a_1}^{a_3} \delta _{a_4}^{a_2} \epsilon ^{i_2i_3} \langle 12\rangle  \langle 34\rangle   & \\
		
		& C_{f_1,f_2f_3,f_4}^{[2]} \delta _{a_1}^{a_3} \delta _{a_4}^{a_2} \epsilon ^{i_2i_3} \langle 13\rangle  \langle 24\rangle   & \\
		
		& C_{f_1,f_2f_3,f_4}^{[1,1]} \delta _{a_1}^{a_3} \delta _{a_4}^{a_2} \epsilon ^{i_2i_3} \langle 12\rangle  \langle 34\rangle   & \\
		
		& C_{f_1,f_2f_3,f_4}^{[1,1]} \delta _{a_1}^{a_3} \delta _{a_4}^{a_2} \epsilon ^{i_2i_3} \langle 13\rangle  \langle 24\rangle   & \\
		
\end{array}\end{align}

\begin{align}\begin{array}{c|ll}
		
		\multirow{3}*{$L Q^3   $}
		
		& C_{f_1,f_2f_3f_4}^{[3]} \epsilon ^{a_2a_3a_4} \epsilon ^{i_1i_3} \epsilon ^{i_2i_4} \langle 12\rangle  \langle 34\rangle   & \\
		
		& C_{f_1,f_2f_3f_4}^{[2,1]} \epsilon ^{a_2a_3a_4} \epsilon ^{i_1i_3} \epsilon ^{i_2i_4} \langle 12\rangle  \langle 34\rangle   & \\
		
		& C_{f_1,f_2f_3f_4}^{[1,1,1]} \epsilon ^{a_2a_3a_4} \epsilon ^{i_1i_3} \epsilon ^{i_2i_4} \langle 12\rangle  \langle 34\rangle   & \\
		
\end{array}\end{align}

$F_{\rm{L}} \psi ^2 \phi$

\begin{align}\begin{array}{c|ll}
		
		\multirow{1}*{$B_{\rm{L}} e_{_\mathbb{C}} L H^{\dagger}{}    $}
		
		& C_{f_2,f_3} \delta _{i_4}^{i_3} \langle 12\rangle  \langle 13\rangle   & \\
		
\end{array}\end{align}

\begin{align}\begin{array}{c|ll}
		
		\multirow{1}*{$B_{\rm{L}} d_{_\mathbb{C}} Q H^{\dagger}{}    $}
		
		& C_{f_2,f_3} \delta _{a_2}^{a_3} \delta _{i_4}^{i_3} \langle 12\rangle  \langle 13\rangle   & \\
		
\end{array}\end{align}

\begin{align}\begin{array}{c|ll}
		
		\multirow{1}*{$B_{\rm{L}} Q u_{_\mathbb{C}} H   $}
		
		& C_{f_2,f_3} \delta _{a_3}^{a_2} \epsilon ^{i_2i_4} \langle 12\rangle  \langle 13\rangle   & \\
		
\end{array}\end{align}

\begin{align}\begin{array}{c|ll}
		
		\multirow{1}*{$W_{\rm{L}} e_{_\mathbb{C}} L H^{\dagger}{}    $}
		
		& C_{f_2,f_3} \left(\tau ^{I_1})_{i_4}^{i_3}\right. \langle 12\rangle  \langle 13\rangle   & \\
		
\end{array}\end{align}

\begin{align}\begin{array}{c|ll}
		
		\multirow{1}*{$W_{\rm{L}} d_{_\mathbb{C}} Q H^{\dagger}{}    $}
		
		& C_{f_2,f_3} \delta _{a_2}^{a_3} \left(\tau ^{I_1})_{i_4}^{i_3}\right. \langle 12\rangle  \langle 13\rangle   & \\
		
\end{array}\end{align}

\begin{align}\begin{array}{c|ll}
		
		\multirow{1}*{$W_{\rm{L}} Q u_{_\mathbb{C}} H   $}
		
		& C_{f_2,f_3} \delta _{a_3}^{a_2} \left(\tau ^{I_1})_k^{i_2}\right. \epsilon ^{i_4k} \langle 12\rangle  \langle 13\rangle   & \\
		
\end{array}\end{align}

\begin{align}\begin{array}{c|ll}
		
		\multirow{1}*{$G_{\rm{L}} d_{_\mathbb{C}} Q H^{\dagger}{}    $}
		
		& C_{f_2,f_3} \delta _{i_4}^{i_3} \left(\lambda ^{A_1})_{a_2}^{a_3}\right. \langle 12\rangle  \langle 13\rangle   & \\
		
\end{array}\end{align}

\begin{align}\begin{array}{c|ll}
		
		\multirow{1}*{$G_{\rm{L}} Q u_{_\mathbb{C}} H   $}
		
		& C_{f_2,f_3} \left(\lambda ^{A_1})_{a_3}^{a_2}\right. \epsilon ^{i_2i_4} \langle 12\rangle  \langle 13\rangle   & \\
		
\end{array}\end{align}

$F_{\rm{L}}^2 \phi ^2$

\begin{align}\begin{array}{c|ll}
		
		\multirow{1}*{$B_{\rm{L}}^2 H H^{\dagger}{}    $}
		
		&   \delta _{i_4}^{i_3} \langle 12\rangle ^2   & \\
		
\end{array}\end{align}

\begin{align}\begin{array}{c|ll}
		
		\multirow{1}*{$B_{\rm{L}} W_{\rm{L}} H H^{\dagger}{}    $}
		
		&   \left(\tau ^{I_2})_{i_4}^{i_3}\right. \langle 12\rangle ^2   & \\
		
\end{array}\end{align}

\begin{align}\begin{array}{c|ll}
		
		\multirow{1}*{$W_{\rm{L}}^2 H H^{\dagger}{}    $}
		
		&   \delta _{i_4}^{i_3} \delta ^{I_1I_2} \langle 12\rangle ^2   & \\
		
\end{array}\end{align}

\begin{align}\begin{array}{c|ll}
		
		\multirow{1}*{$G_{\rm{L}}^2 H H^{\dagger}{}    $}
		
		&   \delta _{i_4}^{i_3} \delta ^{A_1A_2} \langle 12\rangle ^2   & \\
		
\end{array}\end{align}

$\psi ^2 \bar{\psi }^2$

\begin{align}\begin{array}{c|ll}
		
		\multirow{1}*{$e_{_\mathbb{C}}^2 e_{_\mathbb{C}}^{\dagger}{} ^2   $}
		
		& C_{f_1f_2,f_3f_4}^{[2],[2]} \langle 12\rangle [34] & \\
		
\end{array}\end{align}

\begin{align}\begin{array}{c|ll}
		
		\multirow{1}*{$e_{_\mathbb{C}} L e_{_\mathbb{C}}^{\dagger}{}  L^{\dagger}{}    $}
		
		& C_{f_1,f_2,f_3,f_4} \delta _{i_4}^{i_2} \langle 12\rangle [34] & \\
		
\end{array}\end{align}

\begin{align}\begin{array}{c|ll}
		
		\multirow{1}*{$e_{_\mathbb{C}} L d_{_\mathbb{C}}^{\dagger}{}  Q^{\dagger}{}    $}
		
		& C_{f_1,f_2,f_3,f_4} \delta _{a_4}^{a_3} \delta _{i_4}^{i_2} \langle 12\rangle [34] & \\
		
\end{array}\end{align}

\begin{align}\begin{array}{c|ll}
		
		\multirow{1}*{$d_{_\mathbb{C}} L d_{_\mathbb{C}}^{\dagger}{}  L^{\dagger}{}    $}
		
		& C_{f_1,f_2,f_3,f_4} \delta _{a_1}^{a_3} \delta _{i_4}^{i_2} \langle 12\rangle [34] & \\
		
\end{array}\end{align}

\begin{align}\begin{array}{c|ll}
		
		\multirow{1}*{$L u_{_\mathbb{C}} L^{\dagger}{}  u_{_\mathbb{C}}^{\dagger}{}    $}
		
		& C_{f_1,f_2,f_3,f_4} \delta _{a_2}^{a_4} \delta _{i_3}^{i_1} \langle 12\rangle [34] & \\
		
\end{array}\end{align}

\begin{align}\begin{array}{c|ll}
		
		\multirow{1}*{$e_{_\mathbb{C}} Q e_{_\mathbb{C}}^{\dagger}{}  Q^{\dagger}{}    $}
		
		& C_{f_1,f_2,f_3,f_4} \delta _{a_4}^{a_2} \delta _{i_4}^{i_2} \langle 12\rangle [34] & \\
		
\end{array}\end{align}

\begin{align}\begin{array}{c|ll}
		
		\multirow{1}*{$d_{_\mathbb{C}} Q e_{_\mathbb{C}}^{\dagger}{}  L^{\dagger}{}    $}
		
		& C_{f_1,f_2,f_3,f_4} \delta _{a_1}^{a_2} \delta _{i_4}^{i_2} \langle 12\rangle [34] & \\
		
\end{array}\end{align}

\begin{align}\begin{array}{c|ll}
		
		\multirow{2}*{$L^2 L^{\dagger}{} ^2   $}
		
		& C_{f_1f_2,f_3f_4}^{[2],[2]} \delta _{i_3}^{i_1} \delta _{i_4}^{i_2} \langle 12\rangle [34] & \\
		
		& C_{f_1f_2,f_3f_4}^{[1,1],[1,1]} \delta _{i_3}^{i_1} \delta _{i_4}^{i_2} \langle 12\rangle [34] & \\
		
\end{array}\end{align}

\begin{align}\begin{array}{c|ll}
		
		\multirow{1}*{$d_{_\mathbb{C}} e_{_\mathbb{C}} d_{_\mathbb{C}}^{\dagger}{}  e_{_\mathbb{C}}^{\dagger}{}    $}
		
		& C_{f_1,f_2,f_3,f_4} \delta _{a_1}^{a_3} \langle 12\rangle [34] & \\
		
\end{array}\end{align}

\begin{align}\begin{array}{c|ll}
		
		\multirow{1}*{$e_{_\mathbb{C}} u_{_\mathbb{C}} e_{_\mathbb{C}}^{\dagger}{}  u_{_\mathbb{C}}^{\dagger}{}    $}
		
		& C_{f_1,f_2,f_3,f_4} \delta _{a_2}^{a_4} \langle 12\rangle [34] & \\
		
\end{array}\end{align}

\begin{align}\begin{array}{c|ll}
		
		\multirow{2}*{$d_{_\mathbb{C}}^2 d_{_\mathbb{C}}^{\dagger}{} ^2   $}
		
		& C_{f_1f_2,f_3f_4}^{[2],[2]} \delta _{a_1}^{a_3} \delta _{a_2}^{a_4} \langle 12\rangle [34] & \\
		
		& C_{f_1f_2,f_3f_4}^{[1,1],[1,1]} \delta _{a_1}^{a_3} \delta _{a_2}^{a_4} \langle 12\rangle [34] & \\
		
\end{array}\end{align}

\begin{align}\begin{array}{c|ll}
		
		\multirow{2}*{$d_{_\mathbb{C}} u_{_\mathbb{C}} d_{_\mathbb{C}}^{\dagger}{}  u_{_\mathbb{C}}^{\dagger}{}    $}
		
		& C_{f_1,f_2,f_3,f_4} \delta _{a_1}^{a_3} \delta _{a_2}^{a_4} \langle 12\rangle [34] & \\
		
		& C_{f_1,f_2,f_3,f_4} \delta _{a_1}^{a_4} \delta _{a_2}^{a_3} \langle 12\rangle [34] & \\
		
\end{array}\end{align}

\begin{align}\begin{array}{c|ll}
		
		\multirow{2}*{$u_{_\mathbb{C}}^2 u_{_\mathbb{C}}^{\dagger}{} ^2   $}
		
		& C_{f_1f_2,f_3f_4}^{[2],[2]} \delta _{a_1}^{a_3} \delta _{a_2}^{a_4} \langle 12\rangle [34] & \\
		
		& C_{f_1f_2,f_3f_4}^{[1,1],[1,1]} \delta _{a_1}^{a_3} \delta _{a_2}^{a_4} \langle 12\rangle [34] & \\
		
\end{array}\end{align}

\begin{align}\begin{array}{c|ll}
		
		\multirow{1}*{$e_{_\mathbb{C}} u_{_\mathbb{C}} Q^{\dagger}{} ^2   $}
		
		& C_{f_1,f_2,f_3f_4}^{[2]} \epsilon _{a_2a_3a_4} \epsilon _{i_3i_4} \langle 12\rangle [34] & \\
		
\end{array}\end{align}

\begin{align}\begin{array}{c|ll}
		
		\multirow{1}*{$d_{_\mathbb{C}} u_{_\mathbb{C}} L^{\dagger}{}  Q^{\dagger}{}    $}
		
		& C_{f_1,f_2,f_3,f_4} \epsilon _{a_1a_2a_4} \epsilon _{i_3i_4} \langle 12\rangle [34] & \\
		
\end{array}\end{align}

\begin{align}\begin{array}{c|ll}
		
		\multirow{1}*{$L Q d_{_\mathbb{C}}^{\dagger}{}  u_{_\mathbb{C}}^{\dagger}{}    $}
		
		& C_{f_1,f_2,f_3,f_4} \epsilon ^{a_2a_3a_4} \epsilon ^{i_1i_2} \langle 12\rangle [34] & \\
		
\end{array}\end{align}

\begin{align}\begin{array}{c|ll}
		
		\multirow{1}*{$Q^2 e_{_\mathbb{C}}^{\dagger}{}  u_{_\mathbb{C}}^{\dagger}{}    $}
		
		& C_{f_1f_2,f_3,f_4}^{[2]} \epsilon ^{a_1a_2a_4} \epsilon ^{i_1i_2} \langle 12\rangle [34] & \\
		
\end{array}\end{align}

\begin{align}\begin{array}{c|ll}
		
		\multirow{2}*{$L Q L^{\dagger}{}  Q^{\dagger}{}    $}
		
		& C_{f_1,f_2,f_3,f_4} \delta _{a_4}^{a_2} \delta _{i_3}^{i_1} \delta _{i_4}^{i_2} \langle 12\rangle [34] & \\
		
		& C_{f_1,f_2,f_3,f_4} \delta _{a_4}^{a_2} \delta _{i_3}^{i_2} \delta _{i_4}^{i_1} \langle 12\rangle [34] & \\
		
\end{array}\end{align}

\begin{align}\begin{array}{c|ll}
		
		\multirow{2}*{$d_{_\mathbb{C}} Q d_{_\mathbb{C}}^{\dagger}{}  Q^{\dagger}{}    $}
		
		& C_{f_1,f_2,f_3,f_4} \delta _{a_1}^{a_3} \delta _{a_4}^{a_2} \delta _{i_4}^{i_2} \langle 12\rangle [34] & \\
		
		& C_{f_1,f_2,f_3,f_4} \delta _{a_1}^{a_2} \delta _{a_4}^{a_3} \delta _{i_4}^{i_2} \langle 12\rangle [34] & \\
		
\end{array}\end{align}

\begin{align}\begin{array}{c|ll}
		
		\multirow{2}*{$Q u_{_\mathbb{C}} Q^{\dagger}{}  u_{_\mathbb{C}}^{\dagger}{}    $}
		
		& C_{f_1,f_2,f_3,f_4} \delta _{a_2}^{a_1} \delta _{a_3}^{a_4} \delta _{i_3}^{i_1} \langle 12\rangle [34] & \\
		
		& C_{f_1,f_2,f_3,f_4} \delta _{a_2}^{a_4} \delta _{a_3}^{a_1} \delta _{i_3}^{i_1} \langle 12\rangle [34] & \\
		
\end{array}\end{align}

\begin{align}\begin{array}{c|ll}
		
		\multirow{4}*{$Q^2 Q^{\dagger}{} ^2   $}
		
		& C_{f_1f_2,f_3f_4}^{[2],[2]} \delta _{a_3}^{a_1} \delta _{a_4}^{a_2} \delta _{i_3}^{i_1} \delta _{i_4}^{i_2} \langle 12\rangle [34] & \\
		
		& C_{f_1f_2,f_3f_4}^{[2],[2]} \delta _{a_3}^{a_1} \delta _{a_4}^{a_2} \delta _{i_3}^{i_2} \delta _{i_4}^{i_1} \langle 12\rangle [34] & \\
		
		& C_{f_1f_2,f_3f_4}^{[1,1],[1,1]} \delta _{a_3}^{a_1} \delta _{a_4}^{a_2} \delta _{i_3}^{i_1} \delta _{i_4}^{i_2} \langle 12\rangle [34] & \\
		
		& C_{f_1f_2,f_3f_4}^{[1,1],[1,1]} \delta _{a_3}^{a_1} \delta _{a_4}^{a_2} \delta _{i_3}^{i_2} \delta _{i_4}^{i_1} \langle 12\rangle [34] & \\
		
\end{array}\end{align}

$D \psi  \phi ^2 \bar{\psi }$

\begin{align}\begin{array}{c|ll}
		
		\multirow{1}*{$e_{_\mathbb{C}} H H^{\dagger}{}  e_{_\mathbb{C}}^{\dagger}{}  D $}
		
		& C_{f_1,f_4} \delta _{i_3}^{i_2} \langle 13\rangle [34] & \\
		
\end{array}\end{align}

\begin{align}\begin{array}{c|ll}
		
		\multirow{2}*{$L H H^{\dagger}{}  L^{\dagger}{}  D $}
		
		& C_{f_1,f_4} \delta _{i_3}^{i_1} \delta _{i_4}^{i_2} \langle 13\rangle [34] & \\
		
		& C_{f_1,f_4} \delta _{i_3}^{i_2} \delta _{i_4}^{i_1} \langle 13\rangle [34] & \\
		
\end{array}\end{align}

\begin{align}\begin{array}{c|ll}
		
		\multirow{1}*{$d_{_\mathbb{C}} H^{\dagger}{} ^2 u_{_\mathbb{C}}^{\dagger}{}  D $}
		
		& C_{f_1,f_4} \epsilon _{i_2i_3} \delta _{a_1}^{a_4} \langle 13\rangle [34] & \\
		
\end{array}\end{align}

\begin{align}\begin{array}{c|ll}
		
		\multirow{1}*{$d_{_\mathbb{C}} H H^{\dagger}{}  d_{_\mathbb{C}}^{\dagger}{}  D $}
		
		& C_{f_1,f_4} \delta _{a_1}^{a_4} \delta _{i_3}^{i_2} \langle 13\rangle [34] & \\
		
\end{array}\end{align}

\begin{align}\begin{array}{c|ll}
		
		\multirow{1}*{$u_{_\mathbb{C}} H H^{\dagger}{}  u_{_\mathbb{C}}^{\dagger}{}  D $}
		
		& C_{f_1,f_4} \delta _{a_1}^{a_4} \delta _{i_3}^{i_2} \langle 13\rangle [34] & \\
		
\end{array}\end{align}

\begin{align}\begin{array}{c|ll}
		
		\multirow{1}*{$u_{_\mathbb{C}} H^2 d_{_\mathbb{C}}^{\dagger}{}  D $}
		
		& C_{f_1,f_4} \delta _{a_1}^{a_4} \epsilon ^{i_2i_3} \langle 13\rangle [34] & \\
		
\end{array}\end{align}

\begin{align}\begin{array}{c|ll}
		
		\multirow{2}*{$Q H H^{\dagger}{}  Q^{\dagger}{}  D $}
		
		& C_{f_1,f_4} \delta _{a_4}^{a_1} \delta _{i_3}^{i_1} \delta _{i_4}^{i_2} \langle 13\rangle [34] & \\
		
		& C_{f_1,f_4} \delta _{a_4}^{a_1} \delta _{i_3}^{i_2} \delta _{i_4}^{i_1} \langle 13\rangle [34] & \\
		
\end{array}\end{align}

$D^2 \phi ^4$

\begin{align}\begin{array}{c|ll}
		
		\multirow{2}*{$H^2 H^{\dagger}{} ^2 D^2 $}
		
		&-  \delta _{i_3}^{i_1} \delta _{i_4}^{i_2} s_{34} & \\
		
		&   \delta _{i_3}^{i_1} \delta _{i_4}^{i_2} s_{24} & \\
		
\end{array}\end{align}

$\psi ^2 \phi ^3$

\begin{align}\begin{array}{c|ll}
		
		\multirow{1}*{$e_{_\mathbb{C}} L H H^{\dagger}{} ^2   $}
		
		& C_{f_1,f_2} \delta _{i_4}^{i_2} \delta _{i_5}^{i_3} \langle 12\rangle   & \\
		
\end{array}\end{align}

\begin{align}\begin{array}{c|ll}
		
		\multirow{1}*{$d_{_\mathbb{C}} Q H H^{\dagger}{} ^2   $}
		
		& C_{f_1,f_2} \delta _{a_1}^{a_2} \delta _{i_4}^{i_2} \delta _{i_5}^{i_3} \langle 12\rangle   & \\
		
\end{array}\end{align}

\begin{align}\begin{array}{c|ll}
		
		\multirow{1}*{$Q u_{_\mathbb{C}} H^2 H^{\dagger}{}    $}
		
		& C_{f_1,f_2} \delta _{a_2}^{a_1} \delta _{i_5}^{i_4} \epsilon ^{i_1i_3} \langle 12\rangle   & \\
		
\end{array}\end{align}

$\phi ^6$

\begin{align}\begin{array}{c|ll}
		
		\multirow{1}*{$H^3 H^{\dagger}{} ^3   $}
		
		&   \delta _{i_4}^{i_1} \delta _{i_5}^{i_2} \delta _{i_6}^{i_3}   & \\
		
\end{array}\end{align}

\subsection{Dimension-7 On-shell Amplitude Basis}

$D \psi ^3 \bar{\psi }$

\begin{align}\begin{array}{c|ll}
		
		\multirow{1}*{$d_{_\mathbb{C}} L^2 u_{_\mathbb{C}}^{\dagger}{}  D $}
		
		& C_{f_1,f_2f_3,f_4}^{[2]} \delta _{a_1}^{a_4} \epsilon ^{i_2i_3} \langle 13\rangle  \langle 23\rangle [34] & \\
		
\end{array}\end{align}

\begin{align}\begin{array}{c|ll}
		
		\multirow{1}*{$d_{_\mathbb{C}}^2 L Q^{\dagger}{}  D $}
		
		& C_{f_1f_2,f_3,f_4}^{[2]} \epsilon _{a_1a_2a_4} \delta _{i_4}^{i_3} \langle 13\rangle  \langle 23\rangle [34] & \\
		
\end{array}\end{align}

\begin{align}\begin{array}{c|ll}
		
		\multirow{1}*{$d_{_\mathbb{C}}^3 e_{_\mathbb{C}}^{\dagger}{}  D $}
		
		& C_{f_1f_2f_3,f_4}^{[3]} \epsilon _{a_1a_2a_3} \langle 13\rangle  \langle 23\rangle [34] & \\
		
\end{array}\end{align}

$D^2 \psi ^2 \phi ^2$

\begin{align}\begin{array}{c|ll}
		
		\multirow{2}*{$L^2 H^2 D^2 $}
		
		&-C_{f_1f_2}^{[2]} \epsilon ^{i_1i_3} \epsilon ^{i_2i_4} \langle 12\rangle s_{34} & \\
		
		& C_{f_1f_2}^{[2]} \epsilon ^{i_1i_3} \epsilon ^{i_2i_4} \langle 13\rangle  \langle 24\rangle [34] & \\
		
\end{array}\end{align}

$\psi ^4 \phi$

\begin{align}\begin{array}{c|ll}
		
		\multirow{3}*{$e_{_\mathbb{C}} L^3 H   $}
		
		& C_{f_1,f_2f_3f_4}^{[3]} \epsilon ^{i_2i_4} \epsilon ^{i_3i_5} \langle 12\rangle  \langle 34\rangle   & \\
		
		& C_{f_1,f_2f_3f_4}^{[2,1]} \epsilon ^{i_2i_4} \epsilon ^{i_3i_5} \langle 12\rangle  \langle 34\rangle   & \\
		
		& C_{f_1,f_2f_3f_4}^{[1,1,1]} \epsilon ^{i_2i_4} \epsilon ^{i_3i_5} \langle 12\rangle  \langle 34\rangle   & \\
		
\end{array}\end{align}

\begin{align}\begin{array}{c|ll}
		
		\multirow{1}*{$d_{_\mathbb{C}}^3 L H^{\dagger}{}    $}
		
		& C_{f_1f_2f_3,f_4}^{[2,1]} \epsilon _{a_1a_2a_3} \delta _{i_5}^{i_4} \langle 12\rangle  \langle 34\rangle   & \\
		
\end{array}\end{align}

\begin{align}\begin{array}{c|ll}
		
		\multirow{2}*{$d_{_\mathbb{C}}^2 L u_{_\mathbb{C}} H   $}
		
		& C_{f_1f_2,f_3,f_4}^{[2]} \epsilon _{a_1a_2a_4} \epsilon ^{i_3i_5} \langle 13\rangle  \langle 24\rangle   & \\
		
		& C_{f_1f_2,f_3,f_4}^{[1,1]} \epsilon _{a_1a_2a_4} \epsilon ^{i_3i_5} \langle 12\rangle  \langle 34\rangle   & \\
		
\end{array}\end{align}

\begin{align}\begin{array}{c|ll}
		
		\multirow{4}*{$d_{_\mathbb{C}} L^2 Q H   $}
		
		& C_{f_1,f_2f_3,f_4}^{[2]} \delta _{a_1}^{a_4} \epsilon ^{i_2i_4} \epsilon ^{i_3i_5} \langle 12\rangle  \langle 34\rangle   & \\
		
		& C_{f_1,f_2f_3,f_4}^{[2]} \delta _{a_1}^{a_4} \epsilon ^{i_2i_4} \epsilon ^{i_3i_5} \langle 13\rangle  \langle 24\rangle   & \\
		
		& C_{f_1,f_2f_3,f_4}^{[1,1]} \delta _{a_1}^{a_4} \epsilon ^{i_2i_4} \epsilon ^{i_3i_5} \langle 12\rangle  \langle 34\rangle   & \\
		
		& C_{f_1,f_2f_3,f_4}^{[1,1]} \delta _{a_1}^{a_4} \epsilon ^{i_2i_4} \epsilon ^{i_3i_5} \langle 13\rangle  \langle 24\rangle   & \\
		
\end{array}\end{align}

$F_{\rm{L}} \psi ^2 \phi ^2$

\begin{align}\begin{array}{c|ll}
		
		\multirow{1}*{$B_{\rm{L}} L^2 H^2   $}
		
		& C_{f_2f_3}^{[1,1]} \epsilon ^{i_2i_4} \epsilon ^{i_3i_5} \langle 12\rangle  \langle 13\rangle   & \\
		
\end{array}\end{align}

\begin{align}\begin{array}{c|ll}
		
		\multirow{2}*{$W_{\rm{L}} L^2 H^2   $}
		
		& C_{f_2f_3}^{[2]} \left(\tau ^{I_1})_m^{i_3}\right. \epsilon ^{i_2i_5} \epsilon ^{i_4m} \langle 12\rangle  \langle 13\rangle   & \\
		
		& C_{f_2f_3}^{[1,1]} \left(\tau ^{I_1})_m^{i_3}\right. \epsilon ^{i_2i_5} \epsilon ^{i_4m} \langle 12\rangle  \langle 13\rangle   & \\
		
\end{array}\end{align}

$\psi ^2 \phi  \bar{\psi }^2$

\begin{align}\begin{array}{c|ll}
		
		\multirow{2}*{$L^2 H Q^{\dagger}{}  u_{_\mathbb{C}}^{\dagger}{}    $}
		
		& C_{f_1f_2,f_4,f_5}^{[2]} \delta _{a_4}^{a_5} \delta _{i_4}^{i_1} \epsilon ^{i_3i_2} \langle 12\rangle [45] & \\
		
		& C_{f_1f_2,f_4,f_5}^{[1,1]} \delta _{a_4}^{a_5} \delta _{i_4}^{i_1} \epsilon ^{i_3i_2} \langle 12\rangle [45] & \\
		
\end{array}\end{align}

\begin{align}\begin{array}{c|ll}
		
		\multirow{1}*{$d_{_\mathbb{C}} L H e_{_\mathbb{C}}^{\dagger}{}  u_{_\mathbb{C}}^{\dagger}{}    $}
		
		& C_{f_1,f_2,f_4,f_5} \delta _{a_1}^{a_5} \epsilon ^{i_2i_3} \langle 12\rangle [45] & \\
		
\end{array}\end{align}

\begin{align}\begin{array}{c|ll}
		
		\multirow{2}*{$d_{_\mathbb{C}} L H Q^{\dagger}{} ^2   $}
		
		& C_{f_1,f_2,f_4f_5}^{[2]} \epsilon _{a_1a_4a_5} \delta _{i_4}^{i_2} \delta _{i_5}^{i_3} \langle 12\rangle [45] & \\
		
		& C_{f_1,f_2,f_4f_5}^{[1,1]} \epsilon _{a_1a_4a_5} \delta _{i_4}^{i_2} \delta _{i_5}^{i_3} \langle 12\rangle [45] & \\
		
\end{array}\end{align}

\begin{align}\begin{array}{c|ll}
		
		\multirow{1}*{$d_{_\mathbb{C}}^2 H e_{_\mathbb{C}}^{\dagger}{}  Q^{\dagger}{}    $}
		
		& C_{f_1f_2,f_4,f_5}^{[2]} \epsilon _{a_1a_2a_5} \delta _{i_5}^{i_3} \langle 12\rangle [45] & \\
		
\end{array}\end{align}

\begin{align}\begin{array}{c|ll}
		
		\multirow{1}*{$e_{_\mathbb{C}} u_{_\mathbb{C}} H^{\dagger}{}  d_{_\mathbb{C}}^{\dagger}{}  L^{\dagger}{}    $}
		
		& C_{f_1,f_2,f_4,f_5} \epsilon _{i_3i_5} \delta _{a_2}^{a_4} \langle 12\rangle [45] & \\
		
\end{array}\end{align}

\begin{align}\begin{array}{c|ll}
		
		\multirow{1}*{$e_{_\mathbb{C}} Q H^{\dagger}{}  d_{_\mathbb{C}}^{\dagger}{} ^2   $}
		
		& C_{f_1,f_2,f_4f_5}^{[2]} \delta _{i_3}^{i_2} \epsilon ^{a_2a_4a_5} \langle 12\rangle [45] & \\
		
\end{array}\end{align}

\begin{align}\begin{array}{c|ll}
		
		\multirow{2}*{$Q u_{_\mathbb{C}} H^{\dagger}{}  L^{\dagger}{} ^2   $}
		
		& C_{f_1,f_2,f_4f_5}^{[2]} \epsilon _{i_5i_3} \delta _{a_2}^{a_1} \delta _{i_4}^{i_1} \langle 12\rangle [45] & \\
		
		& C_{f_1,f_2,f_4f_5}^{[1,1]} \epsilon _{i_5i_3} \delta _{a_2}^{a_1} \delta _{i_4}^{i_1} \langle 12\rangle [45] & \\
		
\end{array}\end{align}

\begin{align}\begin{array}{c|ll}
		
		\multirow{2}*{$Q^2 H^{\dagger}{}  d_{_\mathbb{C}}^{\dagger}{}  L^{\dagger}{}    $}
		
		& C_{f_1f_2,f_4,f_5}^{[2]} \delta _{i_3}^{i_1} \delta _{i_5}^{i_2} \epsilon ^{a_1a_2a_4} \langle 12\rangle [45] & \\
		
		& C_{f_1f_2,f_4,f_5}^{[1,1]} \delta _{i_3}^{i_1} \delta _{i_5}^{i_2} \epsilon ^{a_1a_2a_4} \langle 12\rangle [45] & \\
		
\end{array}\end{align}

$D \psi  \phi ^3 \bar{\psi }$

\begin{align}\begin{array}{c|ll}
		
		\multirow{1}*{$e_{_\mathbb{C}} H^{\dagger}{} ^3 L^{\dagger}{}  D $}
		
		& C_{f_1,f_5} \epsilon _{i_2i_4} \epsilon _{i_3i_5} \langle 14\rangle [45] & \\
		
\end{array}\end{align}

\begin{align}\begin{array}{c|ll}
		
		\multirow{1}*{$L H^3 e_{_\mathbb{C}}^{\dagger}{}  D $}
		
		& C_{f_1,f_5} \epsilon ^{i_1i_3} \epsilon ^{i_2i_4} \langle 14\rangle [45] & \\
		
\end{array}\end{align}

$\psi ^2 \phi ^4$

\begin{align}\begin{array}{c|ll}
		
		\multirow{1}*{$L^2 H^3 H^{\dagger}{}    $}
		
		& C_{f_1f_2}^{[2]} \delta _{i_6}^{i_4} \epsilon ^{i_3i_1} \epsilon ^{i_5i_2} \langle 12\rangle   & \\
		
\end{array}\end{align}

\subsection{Dimension-8 On-shell Amplitude Basis involving only Bosons}

$F_{\rm{L}}^4$

\begin{align}\begin{array}{c|ll}
		
		\multirow{1}*{$B_{\rm{L}}^4   $}
		
		&   \langle 12\rangle ^2 \langle 34\rangle ^2   & \\
		
\end{array}\end{align}

\begin{align}\begin{array}{c|ll}
		
		\multirow{2}*{$B_{\rm{L}}^2 W_{\rm{L}}^2   $}
		
		&   \delta ^{I_3I_4} \langle 12\rangle ^2 \langle 34\rangle ^2   & \\
		
		&   \delta ^{I_3I_4} \langle 13\rangle ^2 \langle 24\rangle ^2   & \\
		
\end{array}\end{align}

\begin{align}\begin{array}{c|ll}
		
		\multirow{2}*{$W_{\rm{L}}^4   $}
		
		&   \delta ^{I_1I_3} \delta ^{I_2I_4} \langle 12\rangle ^2 \langle 34\rangle ^2   & \\
		
		&   \delta ^{I_1I_3} \delta ^{I_2I_4} \langle 13\rangle ^2 \langle 24\rangle ^2   & \\
		
\end{array}\end{align}

\begin{align}\begin{array}{c|ll}
		
		\multirow{2}*{$B_{\rm{L}}^2 G_{\rm{L}}^2   $}
		
		&   \delta ^{A_3A_4} \langle 12\rangle ^2 \langle 34\rangle ^2   & \\
		
		&   \delta ^{A_3A_4} \langle 13\rangle ^2 \langle 24\rangle ^2   & \\
		
\end{array}\end{align}

\begin{align}\begin{array}{c|ll}
		
		\multirow{2}*{$G_{\rm{L}}^2 W_{\rm{L}}^2   $}
		
		&   \delta ^{A_1A_2} \delta ^{I_3I_4} \langle 12\rangle ^2 \langle 34\rangle ^2   & \\
		
		&   \delta ^{A_1A_2} \delta ^{I_3I_4} \langle 13\rangle ^2 \langle 24\rangle ^2   & \\
		
\end{array}\end{align}

\begin{align}\begin{array}{c|ll}
		
		\multirow{1}*{$B_{\rm{L}} G_{\rm{L}}^3   $}
		
		&   d^{A_2A_3A_4} \langle 12\rangle ^2 \langle 34\rangle ^2   & \\
		
\end{array}\end{align}

\begin{align}\begin{array}{c|ll}
		
		\multirow{3}*{$G_{\rm{L}}^4   $}
		
		&   d^{EA_1A_2} d^{EA_3A_4} \langle 12\rangle ^2 \langle 34\rangle ^2   & \\
		
		&   d^{EA_1A_2} d^{EA_3A_4} \langle 13\rangle ^2 \langle 24\rangle ^2   & \\
		
		&   f^{EA_1A_2} f^{EA_3A_4} \langle 13\rangle ^2 \langle 24\rangle ^2   & \\
		
\end{array}\end{align}

$D^2 F_{\rm{L}}^2 \phi ^2$

\begin{align}\begin{array}{c|ll}
		
		\multirow{1}*{$B_{\rm{L}}^2 H H^{\dagger}{}    $}
		
		&-  \delta _{i_4}^{i_3} \langle 12\rangle ^2 s_{34} & \\
		
\end{array}\end{align}

\begin{align}\begin{array}{c|ll}
		
		\multirow{2}*{$B_{\rm{L}} W_{\rm{L}} H H^{\dagger}{}    $}
		
		&-  \left(\tau ^{I_2})_{i_4}^{i_3}\right. \langle 12\rangle ^2 s_{34} & \\
		
		&   \left(\tau ^{I_2})_{i_4}^{i_3}\right. \langle 12\rangle  \langle 13\rangle  \langle 24\rangle [34] & \\
		
\end{array}\end{align}

\begin{align}\begin{array}{c|ll}
		
		\multirow{2}*{$W_{\rm{L}}^2 H H^{\dagger}{}    $}
		
		&-  \delta _{i_4}^{i_3} \delta ^{I_1I_2} \langle 12\rangle ^2 s_{34} & \\
		
		&   \left(\tau ^K)_{i_4}^{i_3}\right. \epsilon ^{I_1I_2K} \langle 12\rangle  \langle 13\rangle  \langle 24\rangle [34] & \\
		
\end{array}\end{align}

\begin{align}\begin{array}{c|ll}
		
		\multirow{1}*{$G_{\rm{L}}^2 H H^{\dagger}{}    $}
		
		&-  \delta _{i_4}^{i_3} \delta ^{A_1A_2} \langle 12\rangle ^2 s_{34} & \\
		
\end{array}\end{align}

$F_{\rm{L}}^2 F_{\rm{R}}^2$

\begin{align}\begin{array}{c|ll}
		
		\multirow{1}*{$B_{\rm{L}}^2 B_{\rm{R}}^2   $}
		
		&   \langle 12\rangle ^2 [34]^2 & \\
		
\end{array}\end{align}

\begin{align}\begin{array}{c|ll}
		
		\multirow{1}*{$B_{\rm{L}}^2 W_{\rm{R}}^2   $}
		
		&   \delta ^{I_3I_4} \langle 12\rangle ^2 [34]^2 & \\
		
\end{array}\end{align}

\begin{align}\begin{array}{c|ll}
		
		\multirow{1}*{$B_{\rm{L}} W_{\rm{L}} B_{\rm{R}} W_{\rm{R}}   $}
		
		&   \delta ^{I_2I_4} \langle 12\rangle ^2 [34]^2 & \\
		
\end{array}\end{align}

\begin{align}\begin{array}{c|ll}
		
		\multirow{1}*{$W_{\rm{L}}^2 B_{\rm{R}}^2   $}
		
		&   \delta ^{I_1I_2} \langle 12\rangle ^2 [34]^2 & \\
		
\end{array}\end{align}

\begin{align}\begin{array}{c|ll}
		
		\multirow{1}*{$B_{\rm{L}}^2 G_{\rm{R}}^2   $}
		
		&   \delta ^{A_3A_4} \langle 12\rangle ^2 [34]^2 & \\
		
\end{array}\end{align}

\begin{align}\begin{array}{c|ll}
		
		\multirow{1}*{$B_{\rm{L}} G_{\rm{L}} B_{\rm{R}} G_{\rm{R}}   $}
		
		&   \delta ^{A_2A_4} \langle 12\rangle ^2 [34]^2 & \\
		
\end{array}\end{align}

\begin{align}\begin{array}{c|ll}
		
		\multirow{1}*{$G_{\rm{L}}^2 B_{\rm{R}}^2   $}
		
		&   \delta ^{A_1A_2} \langle 12\rangle ^2 [34]^2 & \\
		
\end{array}\end{align}

\begin{align}\begin{array}{c|ll}
		
		\multirow{1}*{$B_{\rm{L}} W_{\rm{L}} W_{\rm{R}}^2   $}
		
		&   \epsilon ^{I_2I_3I_4} \langle 12\rangle ^2 [34]^2 & \\
		
\end{array}\end{align}

\begin{align}\begin{array}{c|ll}
		
		\multirow{1}*{$W_{\rm{L}}^2 B_{\rm{R}} W_{\rm{R}}   $}
		
		&   \epsilon ^{I_1I_2I_4} \langle 12\rangle ^2 [34]^2 & \\
		
\end{array}\end{align}

\begin{align}\begin{array}{c|ll}
		
		\multirow{2}*{$W_{\rm{L}}^2 W_{\rm{R}}^2   $}
		
		&   \delta ^{I_1I_3} \delta ^{I_2I_4} \langle 12\rangle ^2 [34]^2 & \\
		
		&   \delta ^{I_1I_2} \delta ^{I_3I_4} \langle 12\rangle ^2 [34]^2 & \\
		
\end{array}\end{align}

\begin{align}\begin{array}{c|ll}
		
		\multirow{1}*{$W_{\rm{L}}^2 G_{\rm{R}}^2   $}
		
		&   \delta ^{A_3A_4} \delta ^{I_1I_2} \langle 12\rangle ^2 [34]^2 & \\
		
\end{array}\end{align}

\begin{align}\begin{array}{c|ll}
		
		\multirow{1}*{$G_{\rm{L}} W_{\rm{L}} G_{\rm{R}} W_{\rm{R}}   $}
		
		&   \delta ^{A_1A_3} \delta ^{I_2I_4} \langle 12\rangle ^2 [34]^2 & \\
		
\end{array}\end{align}

\begin{align}\begin{array}{c|ll}
		
		\multirow{1}*{$G_{\rm{L}}^2 W_{\rm{R}}^2   $}
		
		&   \delta ^{A_1A_2} \delta ^{I_3I_4} \langle 12\rangle ^2 [34]^2 & \\
		
\end{array}\end{align}

\begin{align}\begin{array}{c|ll}
		
		\multirow{1}*{$B_{\rm{L}} G_{\rm{L}} G_{\rm{R}}^2   $}
		
		&   d^{A_2A_3A_4} \langle 12\rangle ^2 [34]^2 & \\
		
\end{array}\end{align}

\begin{align}\begin{array}{c|ll}
		
		\multirow{1}*{$G_{\rm{L}}^2 B_{\rm{R}} G_{\rm{R}}   $}
		
		&   d^{A_1A_2A_4} \langle 12\rangle ^2 [34]^2 & \\
		
\end{array}\end{align}

\begin{align}\begin{array}{c|ll}
		
		\multirow{3}*{$G_{\rm{L}}^2 G_{\rm{R}}^2   $}
		
		&   d^{EA_1A_2} d^{EA_3A_4} \langle 12\rangle ^2 [34]^2 & \\
		
		&   \delta ^{A_1A_2} \delta ^{A_3A_4} \langle 12\rangle ^2 [34]^2 & \\
		
		&   \delta ^{A_1A_3} \delta ^{A_2A_4} \langle 12\rangle ^2 [34]^2 & \\
		
\end{array}\end{align}

$D^2 F_{\rm{L}} F_{\rm{R}} \phi ^2$

\begin{align}\begin{array}{c|ll}
		
		\multirow{1}*{$B_{\rm{L}} H H^{\dagger}{}  B_{\rm{R}}   $}
		
		&   \delta _{i_3}^{i_2} \langle 13\rangle ^2 [34]^2 & \\
		
\end{array}\end{align}

\begin{align}\begin{array}{c|ll}
		
		\multirow{1}*{$B_{\rm{L}} H H^{\dagger}{}  W_{\rm{R}}   $}
		
		&   \left(\tau ^{I_4})_{i_3}^{i_2}\right. \langle 13\rangle ^2 [34]^2 & \\
		
\end{array}\end{align}

\begin{align}\begin{array}{c|ll}
		
		\multirow{1}*{$W_{\rm{L}} H H^{\dagger}{}  B_{\rm{R}}   $}
		
		&   \left(\tau ^{I_1})_{i_3}^{i_2}\right. \langle 13\rangle ^2 [34]^2 & \\
		
\end{array}\end{align}

\begin{align}\begin{array}{c|ll}
		
		\multirow{2}*{$W_{\rm{L}} H H^{\dagger}{}  W_{\rm{R}}   $}
		
		&   \left(\tau ^K)_{i_3}^{i_2}\right. \epsilon ^{I_1I_4K} \langle 13\rangle ^2 [34]^2 & \\
		
		&   \delta _{i_3}^{i_2} \delta ^{I_4I_1} \langle 13\rangle ^2 [34]^2 & \\
		
\end{array}\end{align}

\begin{align}\begin{array}{c|ll}
		
		\multirow{1}*{$G_{\rm{L}} H H^{\dagger}{}  G_{\rm{R}}   $}
		
		&   \delta _{i_3}^{i_2} \delta ^{A_1A_4} \langle 13\rangle ^2 [34]^2 & \\
		
\end{array}\end{align}

$D^4 \phi ^4$

\begin{align}\begin{array}{c|ll}
		
		\multirow{3}*{$H^2 H^{\dagger}{} ^2   $}
		
		&   \delta _{i_3}^{i_1} \delta _{i_4}^{i_2} s_{34}^2 & \\
		
		&   \delta _{i_3}^{i_1} \delta _{i_4}^{i_2} s_{24}^2 & \\
		
		&-  \delta _{i_3}^{i_1} \delta _{i_4}^{i_2} s_{24} s_{34} & \\
		
\end{array}\end{align}

$F_{\rm{L}}^3 \phi ^2$

\begin{align}\begin{array}{c|ll}
		
		\multirow{1}*{$B_{\rm{L}}^3 H H^{\dagger}{}    $}
		
		&   \delta _{i_5}^{i_4} \langle 12\rangle  \langle 13\rangle  \langle 23\rangle   & \\
		
\end{array}\end{align}

\begin{align}\begin{array}{c|ll}
		
		\multirow{1}*{$B_{\rm{L}}^2 W_{\rm{L}} H H^{\dagger}{}    $}
		
		&   \left(\tau ^{I_3})_{i_5}^{i_4}\right. \langle 12\rangle  \langle 13\rangle  \langle 23\rangle   & \\
		
\end{array}\end{align}

\begin{align}\begin{array}{c|ll}
		
		\multirow{1}*{$B_{\rm{L}} W_{\rm{L}}^2 H H^{\dagger}{}    $}
		
		&   \left(\tau ^K)_{i_5}^{i_4}\right. \epsilon ^{I_2I_3K} \langle 12\rangle  \langle 13\rangle  \langle 23\rangle   & \\
		
\end{array}\end{align}

\begin{align}\begin{array}{c|ll}
		
		\multirow{1}*{$W_{\rm{L}}^3 H H^{\dagger}{}    $}
		
		&   \delta _{i_5}^{i_4} \delta ^{I_2L} \epsilon ^{I_3I_1L} \langle 12\rangle  \langle 13\rangle  \langle 23\rangle   & \\
		
\end{array}\end{align}

\begin{align}\begin{array}{c|ll}
		
		\multirow{1}*{$B_{\rm{L}} G_{\rm{L}}^2 H H^{\dagger}{}    $}
		
		&   \delta _{i_5}^{i_4} \delta ^{A_2A_3} \langle 12\rangle  \langle 13\rangle  \langle 23\rangle   & \\
		
\end{array}\end{align}

\begin{align}\begin{array}{c|ll}
		
		\multirow{1}*{$G_{\rm{L}}^2 W_{\rm{L}} H H^{\dagger}{}    $}
		
		&   \delta ^{A_1A_2} \left(\tau ^{I_3})_{i_5}^{i_4}\right. \langle 12\rangle  \langle 13\rangle  \langle 23\rangle   & \\
		
\end{array}\end{align}

\begin{align}\begin{array}{c|ll}
		
		\multirow{1}*{$G_{\rm{L}}^3 H H^{\dagger}{}    $}
		
		&   \delta _{i_5}^{i_4} f^{A_1A_2A_3} \langle 12\rangle  \langle 13\rangle  \langle 23\rangle   & \\
		
\end{array}\end{align}

$D^2 F_{\rm{L}} \phi ^4$

\begin{align}\begin{array}{c|ll}
		
		\multirow{1}*{$B_{\rm{L}} H^2 H^{\dagger}{} ^2   $}
		
		&-  \delta _{i_4}^{i_2} \delta _{i_5}^{i_3} \langle 13\rangle  \langle 15\rangle [35] & \\
		
\end{array}\end{align}

\begin{align}\begin{array}{c|ll}
		
		\multirow{2}*{$W_{\rm{L}} H^2 H^{\dagger}{} ^2   $}
		
		&   \delta _{i_5}^{i_2} \left(\tau ^{I_1})_{i_4}^{i_3}\right. \langle 14\rangle  \langle 15\rangle [45] & \\
		
		&-  \delta _{i_5}^{i_2} \left(\tau ^{I_1})_{i_4}^{i_3}\right. \langle 13\rangle  \langle 15\rangle [35] & \\
		
\end{array}\end{align}

$F_{\rm{L}}^2 \phi ^4$

\begin{align}\begin{array}{c|ll}
		
		\multirow{1}*{$B_{\rm{L}}^2 H^2 H^{\dagger}{} ^2   $}
		
		&   \delta _{i_5}^{i_3} \delta _{i_6}^{i_4} \langle 12\rangle ^2   & \\
		
\end{array}\end{align}

\begin{align}\begin{array}{c|ll}
		
		\multirow{1}*{$B_{\rm{L}} W_{\rm{L}} H^2 H^{\dagger}{} ^2   $}
		
		&   \delta _{i_6}^{i_3} \left(\tau ^{I_2})_{i_5}^{i_4}\right. \langle 12\rangle ^2   & \\
		
\end{array}\end{align}

\begin{align}\begin{array}{c|ll}
		
		\multirow{2}*{$W_{\rm{L}}^2 H^2 H^{\dagger}{} ^2   $}
		
		&   \left(\tau ^{I_1})_{i_5}^{i_3}\right. \left(\tau ^{I_2})_{i_6}^{i_4}\right. \langle 12\rangle ^2   & \\
		
		&   \delta _{i_5}^{i_4} \delta _{i_6}^{i_3} \delta ^{I_1I_2} \langle 12\rangle ^2   & \\
		
\end{array}\end{align}

\begin{align}\begin{array}{c|ll}
		
		\multirow{1}*{$G_{\rm{L}}^2 H^2 H^{\dagger}{} ^2   $}
		
		&   \delta _{i_5}^{i_3} \delta _{i_6}^{i_4} \delta ^{A_1A_2} \langle 12\rangle ^2   & \\
		
\end{array}\end{align}

$D^2 \phi ^6$

\begin{align}\begin{array}{c|ll}
		
		\multirow{2}*{$H^3 H^{\dagger}{} ^3   $}
		
		&-  \delta _{i_4}^{i_1} \delta _{i_5}^{i_2} \delta _{i_6}^{i_3} s_{56} & \\
		
		&-  \delta _{i_4}^{i_1} \delta _{i_5}^{i_2} \delta _{i_6}^{i_3} s_{36} & \\
		
\end{array}\end{align}

$\phi ^8$

\begin{align}\begin{array}{c|ll}
		
		\multirow{1}*{$H^4 H^{\dagger}{} ^4   $}
		
		&   \delta _{i_5}^{i_1} \delta _{i_6}^{i_2} \delta _{i_7}^{i_3} \delta _{i_8}^{i_4}   & \\
		
\end{array}\end{align}

%\section{References}
\bibliographystyle{JHEP}
\bibliography{ref}

\providecommand{\href}[2]{#2}\begingroup\raggedright\begin{thebibliography}{10}

\bibitem{Weinberg:1979sa}
S.~Weinberg, {\it {Baryon and Lepton Nonconserving Processes}},  {\em Phys.
  Rev. Lett.} {\bf 43} (1979) 1566--1570.

\bibitem{Buchmuller:1985jz}
W.~Buchmuller and D.~Wyler, {\it {Effective Lagrangian Analysis of New
  Interactions and Flavor Conservation}},  {\em Nucl. Phys. B} {\bf 268} (1986)
  621--653.

\bibitem{Grzadkowski:2010es}
B.~Grzadkowski, M.~Iskrzynski, M.~Misiak, and J.~Rosiek, {\it {Dimension-Six
  Terms in the Standard Model Lagrangian}},  {\em JHEP} {\bf 10} (2010) 085,
  [\href{http://arxiv.org/abs/1008.4884}{{\tt arXiv:1008.4884}}].

\bibitem{Lehman:2014jma}
L.~Lehman, {\it {Extending the Standard Model Effective Field Theory with the
  Complete Set of Dimension-7 Operators}},  {\em Phys. Rev. D} {\bf 90} (2014),
  no.~12 125023, [\href{http://arxiv.org/abs/1410.4193}{{\tt
  arXiv:1410.4193}}].

\bibitem{Li:2020gnx}
H.-L. Li, Z.~Ren, J.~Shu, M.-L. Xiao, J.-H. Yu, and Y.-H. Zheng, {\it {Complete
  set of dimension-eight operators in the standard model effective field
  theory}},  {\em Phys. Rev. D} {\bf 104} (2021), no.~1 015026,
  [\href{http://arxiv.org/abs/2005.00008}{{\tt arXiv:2005.00008}}].

\bibitem{Murphy:2020rsh}
C.~W. Murphy, {\it {Dimension-8 operators in the Standard Model Eective Field
  Theory}},  {\em JHEP} {\bf 10} (2020) 174,
  [\href{http://arxiv.org/abs/2005.00059}{{\tt arXiv:2005.00059}}].

\bibitem{Li:2020xlh}
H.-L. Li, Z.~Ren, M.-L. Xiao, J.-H. Yu, and Y.-H. Zheng, {\it {Complete set of
  dimension-nine operators in the standard model effective field theory}},
  {\em Phys. Rev. D} {\bf 104} (2021), no.~1 015025,
  [\href{http://arxiv.org/abs/2007.07899}{{\tt arXiv:2007.07899}}].

\bibitem{Liao:2020jmn}
Y.~Liao and X.-D. Ma, {\it {An explicit construction of the dimension-9
  operator basis in the standard model effective field theory}},  {\em JHEP}
  {\bf 11} (2020) 152, [\href{http://arxiv.org/abs/2007.08125}{{\tt
  arXiv:2007.08125}}].

\bibitem{Liao:2016hru}
Y.~Liao and X.-D. Ma, {\it {Renormalization Group Evolution of Dimension-seven
  Baryon- and Lepton-number-violating Operators}},  {\em JHEP} {\bf 11} (2016)
  043, [\href{http://arxiv.org/abs/1607.07309}{{\tt arXiv:1607.07309}}].

\bibitem{Jenkins:2017jig}
E.~E. Jenkins, A.~V. Manohar, and P.~Stoffer, {\it {Low-Energy Effective Field
  Theory below the Electroweak Scale: Operators and Matching}},  {\em JHEP}
  {\bf 03} (2018) 016, [\href{http://arxiv.org/abs/1709.04486}{{\tt
  arXiv:1709.04486}}].

\bibitem{Liao:2020zyx}
Y.~Liao, X.-D. Ma, and Q.-Y. Wang, {\it {Extending low energy effective field
  theory with a complete set of dimension-7 operators}},  {\em JHEP} {\bf 08}
  (2020) 162, [\href{http://arxiv.org/abs/2005.08013}{{\tt arXiv:2005.08013}}].

\bibitem{Li:2020tsi}
H.-L. Li, Z.~Ren, M.-L. Xiao, J.-H. Yu, and Y.-H. Zheng, {\it {Low energy
  effective field theory operator basis at d \ensuremath{\leq} 9}},  {\em JHEP}
  {\bf 06} (2021) 138, [\href{http://arxiv.org/abs/2012.09188}{{\tt
  arXiv:2012.09188}}].

\bibitem{Murphy:2020cly}
C.~W. Murphy, {\it {Low-Energy Effective Field Theory below the Electroweak
  Scale: Dimension-8 Operators}},  {\em JHEP} {\bf 04} (2021) 101,
  [\href{http://arxiv.org/abs/2012.13291}{{\tt arXiv:2012.13291}}].

\bibitem{delAguila:2008ir}
F.~del Aguila, S.~Bar-Shalom, A.~Soni, and J.~Wudka, {\it {Heavy Majorana
  Neutrinos in the Effective Lagrangian Description: Application to Hadron
  Colliders}},  {\em Phys. Lett. B} {\bf 670} (2009) 399--402,
  [\href{http://arxiv.org/abs/0806.0876}{{\tt arXiv:0806.0876}}].

\bibitem{Aparici:2009fh}
A.~Aparici, K.~Kim, A.~Santamaria, and J.~Wudka, {\it {Right-handed neutrino
  magnetic moments}},  {\em Phys. Rev. D} {\bf 80} (2009) 013010,
  [\href{http://arxiv.org/abs/0904.3244}{{\tt arXiv:0904.3244}}].

\bibitem{Bhattacharya:2015vja}
S.~Bhattacharya and J.~Wudka, {\it {Dimension-seven operators in the standard
  model with right handed neutrinos}},  {\em Phys. Rev. D} {\bf 94} (2016),
  no.~5 055022, [\href{http://arxiv.org/abs/1505.05264}{{\tt
  arXiv:1505.05264}}]. [Erratum: Phys.Rev.D 95, 039904 (2017)].

\bibitem{Liao:2016qyd}
Y.~Liao and X.-D. Ma, {\it {Operators up to Dimension Seven in Standard Model
  Effective Field Theory Extended with Sterile Neutrinos}},  {\em Phys. Rev. D}
  {\bf 96} (2017), no.~1 015012, [\href{http://arxiv.org/abs/1612.04527}{{\tt
  arXiv:1612.04527}}].

\bibitem{Li:2021tsq}
H.-L. Li, Z.~Ren, M.-L. Xiao, J.-H. Yu, and Y.-H. Zheng, {\it {Operator bases
  in effective field theories with sterile neutrinos: d \ensuremath{\leq} 9}},
  {\em JHEP} {\bf 11} (2021) 003, [\href{http://arxiv.org/abs/2105.09329}{{\tt
  arXiv:2105.09329}}].

\bibitem{Chala:2020vqp}
M.~Chala and A.~Titov, {\it {One-loop matching in the SMEFT extended with a
  sterile neutrino}},  {\em JHEP} {\bf 05} (2020) 139,
  [\href{http://arxiv.org/abs/2001.07732}{{\tt arXiv:2001.07732}}].

\bibitem{Li:2020lba}
T.~Li, X.-D. Ma, and M.~A. Schmidt, {\it {General neutrino interactions with
  sterile neutrinos in light of coherent neutrino-nucleus scattering and meson
  invisible decays}},  {\em JHEP} {\bf 07} (2020) 152,
  [\href{http://arxiv.org/abs/2005.01543}{{\tt arXiv:2005.01543}}].

\bibitem{Shadmi:2018xan}
Y.~Shadmi and Y.~Weiss, {\it {Effective Field Theory Amplitudes the On-Shell
  Way: Scalar and Vector Couplings to Gluons}},  {\em JHEP} {\bf 02} (2019)
  165, [\href{http://arxiv.org/abs/1809.09644}{{\tt arXiv:1809.09644}}].

\bibitem{Ma:2019gtx}
T.~Ma, J.~Shu, and M.-L. Xiao, {\it {Standard Model Effective Field Theory from
  On-shell Amplitudes}},  \href{http://arxiv.org/abs/1902.06752}{{\tt
  arXiv:1902.06752}}.

\bibitem{Durieux:2019siw}
G.~Durieux and C.~S. Machado, {\it {Enumerating higher-dimensional operators
  with on-shell amplitudes}},  {\em Phys. Rev. D} {\bf 101} (2020), no.~9
  095021, [\href{http://arxiv.org/abs/1912.08827}{{\tt arXiv:1912.08827}}].

\bibitem{AccettulliHuber:2021uoa}
M.~Accettulli~Huber and S.~De~Angelis, {\it {Standard Model EFTs via on-shell
  methods}},  {\em JHEP} {\bf 11} (2021) 221,
  [\href{http://arxiv.org/abs/2108.03669}{{\tt arXiv:2108.03669}}].

\bibitem{Balkin:2021dko}
R.~Balkin, G.~Durieux, T.~Kitahara, Y.~Shadmi, and Y.~Weiss, {\it {On-shell
  Higgsing for EFTs}},  \href{http://arxiv.org/abs/2112.09688}{{\tt
  arXiv:2112.09688}}.

\bibitem{Durieux:2019eor}
G.~Durieux, T.~Kitahara, Y.~Shadmi, and Y.~Weiss, {\it {The electroweak
  effective field theory from on-shell amplitudes}},  {\em JHEP} {\bf 01}
  (2020) 119, [\href{http://arxiv.org/abs/1909.10551}{{\tt arXiv:1909.10551}}].

\bibitem{Durieux:2020gip}
G.~Durieux, T.~Kitahara, C.~S. Machado, Y.~Shadmi, and Y.~Weiss, {\it
  {Constructing massive on-shell contact terms}},  {\em JHEP} {\bf 12} (2020)
  175, [\href{http://arxiv.org/abs/2008.09652}{{\tt arXiv:2008.09652}}].

\bibitem{Dong:2021yak}
Z.-Y. Dong, T.~Ma, and J.~Shu, {\it {Constructing on-shell operator basis for
  all masses and spins}},  \href{http://arxiv.org/abs/2103.15837}{{\tt
  arXiv:2103.15837}}.

\bibitem{Li:2020zfq}
H.-L. Li, J.~Shu, M.-L. Xiao, and J.-H. Yu, {\it {Depicting the Landscape of
  Generic Effective Field Theories}},
  \href{http://arxiv.org/abs/2012.11615}{{\tt arXiv:2012.11615}}.

\bibitem{Henning:2019enq}
B.~Henning and T.~Melia, {\it {Constructing effective field theories via their
  harmonics}},  {\em Phys. Rev. D} {\bf 100} (2019), no.~1 016015,
  [\href{http://arxiv.org/abs/1902.06754}{{\tt arXiv:1902.06754}}].

\bibitem{Jiang:2020rwz}
M.~Jiang, J.~Shu, M.-L. Xiao, and Y.-H. Zheng, {\it {Partial Wave Amplitude
  Basis and Selection Rules in Effective Field Theories}},  {\em Phys. Rev.
  Lett.} {\bf 126} (2021), no.~1 011601,
  [\href{http://arxiv.org/abs/2001.04481}{{\tt arXiv:2001.04481}}].

\bibitem{ma2007group}
Z.~Ma, {\em Group Theory for Physicists}.
\newblock World Scientific, 2007.

\bibitem{Hagiwara:1993ck}
K.~Hagiwara, S.~Ishihara, R.~Szalapski, and D.~Zeppenfeld, {\it {Low-energy
  effects of new interactions in the electroweak boson sector}},  {\em Phys.
  Rev. D} {\bf 48} (1993) 2182--2203.

\bibitem{Giudice:2007fh}
G.~F. Giudice, C.~Grojean, A.~Pomarol, and R.~Rattazzi, {\it {The
  Strongly-Interacting Light Higgs}},  {\em JHEP} {\bf 06} (2007) 045,
  [\href{http://arxiv.org/abs/hep-ph/0703164}{{\tt hep-ph/0703164}}].

\bibitem{Cheung:2015aba}
C.~Cheung and C.-H. Shen, {\it {Nonrenormalization Theorems without
  Supersymmetry}},  {\em Phys. Rev. Lett.} {\bf 115} (2015), no.~7 071601,
  [\href{http://arxiv.org/abs/1505.01844}{{\tt arXiv:1505.01844}}].

\end{thebibliography}\endgroup

\end{document}